\documentclass[draft,tightenlines,nofootinbib,preprint,aps,eqsecnum,amsmath,amssymb]{revtex4}

\newcommand{\beq}{\begin{equation}}
\newcommand{\eeq}{\end{equation}}
\newcommand{\bea}{\begin{eqnarray}}
\newcommand{\eea}{\end{eqnarray}}

\newcommand{\sgn}{\epsilon}
\newcommand{\eo}{{}^4{\buildrel \circ \over E}}

\begin{document}

\title{The York Map as a Shanmugadhasan Canonical Transformation in Tetrad
Gravity and the Role of Non-Inertial Frames in the Geometrical View
of the Gravitational Field.}

\medskip

\author{David Alba}

\affiliation{Dipartimento di Fisica\\
Universita' di Firenze\\Polo Scientifico, via Sansone 1\\
 50019 Sesto Fiorentino, Italy\\
 E-mail ALBA@FI.INFN.IT}

\author{Luca Lusanna}

\affiliation{ Sezione INFN di Firenze\\ Polo Scientifico\\ Via Sansone 1\\
50019 Sesto Fiorentino (FI), Italy\\ Phone: 0039-055-4572334\\
FAX: 0039-055-4572364\\ E-mail: lusanna@fi.infn.it}

\begin{abstract}

A new parametrization of the 3-metric allows to  find explicitly a
York map by means of a partial Shanmugadhasan canonical
transformation in canonical ADM tetrad gravity. This allows to
identify the two pairs of physical tidal degrees of freedom (the
Dirac observables of the gravitational field have to be built in
term of them) and 14  gauge variables. These gauge quantities, whose
role in describing generalized inertial effects is clarified, are
all configurational except one, the York time, i.e. the trace
${}^3K(\tau ,\vec \sigma )$ of the extrinsic curvature of the
instantaneous 3-spaces $\Sigma_{\tau}$ (corresponding to a clock
synchronization convention) of a non-inertial frame centered on an
arbitrary observer. In $\Sigma_{\tau}$ the Dirac Hamiltonian is the
sum of the weak ADM energy $E_{ADM} = \int d^3\sigma\, {\cal
E}_{ADM}(\tau ,\vec \sigma )$ (whose density ${\cal E}_{ADM}(\tau
,\vec \sigma )$ is coordinate-dependent, containing the  inertial
potentials) and of the first-class constraints.

The main results of the paper, deriving from a coherent use of
constraint theory, are:

i) The explicit form of the Hamilton equations for the two tidal
degrees of freedom of the gravitational field in an arbitrary gauge:
a deterministic evolution can be defined only in a completely fixed
gauge, i.e. in a non-inertial frame with its pattern of inertial
forces. The simplest such gauge is the 3-orthogonal one, but other
gauges are discussed and the Hamiltonian interpretation of the
harmonic gauges is given. This frame-dependence derives from the
geometrical view of the gravitational field and is lost when the
theory is reduced to a linear spin 2 field on a background
space-time.

ii) A general solution of the super-momentum constraints, which
shows the existence of a generalized Gribov ambiguity associated to
the 3-diffeomorphism gauge group. It influences: a) the explicit
form of the solution of the super-momentum constraint and then of
the Dirac Hamiltonian; b) the determination of the shift functions
and then of the lapse one.

iii) The dependence of the Hamilton equations for the two pairs of
dynamical gravitational degrees of freedom (the generalized tidal
effects) and for the matter, written in a completely fixed
3-orthogonal Schwinger time gauge, upon the gauge variable
${}^3K(\tau ,\vec \sigma )$, determining the convention of clock
synchronization. The associated relativistic inertial effects,
absent in Newtonian gravity and implying inertial forces changing
from attractive to repulsive in regions with different sign of
${}^3K(\tau ,\vec \sigma )$, are completely unexplored and may have
astrophysical relevance in the interpretation of the dark side of
the universe.

\end{abstract}

\maketitle

\vfill\eject

\section{Introduction}

In a series of papers a new formulation of canonical metric \cite{1}
and tetrad \cite{2,3} gravity both based on the ADM action
\footnote{Tetrad gravity is more natural for the coupling to the
fermions. This leads to an interpretation of gravity based on a
congruence of time-like observers endowed with orthonormal tetrads:
in each point of space-time the time-like axis is the  unit
4-velocity of the observer, while the spatial axes are a (gauge)
convention for observer's gyroscopes. Tetrad gravity has 10 primary
first class constraints and 4 secondary first class ones. Six of the
primary constraints describe the extra freedom in the choice of the
tetrads. The other 4 primary (the vanishing of the momenta of the
lapse and shift functions) and 4 secondary (the super-Hamiltonian
and super-momentum constraints) constraints are the same as in
metric gravity. In Ref.\cite{3} 13 of the 14 constraints were
solved: the super-Hamiltonian one can be solved only after
linearization.} was given with the aim \cite{4} to identify the
Dirac observables of the gravitational field (the generalized {\it
tidal} effects) after having separated them from the gauge variables
(the generalized {\it inertial} effects) by using the Shanmugadhasan
canonical transformation \cite{5} adapted to the first class
constraints of the theory.

\bigskip

The formulation was given in a family of {\it non-compact}
space-times $M^4$ with the following properties:\hfill\break
 i) {\it globally hyperbolic} and {\it topologically trivial},
so that they can be foliated with space-like hyper-surfaces
$\Sigma_{\tau}$ diffeomorphic to $R^3$ (3+1 splitting of space-time
with $\tau$, the scalar parameter labeling the leaves, as a {\it
mathematical time});\hfill\break
 ii) {\it asymptotically flat at spatial infinity} and with
boundary conditions at spatial infinity independent from the
direction, so that the {\it spi group} of asymptotic symmetries is
reduced to the Poincare' group with the ADM Poincare' charges as
generators. In this way we can eliminate the {\it
super-translations}, namely the obstruction to define angular
momentum in general relativity, and we have the same type of
boundary conditions which are needed to get well defined non-Abelian
charges in Yang-Mills theory, opening the possibility of a unified
description of the four interactions with all the fields belonging
to same function space \cite{1,6}. All these requirements imply that
the {\it admissible foliations} of space-time must have the
space-like hyper-surfaces tending in a direction-independent way to
Minkowski space-like hyper-planes at spatial infinity, which
moreover must be orthogonal there to the ADM 4-momentum. Therefore,
$M^4$ is {\it asymptotically Minkowskian} \cite{8}. Moreover the
simultaneity 3-surfaces must admit an involution (Lichnerowicz
3-manifolds \cite{9}) allowing the definition of a generalized
Fourier transform with its associated concepts of positive and
negative energy, so to avoid the claimed impossibility to define
particles in curved space-times. \hfill\break
 iv) All the fields have to belong to suitable {\it weighted
Sobolev spaces} so that; i) the admissible space-like hyper-surfaces
are Riemannian 3-manifolds without asymptotically vanishing Killing
vectors \cite{8,10} (we furthermore assume the absence of any
Killing vector); ii) the inclusion of particle physics leads to a
formulation without Gribov ambiguity \cite{11,12}.
\medskip

In absence of matter the class of Christodoulou-Klainermann
space-times \cite{7}, admitting asymptotic ADM Poincare' charges and
an asymptotic flat metric is selected.

\bigskip

This formulation, the {\it rest-frame instant form of metric and
tetrad gravity}, emphasizes the role of {\it non-inertial frames}
(the only ones existing in general relativity due to the global
interpretation of the {\it equivalence principle}; see Ref.\cite{4}
for this viewpoint) and {\it deparametrizes} to the rest-frame
instant form of dynamics in Minkowski space-time \cite{4,6} when
matter is present if the Newton constant is switched off. The
non-inertial frames are the 3+1 splittings admissible for the given
space-time, after having chosen an arbitrary time-like observer as
the origin of the 3-coordinates on the leaves $\Sigma_{\tau}$, which
are both \cite{4} Cauchy surfaces and instantaneous 3-spaces
corresponding to a convention for the synchronization of distant
clocks \footnote{See Ref.\cite{13} for the special relativistic case
and Ref.\cite{14} for the quantization of particles in non-inertial
frames.}. As a consequence the 3+1 splitting identifies a global
non-inertial frame centered on the observer, namely a possible
extended physical laboratory with its metrological
conventions.\medskip

As shown in Refs.\cite{1,3} in this way one gets the {\it rest-frame
instant form of metric and tetrad gravity} with the weak ADM energy
$E_{ADM} = \int d^3\sigma \, {\cal E}_{ADM}(\tau ,\vec \sigma )$ as
the effective Hamiltonian (in accord with Refs.\cite{8,15})
\footnote{Therefore the formulations with a frozen reduced phase
space are avoided \cite{4}. The super-hamiltonian constraint
generates {\it normal} deformations of the space-like
hyper-surfaces, which are {\it not} interpreted as a time evolution
(like in the Wheeler-DeWitt approach) but as the Hamiltonian gauge
transformations ensuring that the description of gravity is
independent from the 3+1 splittings of space-time (i.e. from the
clock synchronization convention) like it happens in parametrized
Minkowski theories.}. The $\Gamma$-$\Gamma$ term in the ADM energy
density ${\cal E}_{ADM}(\tau ,\vec \sigma )$ is coordinate-dependent
(the problem of energy in general relativity) because it contains
the {\it inertial potentials} giving rise to the generalized
inertial effects in the non-inertial frame associated to the chosen
3+1 splitting of the space-time.
\medskip

In Ref.\cite{16} there is the study of the Hamiltonian linearization
of tetrad gravity without matter in these space-times, where the
existence of an asymptotic flat metric at spatial infinity ({\it
asymptotic background} with the presence of asymptotic inertial
observers to be identified with the fixed stars) allows to avoid the
splitting of the 4-metric into the flat one plus a perturbation. In
this way we have obtained {\it post-Minkowskian background-
independent gravitational waves} in a special non-harmonic
3-orthogonal gauge where the 3-metric is diagonal. As a consequence
these space-times, after the inclusion of matter, are candidates for
a general relativistic model of the solar system or of the galaxy.
Maybe they can also be used in the cosmological context if the
asymptotic inertial observers are identified with the preferred
observers of the cosmic background radiation.

\medskip

In Refs.\cite{17,18} there is the description of relativistic fluids
and of the Klein-Gordon field in the framework of parametrized
Minkowski theories. This formalism allows to get the Lagrangian of
these matter systems in the  formulation of tetrad gravity of
Refs.\cite{2,3,16}. The resulting first-class constraints depend
only on the mass density ${\cal M}(\tau ,\vec \sigma )$ (which is
metric-dependent) and the mass-current density ${\cal M}_r(\tau
,\vec \sigma )$ (which is metric-independent) of the matter. For
Dirac fields the situation is more complicated due to the presence
of second class constraints (see Ref.\cite{19} for the case of
parametrized Minkowski theories with fermions). It turns out that
the point Shanmugadhasan  canonical transformation of Ref.\cite{16},
adapted to 13 of the 14 first class constraints is not suited for
the inclusion of matter due to its {\it non-locality}. Therefore in
this paper we will look for a local point Shanmugadhasan
transformation adapted only to 10 of the 14 constraints.\bigskip

The new insight comes from the so-called York - Lichnerowicz
conformal approach \cite{20,21,22} (see also the book \cite{23} for
a review and more bibliographical information) to metric gravity in
globally hyperbolic ({\it but spatially compact} \footnote{This is
due to the influence of Mach principle, see for instance Chapter 5
of Ref.\cite{23}. However, let us remark that the non-locality of
the Dirac observables of the non-compact case (all the instantaneous
3-space is needed for their determination) has a Machian flavor.})
space-times. The starting point is the decomposition ${}^3g_{ij} =
\phi^4\, {}^3{\hat g}_{ij}$ of the 3-metric on an instantaneous
3-space $\Sigma_o$ of a 3+1 splitting of space-time in the product
of a {\it conformal factor} $\phi = (det\, {}^3g)^{1/12}$ and a {\it
conformal 3-metric} ${}^3{\hat g}_{ij}$ with $det\, {}^3{\hat
g}_{ij} = 1$ (${}^3{\hat g}_{ij}$ contains 5 of the 6 degrees of
freedom of ${}^3g_{ij}$). The extrinsic curvature 3-tensor
${}^3K_{ij}$ of $\Sigma_o$ (determining the ADM momentum) is
decomposed in its trace ${}^3K$ (the {\it York time}) plus the {\it
distorsion tensor}, which is the sum of a TT \footnote{Traceless and
transverse with respect to the conformal 3-metric.} symmetric
2-tensor ${}^3A_{ij}$ (2 degrees of freedom) plus the 3-tensor
${}^3W_{i;j} + {}^3W_{j;i} - {2\over 3}\, {}^3g_{ij}\,
{}^3W^k{}_{;k}$ depending on a covariant 3-vector ${}^3W_i$ ({\it
York gravitomagnetic vector potential}; 3 degrees of freedom).
Having fixed the lapse and shift functions of the 3+1 splitting and
having put ${}^3K = const.$, one assigns ${}^3{\hat g}_{ij}$ and
${}^3A_{ij}$ on the Cauchy surface $\Sigma_o$. Then, ${}^3W_i$ is
determined by the super-momentum constraints on $\Sigma_o$ and
$\phi$ is determined by the super-Hamiltonian constraint on
$\Sigma_o$. Then, the remaining Einstein's equations (see
Refs.\cite{7,10,20} for the existence and unicity of solutions)
determine the time derivatives of ${}^3g_{ij}$ and of ${}^3K_{ij}$,
allowing to find the time development from the initial data on
$\Sigma_o$.\bigskip

However, a canonical basis adapted the the previous splittings was
never found. The only result is contained in Ref.\cite{24}, where it
was shown that, having fixed ${}^3K$, the transition from the
non-canonical variables ${}^3{\hat g}_{ij}$, ${}^3A_{ij}$, ${}^3W_i$
to the space of the gravitational initial data satisfying the
constraints is a canonical transformation, named {\it York
map}.\bigskip

In this paper we will show that a new parametrization of the
original 3-metric ${}^3g_{ij}$ allows to find local point
Shanmugadhasan canonical transformation adapted to 10 of the 14
constraints of tetrad gravity, which implements a York map. In
particular one of the new momenta (a gauge variable) will be the
York time  ${}^3K$. The use of Dirac theory of constraints
introduces a different point of view on the gauge-fixing and the
Cauchy problem. While the gauge fixing to the extra 6 primary
constraints fixes the tetrads (i.e. the spatial gyroscopes and their
transport law), the gauge fixing to the 4 primary plus 4 secondary
constraints follows a different scheme from the one used in the
York-Lichnerowicz approach, which influenced contemporary numerical
gravity. Firstly one adds the 4 gauge fixings to the secondary
constraints (the super-Hamiltonian and super-momentum ones), i.e.
one fixes ${}^3K$, i.e.the simultaneity 3-surface, and the
3-coordinates on it (namely 3 of the 5 degrees of freedom of the
conformal 3-metric ${}^3{\hat g}_{ij}$). The preservation in time of
these 4 gauge fixings generates other 4 gauge fixing constraints
determining the lapse and shift functions consistently with the
shape of the simultaneity 3-surface and with the choice of
3-coordinates on it (here is the main difference with the conformal
approach and most of the approaches to numerical gravity). While the
super-Hamiltonian constraint determines the conformal factor $\phi$
\footnote{The only role of the conformal decomposition ${}^3g_{ij} =
\phi^4\, {}^3{\hat g}_{ij}$ is to identify the conformal factor
$\phi$ as the natural unknown in the super-Hamiltonian constraint,
which becomes the {\it Lichnerowicz equation}. See Ref.\cite{1} for
a different justification of this result based on constraint theory
and the two notions of strong and weak ADM energy.}, the
super-momentum constraint determines 3 momenta (replacing the York
gravitomagnetic potential ${}^3W_i$). The remaining 2+2 degrees of
freedom (the genuine tidal effects) are the other two degrees of
freedom in ${}^3{\hat g}_{ij}$ and the two ones inside ${}^3A_{ij}$.
On the Cauchy surface the 2+2 tidal degrees of freedom are assigned
and we have consistency with the initial data of the
York-Lichnerowicz approach.

\bigskip
This is the natural procedure of fixing the gauge and of getting
deterministic Hamilton equations for the tidal degrees of freedom
according to Dirac theory of constraints. Since a completely fixed
gauge is equivalent to give a non-inertial frame centered on some
time-like observer, the gauge-fixed gauge variables will describe
the inertial effects (the {\it appearances} of phenomena) present in
this non-inertial frame, where the Dirac observables describe the
tidal effects of the gravitational field. In particular, the gauge
variable ${}^3K(\tau ,\vec \sigma )$ (York time) describes the
freedom in the choice of the clock synchronization convention, i.e.
in the definition of the instantaneous 3-spaces $\Sigma_{\tau}$.

\bigskip

In Section II we will find the  York map and we discuss some classes
of  Hamiltonian gauges. In the York canonical basis it is possible
to express both the gauge variables (inertial effects) and the tidal
degrees of freedom in terms of the original variables.

In Section III we give the general solution of the super-momentum
constraints in the York canonical basis: it is defined modulo the
{\it zero modes} of the covariant derivative.

In Section IV there is the form of the super-Hamiltonian constraint
and the weak ADM energy in a family of completely fixed 3-orthogonal
Schwinger time gauges parametrized by the gauge variable ${}^3K(\tau
,\vec \sigma )$ (it is a family of non-inertial frames with a fixed
pattern of inertial effects) defined by suitable set of primary
gauge-fixing constraints.

In Section V there are the equations determining the lapse and shift
functions of the 3-orthogonal gauges gauges: they arise from the
preservation in time of the primary gauge-fixing constraints. It is
shown that, like in Yang-Mills theories, a generalized Gribov
ambiguity arises in the determination of the shift functions and, as
a consequence, also of the lapse function. It is induced by the zero
modes of the covariant derivative.

In Section VI there are the final Hamilton equations describing the
deterministic evolution of the dynamical gravitational degrees of
freedom (the generalized tidal effects) both in an arbitrary gauge
and in the completely fixed ones. It is shown that only in a
completely fixed gauge we can obtain a deterministic evolution,
which, however depends upon the chosen non-inertial frame with its
pattern of relativistic inertial forces. This frame-dependence
derives from the geometrical view of the gravitational field and is
lost when the theory is reduced to a linear spin 2 field on a
background space-time.

In the Conclusions we make a summary of the results, emphasizing the
methodological and interpretational insights induced by a correct
use of constraint's theory. We also make some comments on the
perspectives of technical developments (for instance the weak field
limit but with relativistic motion) in connection with physical
problems connected with space experiments in the solar system and
with astrophysics.

Finally there are four Appendices: A) with the notations for tetrad
gravity; B) with the calculations for the canonical transformation
of Section III; C) with the 3-geometry in 3-orthogonal gauges; D)
with Green functions.

\bigskip

In Ref.\cite{25} there is an expanded version of this paper,
containing many Appendices with the explicit expression of many
quantities in the York canonical basis and in the 3-orthogonal
gauge. To have an exposition concentrated on the main theoretical
aspects implied by a coherent and systematic use of constraint's
theory and on the interpretational issues, we have not given the
explicit expression of heavy calculations and of some cumbersome
formulas. They can be found in Ref.\cite{25} at the quoted
positions.

\section{The York Map from a Shanmugadhasan Canonical
Transformation adapted only to the Rotation Constraints.}

Let us look for a Shanmugadhasan canonical transformation
interpretable as a York map. It can be obtained starting from a
natural parametrization of the 3-metric and then by making an
adaptation only to the rotation constraints (and not also the the
super-momentum ones like in Refs.\cite{3,16}).\bigskip

\subsection{Diagonalization of the 3-Metric.}

The 3-metric ${}^3g_{rs}$ may be diagonalized with an {\it
orthogonal} matrix $V(\theta^r)$, $V^{-1} = V^T$, $det\, V = 1$,
depending on 3 Euler angles $\theta^r$ \footnote{Due to the positive
signature of the 3-metric, we define the matrix $V$ with the
following indices: $V_{ru}$. Since the choice of Shanmugadhasan
canonical bases breaks manifest covariance, we will use the notation
$V_{ua} = \sum_v\, V_{uv}\, \delta_{v(a)}$ instead of $V_{u(a)}$. We
use the following types of indices: $a = 1,2,3$ and $\bar a = 1,2$.}

\bea
 {}^3g_{rs} &=& \sum_{uv}\, V_{ru}(\theta^n)\, \lambda_u\, \delta_{uv}\,
 V^T_{vs}(\theta^n) = \sum_a\, \Big(V_{ra}(\theta^n)\, \Lambda^a\Big)\,
 \Big(V_{sa}(\theta^n)\, \Lambda^a\Big) =\nonumber \\
 &=& \sum_a\, {}^3{\bar e}_{(a)r}\, {}^3{\bar e}_{(a)s} = \sum_a\, {}^3e_{(a)r}\,
 {}^3e_{(a)s} = \phi^4\, {}^3{\hat g}_{rs} {\buildrel {def}\over
 =}\, \phi^4\, \sum_a\, Q^2_a\, V_{ra}(\theta^n)\, V_{sa}(\theta^n),\nonumber \\
 &&{}\nonumber \\
 \Lambda_a(\tau ,\vec \sigma ) &{\buildrel {def}\over =}&
 \sum_u\, \delta_{au}\, \sqrt{\lambda_u(\tau ,\vec \sigma )}\,
 {\buildrel {def}\over =}\, \phi^2(\tau ,\vec \sigma )\, Q_a(\tau
 ,\vec \sigma ),\nonumber \\
 &&{}\nonumber \\
 Q_a &{\buildrel {def}\over =}& e^{\sum_{\bar a}^{1,2}\, \gamma_{\bar aa}\, R_{\bar a}},
 \qquad R_{\bar a}  = \sum_b\, \gamma_{\bar ab}\, ln\, {{\Lambda_b}\over
 {(\Lambda_1\, \Lambda_2\, \Lambda_3)^{1/3}}},\nonumber \\
 \phi &=& (det\, {}^3g)^{1/12} = ( {}^3e)^{1/6} = (\lambda_1\,
\lambda_2\, \lambda_3)^{1/12} = (\Lambda_1\, \Lambda_2\,
\Lambda_3)^{1/6},
 \label{II1}
 \eea

\noindent where the set of numerical parameters $\gamma_{\bar aa}$
satisfies \cite{1} $\sum_u\, \gamma_{\bar au} = 0$, $\sum_u\,
\gamma_{\bar a u}\, \gamma_{\bar b u} = \delta_{\bar a\bar b}$,
$\sum_{\bar a}\, \gamma_{\bar au}\, \gamma_{\bar av} = \delta_{uv} -
{1\over 3}$. The assumed boundary conditions imply $\Lambda_a(\tau
,\vec \sigma )\,  \rightarrow_{r \rightarrow \infty}\,\, 1 + {M\over
{4r}} + {{a_a}\over {r^{3/2}}} + O(r^{-3})$ and $\phi (\tau ,\vec
\sigma )\, \rightarrow_{r \rightarrow \infty}\,\, 1 +  O(r^{-1})$.

 \bigskip

Cotriads and triads are defined modulo rotations $R(\alpha_{(a)})$
on the flat 3-index $(a)$

 \bea
 {}^3e_{(a)r} &=& R_{(a)(b)}(\alpha_{(c)})\, {}^3{\bar e}_{(b)r},
 \nonumber \\
 &&{}\nonumber \\
 {}^3{\bar e}_{(a)r} &{\buildrel {def}\over =}& \sum_u\, \sqrt{\lambda_u}\, \delta_{u(a)}\,
 V^T_{ur}(\theta^n) = V_{ra}(\theta^n)\, \Lambda_a = \phi^2\,
 Q_a\, V_{ra}(\theta^n),\nonumber \\
 &&{}\nonumber \\
 {}^3{\bar e}^r_{(a)} &=& \sum_u\, {{\delta_{u(a)}}\over {\sqrt{\lambda_u}}}\,
 V_{ru} = {{V_{ra}(\theta^n)}\over {\Lambda_a}} = \phi^{-2}\, Q_a^{-1}\,
 V_{ra}(\theta^n).
 \label{II2}
 \eea

The gauge Euler angles $\theta^r$ give a  description of the
3-coordinate systems on $\Sigma_{\tau}$ from a local point of view,
because they give the orientation of the tangents to the 3
coordinate lines through each point (their conjugate momenta are
determined by the super-momentum constraints), $\phi$ is the
conformal factor of the 3-metric, i.e. the unknown in the
super-hamiltonian constraint (its conjugate momentum is a gauge
variable, describing the form of the simultaneity surfaces
$\Sigma_{\tau}$), while the two independent eigenvalues of the
conformal 3-metric ${}^3{\hat g}_{rs}$ (with determinant equal to 1)
describe the genuine {\it tidal} effects of general relativity (the
non-linear "graviton").

\bigskip

\subsection{An Intermediate Point Shanmugadhasan Canonical
Transformation.}

Let us consider the following point canonical transformation
(realized in two steps)

\bea
 &&\begin{minipage}[t]{3cm}
\begin{tabular}{|l|l|l|l|} \hline
$\varphi_{(a)}$ & $n$ & $n_{(a)}$ & ${}^3e_{(a)r}$ \\ \hline
$\approx 0$ & $\approx 0$ & $  \approx 0 $ & ${}^3{ \pi}^r_{(a)}$
\\ \hline
\end{tabular}
\end{minipage} \hspace{1cm} {\longrightarrow \hspace{.2cm}} \
\begin{minipage}[t]{4 cm}
\begin{tabular}{|ll|ll|l|} \hline
$\varphi_{(a)}$ & $\alpha_{(a)}$ & $n$ & ${\bar n}_{(a)}$ &  ${}^3{\bar e}_{(a)r}$\\
\hline $\approx0$ &
 $\approx 0$ & $\approx 0$ & $\approx 0$
& ${}^3{\tilde {\bar \pi}}^r_{(a)}$ \\ \hline
\end{tabular}
\end{minipage} \nonumber \\
 &&{}\nonumber \\
 &&\hspace{1cm} {\longrightarrow \hspace{.2cm}} \
\begin{minipage}[t]{4 cm}
\begin{tabular}{|ll|ll|ll|} \hline
$\varphi_{(a)}$ & $\alpha_{(a)}$ & $n$ & ${\bar n}_{(a)}$ &
$\theta^r$ & $\Lambda_r$\\ \hline $\approx0$ &
 $\approx 0$ & $\approx 0$ & $\approx 0$
& $\pi^{(\theta )}_r$ & $P^{ r}$ \\ \hline
\end{tabular}
\end{minipage}
 \label{II3}
 \eea

 \noindent where ${\bar n}_{(a)} = \sum_b\, n_{(b)}\,
 R_{(b)(a)}(\alpha_{(e)})$ are the shift functions at
 $\alpha_{(a)}(\tau ,\vec \sigma ) = 0$.
\bigskip

This is a {\it Shanmugadhasan canonical transformation adapted also
to the rotation constraints}. It allows to separate the gauge
variables ($\alpha_{(a)}$, $\varphi_{(a)}$) of the Lorentz gauge
group acting on the tetrads.
\bigskip

Being a point transformation, we have

\bea
 {}^3{\tilde \pi}^r_{(a)}(\tau ,\vec \sigma ) &=&
 \sum_b\, K^r_{(a)b}(\tau ,\vec \sigma
 )\, P^b(\tau ,\vec \sigma ) + \sum_i\, G^r_{(a)i}(\tau ,\vec \sigma
 )\, \pi^{(\theta )}_i(\tau ,\vec \sigma ) +\nonumber \\
  &+& \sum_{(c)}\,
 F^r_{(a)(c)}(\tau ,\vec \sigma )\, \pi^{(\alpha )}_{(c)}(\tau ,\vec \sigma )
 \approx \nonumber \\
 &\approx& \sum_b\, K^r_{(a)b}(\tau ,\vec \sigma
 )\, P^b(\tau ,\vec \sigma ) + \sum_i\, G^r_{(a)i}(\tau ,\vec \sigma
 )\, \pi^{(\theta )}_i(\tau ,\vec \sigma ).
 \label{II4}
 \eea

Here $\pi^{(\alpha )}_{(a)}(\tau ,\vec \sigma ) \approx 0$ are the
Abelianized rotation constraints \cite{1,3}, canonically conjugate
to $\alpha_{(a)}(\tau ,\vec \sigma )$.

\bigskip

Let us remark that the Shanmugadhasan canonical transformation
identifying the York map is valid only in the configuration region
where the 3-metric ${}^3g_{rs}(\tau ,\vec \sigma )$ has three
distinct eigenvalues everywhere [i.e. $a_a \not= a_b$ for $a \not=
b$ in the asymptotic behavior of $\Lambda_a$] except at spatial
infinity, where they tend to the common value $1$. The degenerate
cases with two or three equal eigenvalues are {\it singular
configurations} with less configurational degrees of freedom. To
treat these cases we must add by hand extra first class constraints
of the type $\Lambda_a(\tau ,\vec \sigma ) - \Lambda_b(\tau ,\vec
\sigma ) \approx 0$, $a \not= b$, and apply the Dirac algorithm to
the enlarged set of constraints.

\bigskip

 The generating function of the canonical transformation is

 \beq
 \Phi = \int d^3\sigma\, \sum_{ar}\,
 {}^3{\tilde \pi}^r_{(a)}(\tau ,\vec
 \sigma )\, \Big[\sum_b\, R_{(a)(b)}(\alpha_{(e)})\,
 V_{rb}(\theta^n)\, \Lambda_b\Big](\tau ,\vec \sigma ),
 \label{II5}
 \eeq

\noindent so that we get (see Ref.\cite{3} for the O(3) Lie algebra
-valued  matrices $A_{(a)(b)}(\alpha_{(c)})$, $B(\alpha_{(c)}) =
A^{-1}(\alpha_{(c)})$, such that ${{\partial\,
R_{(b)(c)}(\alpha_{(e)})} \over {\partial\, \alpha_{(a)}}} =
\sum_{da}\, \epsilon_{(b)(d)(n)}\, R_{(d)(c)}(\alpha_{(e)})\,
A_{(n)(a)}(\alpha_{(e)})$)

\bea
 \pi^{(\alpha )}_{(c)}(\tau ,\vec \sigma ) &=& {{\delta\, \Phi}\over
 {\delta\, \alpha_{(c)}(\tau ,\vec \sigma )}} =\nonumber \\
 &=& - \sum_{krab}\, \Big[A_{(k)(c)}(\alpha_{(e)})\,
 \epsilon_{(k)(b)(a)}\, {}^3e_{(b)r}\, {}^3{\tilde \pi}^r_{(a)}
\Big](\tau ,\vec \sigma ) =\nonumber \\
 &=& - \sum_k\, \Big[A_{(k)(c)}(\alpha_{(e)})\, M_{(k)}
 \Big](\tau ,\vec \sigma ) \approx 0,\nonumber \\
 P^b(\tau ,\vec \sigma ) &=& {{\delta\, \Phi}\over {\delta\,
 \Lambda_b(\tau ,\vec \sigma )}} =
  \sum_{ar}\, \Big[{}^3{\tilde \pi}^r_{(a)}\, R_{(a)(b)}(\alpha_{(e)})\,
 V_{rb}(\theta^n)\Big](\tau ,\vec \sigma ) =\nonumber \\
 &=& \sum_{ar}\, {{{}^3{\tilde \pi}^r_{(a)}\, R_{(a)(b)}(\alpha_{(e)})\,
 {}^3{\bar e}_{(b)r}}\over {\Lambda_b}}(\tau ,\vec \sigma ),\nonumber \\
 \pi^{(\theta )}_i(\tau ,\vec \sigma ) &=& {{\delta\, \Phi}\over
 {\delta\, \theta^i(\tau ,\vec \sigma )}} =
  - \sum_{lmra}\, \Big[A_{mi}(\theta^n)\, \epsilon_{mlr}\,
 {}^3e_{(a)l}\, {}^3{\tilde \pi}^r_{(a)}\Big](\tau ,\vec \sigma ).
 \label{II6}
 \eea

As a consequence of the calculation of Appendix B we have
[Eqs.(\ref{a10}) are used; ${}^3{\tilde {\bar \pi}}^r_{(a)}$ and
${}^3{\tilde K}_{rs}$ are the cotriad momentum and the extrinsic
curvature of $\Sigma_{\tau}$ after having used the rotation
constraints $M_{(a)}(\tau ,\vec \sigma ) \approx 0$]

\begin{eqnarray*}
 {}^3{\tilde \pi}^r_{(a)} &{\buildrel {def}\over =}& \sum_b\,
 R_{(a)(b)}(\alpha_{(e)})\, {\bar \pi}^r_{(b)},\nonumber \\
 &&{}\nonumber \\
 {}^3{\bar \pi}^r_{(a)} &=& \sum_b\, \pi^r_{(b)}\,
 R_{(b)(a)}(\alpha_{(e)}) =\nonumber \\
 &=& V_{ra}(\theta^n)\,
 P^a + \sum_{b}^{b \not= a}\, \sum_{twi}\, {{V_{rb}(\theta^n)\,
 \epsilon_{abt}\, V_{tw}(\theta^n)}\over {\Lambda_b\,
 \Big({{\Lambda_b}\over {\Lambda_a}} - {{\Lambda_a}\over {\Lambda_b}}\Big)}}\,
 B_{iw}(\theta^n)\, \pi^{(\theta )}_i -\nonumber \\
 &-& \sum_b^{b \not= a}\, \sum_{kti}\, {{V_{rb}(\theta^n)\,
 \epsilon_{bat}\, R_{(t)(k)}(\alpha_{(e)})}\over {\Lambda_b\,
 \Big({{\Lambda_b}\over {\Lambda_a}} - {{\Lambda_a}\over {\Lambda_b}}\Big)}}\,
 B_{(c)(k)}(\alpha_{(e)})\, \pi^{(\alpha )}_{(c)} \approx
 \end{eqnarray*}

 \begin{eqnarray*}
 &\approx& V_{ra}(\theta^n)\,
 P^a + \sum_b^{b \not= a}\, \sum_{twi}\, {{V_{rb}(\theta^n)\,
 \epsilon_{abt}\, V_{tw}(\theta^n)}\over {\Lambda_b\,
 \Big({{\Lambda_b}\over {\Lambda_a}} - {{\Lambda_a}\over {\Lambda_b}}\Big)}}\,
 B_{iw}(\theta^n)\, \pi^{(\theta )}_i =\nonumber \\
 &{\buildrel {def}\over =}& {}^3{\tilde {\bar \pi}}^r_{(a)}\quad
 \rightarrow_{\theta^n \rightarrow 0}\,\, \delta_{ra}\, P^a +
 {{\epsilon_{ari}}\over {\Lambda_r\,
 \Big({{\Lambda_r}\over {\Lambda_a}} - {{\Lambda_a}\over
 {\Lambda_r}}\Big)}}\, \pi_i^{(\theta )}.
 \end{eqnarray*}

\bea
 {}^3K_{rs}  &\approx& {}^3{\tilde K}_{rs} =
 {{\sgn\, 4\pi\, G}\over {c^3\, \Lambda_1\, \Lambda_2\,
 \Lambda_3}}\, \Big[\sum_a\, \Lambda_a^2\, V_{ra}(\theta^n)\,
 V_{sa}(\theta^n)\, (2\, \Lambda_a\, P^a - \sum_b\, \Lambda_b\,
 P^b) +\nonumber \\
 &+& \sum_{ab}^{a \not= b}\, \Lambda_a\, \Lambda_b\, \Big(V_{ra}(\theta^n)\,
 V_{sb}(\theta^n) + V_{rb}(\theta^n)\, V_{sa}(\theta^n)\Big)\,
 \sum_{twi}\, {{\epsilon_{abt}\, V_{tw}(\theta^n)\, B_{iw}(\theta^n)\,
 \pi_i^{(\theta )}}\over {{{\Lambda_b}\over {\Lambda_a}} -
 {{\Lambda_a}\over {\Lambda_b}}}}\Big],\nonumber \\
 &&{}\nonumber \\
 &&{}\nonumber \\
 {}^3K &\approx& {}^3{\tilde K} =
  -  \sgn\, {{4\pi\, G}\over {c^3}}\, {{\sum_b\, \Lambda_b\, P^b}\over
 { \Lambda_1\, \Lambda_2\, \Lambda_3}}.
 \label{II7}
 \eea

\medskip

Since in Ref.\cite{3} it was assumed ${}^3{\tilde \pi}^r_{(a)}(\tau
,\vec \sigma )\, \rightarrow_{r \rightarrow \infty}\,\,
O(r^{-5/2})$, from Eqs.(\ref{II4}) and Appendix B we get $P^a(\tau
,\vec \sigma )\, \rightarrow_{r \rightarrow \infty}\, O(r^{-5/2})$.
However we must have $\pi_i^{(\theta )}(\tau ,\vec \sigma )\,
\rightarrow_{r \rightarrow \infty}\,\, O(r^{-4})$, since the
requirement $\Lambda_a(\tau ,\vec \sigma ) \not= \Lambda_b(\tau
,\vec \sigma )$ for $a \not= b$, needed to avoid singularities,
implies $a_a \not= a_b$ for $a \not= b$ in their asymptotic
behavior, so that we get $\Big({{\Lambda_b}\over {\Lambda_a}} -
{{\Lambda_a}\over {\Lambda_b}}\Big)^{-1}(\tau ,\vec \sigma )\,
\rightarrow_{r \rightarrow \infty}\,\, {{r^{3/2}}\over { 2\, (a_b -
a_a)}}$. As a consequence, consistently with Eqs.(\ref{II6}), we
have $\pi^{(\alpha )}_{(a)}(\tau ,\vec \sigma )\, \rightarrow_{r
\rightarrow \infty}\, O(r^{-5/2})$. Also the angles
$\alpha_{(a)}(\tau ,\vec \sigma )$ and $\theta^i(\tau ,\vec \sigma
)$ must tend to zero in a direction-independent way at spatial
infinity.

\subsection{The York Map.}

Since from Eq.(\ref{II7}) we have ${}^3\tilde K = - \sgn\, {{4\pi\,
G}\over {c^3}}\, {{\sum_b\, \Lambda_b\, P^b}\over {\Lambda_1\,
\Lambda_2\, \Lambda_3}}$, we can introduce the following pair
$\tilde \phi$, $\pi_{\tilde \phi}$ of canonical variables

\bea
 \tilde \phi &=& \phi^6 = \prod_a\, \Lambda_a,\qquad
 \pi_{\tilde \phi} = - \sgn\, {{c^3}\over {12\pi\, G}}\, {}^3K =
 {{\sum_b\, \Lambda_b\, P^b}\over {3\, \Lambda_1\, \Lambda_2\,
 \Lambda_3}},\nonumber \\
 &&{}\nonumber \\
 &&\{ \tilde \phi(\tau ,\vec \sigma ), \pi_{\tilde \phi}(\tau ,{\vec
 \sigma}^{'})\} = \delta^3(\vec \sigma , {\vec \sigma}^{'}),
 \label{II8}
 \eea

\noindent with $\pi_{\tilde \phi}(\tau ,\vec \sigma )\,
\rightarrow_{r \rightarrow \infty}\,\, O(r^{-5/2})$ at spatial
infinity.
\medskip

Let us consider the following point canonical transformation (it is
a family of canonical transformations depending on the set of
numerical parameters $\gamma_{\bar aa}$)

\bea
 &&\begin{minipage}[t]{3cm}
\begin{tabular}{|l|} \hline
 $\Lambda_r$ \\ \hline $P^r$
\\ \hline
\end{tabular}
\end{minipage} \hspace{1cm} {\longrightarrow \hspace{.2cm}} \
\begin{minipage}[t]{4 cm}
\begin{tabular}{|l|l|} \hline
 $ \tilde \phi$ &  $R_{\bar a}$\\
\hline $\pi_{\tilde \phi}$ & $\Pi_{\bar a}$ \\
\hline
\end{tabular}
\end{minipage}
 \label{II9}
 \eea
\medskip

Since the generating function is $\Psi = \int d^3\sigma\,
\Big[\sum_b\, P^b\, {\tilde \phi}^{1/3}\, e^{\sum_{\bar a}\,
\gamma_{\bar ab}\, R_{\bar a}}\Big](\tau ,\vec \sigma )$, we get

\bea
 \pi_{\tilde \phi}(\tau ,\vec \sigma ) &=& {{\delta\, \Psi}\over {\delta \tilde \phi (\tau ,\vec
 \sigma )}} =  {{\sum_b\, \Lambda_b\, P^b}\over  {3\, \Lambda_1\, \Lambda_2\,
 \Lambda_3}}(\tau ,\vec \sigma ),\nonumber \\
 \Pi_{\bar a}(\tau ,\vec \sigma ) &=& {{\delta\, \Psi}\over {\delta\,
 R_{\bar a}(\tau ,\vec \sigma )}} = \Big[(\Lambda_1\, \Lambda_2\,
 \Lambda_3)^{1/3}\, \sum_b\, \gamma_{\bar ab}\, P^b\,
 e^{\sum_{\bar c}\, \gamma_{\bar cb}\, R_{\bar c}}\Big](\tau ,\vec \sigma
 ).
 \label{II10}
 \eea

Therefore, besides the definitions in Eqs.(\ref{II1}), we get

 \begin{eqnarray*}
 P^b &=&  {\tilde \phi}^{-1/3}\, Q^{-1}_b\,
\Big[\tilde \phi\, \pi_{\tilde \phi} +  \sum_{\bar a}\,
\gamma_{\bar bb}\, \Pi_{\bar b}\Big],\nonumber \\
 &&{}\nonumber \\
 \Pi_{\bar a} &=& (\Lambda_1\, \Lambda_2\,
 \Lambda_3)^{1/3}\, \sum_b\, \gamma_{\bar ab}\, P^b\,
 Q_b  = \sum_b\, \gamma_{\bar ab}\, \Lambda_b\, P^b,\nonumber \\
 &&{}\nonumber \\
  &&\sum_b\, \Lambda_b\, P^b = 3\, \tilde \phi\,
 \pi_{\tilde \phi},\qquad
 \sum_{\bar a}\, \gamma_{\bar au}\, \Pi_{\bar a} = \Lambda_u\,
 P^u -  \tilde \phi\, \pi_{\tilde \phi},
 \end{eqnarray*}

\begin{eqnarray*}
 {}^3{\tilde {\bar \pi}}^r_{(a)} &=& {\tilde \phi}^{-1/3}\, \Big[
 V_{ra}(\theta^n)\, Q^{-1}_a\,
 (\tilde \phi\, \pi_{\tilde \phi} +  \sum_{\bar b}\, \gamma_{\bar ba}\,
 \Pi_{\bar b}) +\nonumber \\
 &+& \sum_{l}^{l \not= a}\, \sum_{twi}\, Q^{-1}_l\, {{V_{rl}(\theta^n)\,
 \epsilon_{alt}\, V_{tw}(\theta^n)}\over {Q_l\, Q^{-1}_a - Q_a\, Q^{-1}_l
}}\, B_{iw}(\theta^n)\, \pi^{(\theta )}_i \Big],
 \end{eqnarray*}

\bea
 &&\sum_r\, {}^3{\bar e}_{(b)r}\, {}^3{\tilde {\bar \pi}}^r_{(a)} =
 \delta_{ab}\, [\tilde \phi\, \pi_{\tilde \phi} +  \sum_{\bar
a}\, \gamma_{\bar aa}\, \Pi_{\bar a}] +
  \sum_{twi}\, {{\epsilon_{abt}\, V_{tw}(\theta^n)\, B_{iw}(\theta^n)\,
 \pi^{(\theta )}_i}\over {Q_b\, Q^{-1}_a - Q_a\, Q^{-1}_b}},\nonumber \\
 &&{}\nonumber \\
 &&{}\nonumber \\
 {}^3{\tilde K}_{rs} &=& \sgn\, {{4\pi\, G}\over {c^3}}\, {\tilde \phi}^{-1/3}\, \Big(\sum_a\,
Q^2_a\, V_{ra}(\theta^n)\, V_{sa}(\theta^n)\, [2\, \sum_{\bar b}\,
\gamma_{\bar ba}\, \Pi_{\bar b} -  \tilde \phi\,
\pi_{\tilde \phi}] +\nonumber \\
 &+& \sum_{ab}\, Q_a\, Q_b\, [V_{ra}(\theta^n)\, V_{sb}(\theta^n) +
 V_{rb}(\theta^n)\, V_{sa}(\theta^n)]\, \sum_{twi}\, {{\epsilon_{abt}\,
 V_{tw}(\theta^n)\, B_{iw}(\theta^n)\, \pi_i^{(\theta )}}\over {
 Q_b\, Q^{-1}_a  - Q_a\, Q^{-1}_b}} \Big).\nonumber \\
 &&{}
 \label{II11}
 \eea

See Appendix B of Ref.\cite{25} for the explicit expression of the
3-Christoffel symbols ${}^3\Gamma^r_{uv}$ on $\Sigma_{\tau}$ and for
the $\Gamma$-$\Gamma$ potential ${\cal S}(\tau ,\vec \sigma )$ in
the York canonical basis.
\medskip

The sequence of canonical transformations (\ref{II3}) and
(\ref{II9}) realize a {\it York map} because the gauge variable
$\pi_{\tilde \phi}$ is proportional to {\it York internal extrinsic
time} ${}^3K$. Its conjugate variable, to be determined by the
super-hamiltonian constraint, is $\tilde \phi = {}^3\bar e$, which
is proportional to {\it Misner's internal intrinsic time}; moreover
$\tilde \phi$ is the {\it volume density} on $\Sigma_{\tau}$: $V_R =
\int_R d^3\sigma\, \phi^6$, $R \subset \Sigma_{\tau}$.
\medskip

Eqs.(\ref{II3}), (\ref{II9}) and (\ref{II11}) identify the two pairs
of canonical variables $R_{\bar a}$, $\Pi_{\bar a}$, $\bar a = 1,2$,
as those describing the generalized {\it tidal effects}, namely the
independent degrees of freedom of the gravitational field In
particular the configuration tidal variables $R_{\bar a}$ depend
{\it only on the eigenvalues of the 3-metric}. They are Dirac
observables {\it only} with respect to the gauge transformations
generated by 10 of the 14 first class constraints. Let us remark
that, if we fix completely the gauge and we go to Dirac brackets,
then the only surviving dynamical variables $R_{\bar a}$ and
$\Pi_{\bar a}$ become two pairs of {\it non canonical} Dirac
observables for that gauge: the two pairs of canonical Dirac
observables have to be found as a Darboux basis of the copy of the
reduced phase space identified by the gauge and they will be (in
general non-local) functionals of the $R_{\bar a}$, $\Pi_{\bar a}$
variables. This shows the importance of canonical bases like the
York one: the tidal effects are described by {\it local} functions
of the 3-metric and its conjugate momenta.\medskip

Since the variables $\tilde \phi$ [given in Eq.(\ref{II8})] and
$\pi_i^{(\theta )}$ [given in Eqs.(\ref{II6})] are determined by the
super-Hamiltonian and super-momentum constraints, the {\it arbitrary
gauge variables} are $\alpha_{(a)}$, $\varphi_{(a)}$, $\theta^i$,
$\pi_{\tilde \phi}$, $n$ and ${\bar n}_{(a)}$. As shown in
Refs.\cite{4}, they describe the following generalized {\it inertial
effects}:

a) $\alpha_{(a)}(\tau ,\vec \sigma )$ and $\varphi_{(a)}(\tau ,\vec
\sigma )$  describe the arbitrariness in the choice of a tetrad to
be associated to a time-like observer, whose world-line goes through
the point $(\tau ,\vec \sigma )$. They fix {\it the unit 4-velocity
of the observer and the conventions for the gyroscopes and their
transport along the world-line of the observer}.

b) $\theta^i(\tau ,\vec \sigma )$ [depending only on the 3-metric,
as shown in Eq.(\ref{II1})] describe the arbitrariness in the choice
of the 3-coordinates on the simultaneity surfaces $\Sigma_{\tau}$ of
the chosen non-inertial frame  centered on an arbitrary time-like
observer. Their choice will induce a pattern of {\it relativistic
standard inertial forces} (centrifugal, Coriolis,...), whose
potentials are contained in the term ${\cal S}(\tau ,\vec \sigma )$
of the weak ADM energy $E_{ADM}$ given in Eqs.(\ref{a8}). These
inertial effects are the relativistic counterpart of the
non-relativistic ones (they are present also in the non-inertial
frames of Minkowski space-time).

c) ${\bar n}_{(a)}(\tau ,\vec \sigma )$, the shift functions
appearing in the Dirac Hamiltonian, describe which points on
different simultaneity surfaces have the same numerical value of the
3-coordinates. They are the inertial potentials describing the
effects of the non-vanishing off-diagonal components ${}^4g_{\tau
r}(\tau ,\vec \sigma )$ of the 4-metric, namely they are the {\it
gravito-magnetic potentials} \footnote{In the post-Newtonian
approximation in harmonic gauges they are the counterpart of the
electro-magnetic vector potentials describing magnetic fields
\cite{23}, \cite{16}: A) $N = 1 + n$, $n\, {\buildrel {def}\over
=}\, - {{4\, \sgn}\over {c^2}}\, \Phi_G$ with $\Phi_G$ the {\it
gravito-electric potential}; B) $n_r\, {\buildrel {def}\over =}\,
{{2\, \sgn}\over {c^2}}\, A_{G\, r}$ with $A_{G\, r}$ the {\it
gravito-magnetic} potential; C) $E_{G\, r} = \partial_r\, \Phi_G -
\partial_{\tau}\, ({1\over 2}\, A_{G\, r})$ (the {\it
gravito-electric field}) and $B_{G\, r} = \epsilon_{ruv}\,
\partial_u\, A_{G\, v} = c\, \Omega_{G\, r}$ (the {\it
gravito-magnetic field}). Let us remark that in arbitrary gauges the
analogy with electro-magnetism \cite{23} breaks down.} responsible
of effects like the dragging of inertial frames (Lens-Thirring
effect) \cite{23} in the post-Newtonian approximation.

d) $\pi_{\tilde \phi}(\tau ,\vec \sigma )$, i.e. the York time
${}^3K(\tau ,\vec \sigma )$, describes the arbitrariness in the
shape of the simultaneity surfaces $\Sigma_{\tau}$ of the
non-inertial frame, namely the arbitrariness in the choice of the
convention for the synchronization of distant clocks. Since this
variable is present in the Dirac Hamiltonian \footnote{See
Eqs.(\ref{II12}) for its presence in the super-Hamiltonian
constraint and in the weak ADM energy, and Eqs.(\ref{III1}) for its
presence in the super-momentum constraints.}, it is a  {\it new
inertial potential} connected to the problem of the relativistic
freedom in the choice of the {\it instantaneous 3-space}, which has
no non-relativistic analogue (in Galilei space-time time is absolute
and there is an absolute notion of Euclidean 3-space). Its effects
are completely unexplored. For instance, since the sign of the trace
of the extrinsic curvature may change from a region to another one
on the simultaneity surface $\Sigma_{\tau}$, {\it the associated
inertial force in the Hamilton equations may change from attractive
to repulsive in different regions}.

e) $n(\tau ,\vec \sigma )$, the lapse function appearing in the
Dirac Hamiltonian, describes the arbitrariness in the choice of the
unit of proper time in each point of the simultaneity surfaces
$\Sigma_{\tau}$, namely how these surfaces are packed in the 3+1
splitting.

\bigskip

From Eqs.(\ref{a4}), (\ref{a8}) and Eq.(B4) of Ref.\cite{25}, where
Eq.(B1) gives the expression of the $\Gamma$-$\Gamma$ term ${\cal
S}$, we get the following expression of the super-Hamiltonian
constraint and of weak ADM energy in the York canonical basis
(${}^3\hat R$ and $\hat \triangle$ are the 3-curvature of
$\Sigma_{\tau}$ and the Laplace-Beltrami operator for the conformal
3-metric ${}^3{\hat g}_{rs}$, respectively)

\bea
 {\cal H}(\tau ,\vec \sigma )&=&
 \epsilon\, {{c^3}\over {16\pi\, G}}\, {\tilde \phi}^{1/6} (\tau ,\vec \sigma ) \,
[- 8\,   {\hat \triangle}\, {\tilde \phi}^{1/6} +  {}^3{\hat R}\,
{\tilde \phi}^{1/6}](\tau ,\vec \sigma ) -{{\sgn}\over c}\,
{\cal M}(\tau ,\vec \sigma )-\nonumber \\
 &&{}\nonumber \\
  &-& \sgn\, {{2\pi\, G}\over {c^3}}\, {\tilde \phi}^{-1}\, \Big[
- 3\, (\tilde \phi\, \pi_{\tilde \phi})^2 + 2\, \sum_{\bar b}\,
 \Pi^2_{\bar b} +\nonumber \\
 &+& 2\, \sum_{abtwiuvj}\, {{\epsilon_{abt}\, \epsilon_{abu}\, V_{tw}(\theta^n)\,
 B_{iw}(\theta^n)\, V_{uv}(\theta^n)\, B_{jv}(\theta^n)\, \pi_i^{(\theta )}\,
 \pi_j^{(\theta )}}\over {\Big[Q_a\, Q^{-1}_b - Q_b\,
 Q^{-1}_a \Big]^2}}\Big](\tau ,\vec \sigma )
 \approx 0,\nonumber \\
 &&{}\nonumber \\
 E_{ADM} &=&  \int d^3\sigma\, \Big[{\cal M} -
  {{c^4}\over {16\pi\, G}}\, {\cal S} +
    {{2\pi\, G}\over {c^2}}\, {\tilde \phi}^{-1}\, \Big(
 - 3\, (\tilde \phi\, \pi_{\tilde \phi})^2 + 2\, \sum_{\bar b}\,
 \Pi^2_{\bar b} +\nonumber \\
 &+& 2\, \sum_{abtwiuvj}\, {{\epsilon_{abt}\, \epsilon_{abu}\, V_{tw}(\theta^n)\,
 B_{iw}(\theta^n)\, V_{uv}(\theta^n)\, B_{jv}(\theta^n)\, \pi_i^{(\theta )}\,
 \pi_j^{(\theta )}}\over {\Big[Q_a\, Q^{-1}_b - Q_b\,
 Q^{-1}_a \Big]^2}}\Big)\,\, \Big](\tau ,\vec \sigma ).
 \label{II12}
 \eea

\subsection{Gauges}

Once we are in the York canonical basis, it is useful to restrict
ourselves to the {\it Schwinger time gauges} implied by the gauge
fixing constraints $\varphi_{(a)}(\tau ,\vec \sigma ) \approx 0$,
$\alpha_{(a)}(\tau ,\vec \sigma ) \approx 0$, which imply
$\lambda_{(a)}(\tau ,\vec \sigma ) \approx 0$, $\lambda_{\vec
\varphi (a)}(\tau ,\vec \sigma ) \approx 0$ in Eq.(\ref{a7}). In
this way we can go to Dirac brackets with respect to the primary 6
constraints $\pi_{\vec \varphi\, (a)}(\tau ,\vec \sigma ) \approx
0$, $\pi_{(a)}^{(\alpha )}(\tau ,\vec \sigma ) \approx 0$ [the
Abelianized rotation constraints of Eq.(\ref{II6})] and  these gauge
fixings (in total there are 6 pairs of second class constraints). In
this reduced phase space the York canonical basis is formed by the
pairs: $n(\tau ,\vec \sigma ) $, $\pi_n(\tau ,\vec \sigma ) \approx
0$, ${\bar n}_{(a)}(\tau ,\vec \sigma ) $, $\pi_{\vec n\, (a)}(\tau
,\vec \sigma ) \approx 0$, $\theta^i(\tau ,\vec \sigma ) $,
$\pi_i^{(\theta )}(\tau ,\vec \sigma ) $, $\tilde \phi (\tau ,\vec
\sigma ) $, $\pi_{\tilde \phi}(\tau ,\vec \sigma ) $, $R_{\bar
a}(\tau ,\vec \sigma ) $, $\Pi_{\bar a}(\tau ,\vec \sigma ) $. We
shall ignore global problems about the validity of the gauge fixing
constraints everywhere in $M^4$: our results will in general be
valid only locally.
\medskip

The {\it CMC gauges} ($\Sigma_{\tau}$ has constant mean curvature
${}^3\tilde K(\tau ,\vec \sigma ) = const.$) \cite{23} are those
associated to the gauge fixing $\pi_{\tilde \phi}(\tau ,\vec \sigma
) \approx - \sgn\, {{c^3}\over {12\pi\, G}}\, \times const.$. See
Ref.\cite{26} for the existence of surfaces of prescribed mean
curvature in asymptotically flat space-times.\medskip

The CMC gauge fixing $\pi_{\tilde \phi}(\tau ,\vec \sigma ) = -
\sgn\, {{c^3}\over {12\pi\, G}}\, {}^3\tilde K(\tau ,\vec \sigma )
\approx 0$ identifies the special gauge in which the simultaneity
and Cauchy hyper-surfaces $\Sigma_{\tau}$ are the CMC hyper-surfaces
with ${}^3\tilde K(\tau ,\vec \sigma ) \approx 0$.
\medskip

We shall not use the special CMC gauge $\pi_{\tilde \phi}(\tau ,\vec
\sigma ) \approx 0$, but we shall consider the class of gauges with
given trace of the extrinsic curvature ${}^3\tilde K(\tau ,\vec
\sigma ) \approx \sgn\, K(\tau ,\vec \sigma )$, so that
$\pi_{\phi}(\tau ,\vec \sigma ) \approx - {{c^3}\over {12\pi\, G}}\,
 K(\tau ,\vec \sigma )$, to see the dependance of the dynamics on
the shape of the simultaneity surfaces $\Sigma_{\tau}$, namely on
the convention chosen for clock synchronization.
\medskip

Let us remember that the gauge fixings determining the lapse and
shift functions are obtained by requiring the $\tau$ -constancy of
the gauge fixings determining $\pi_{\tilde \phi}$ and $\theta^n$.

\bigskip

The {\it 3-orthogonal gauges}  correspond to the gauge fixings
$\theta^n(\tau ,\vec \sigma ) \approx 0$ and imply

\begin{eqnarray*}
 {}^3{\bar e}_{(a)r} &=& {\tilde \phi}^{1/3}\, \delta_{ra}\,
 Q_a,\qquad
 {}^3{\bar e}^r_{(a)} = {\tilde \phi}^{-1/3}\, \delta_{ra}\,
 Q_a^{-1},\qquad
 {}^3g_{rs} = {\tilde \phi}^{2/3}\,  Q^2_a \, \delta_{rs},\qquad
 {}^3g^{rs} =  {\tilde \phi}^{-2/3}\, Q_a^{-2}\, \delta_{rs},\nonumber \\
 &&{}\nonumber \\
 {}^3{\tilde {\bar \pi}}^r_{(a)} &=& {\tilde \phi}^{-1/3}\, \Big[
 \delta_{ra}\, Q_a\, (\tilde \phi\, \pi_{\tilde \phi} +  \sum_{\bar b}\,
 \gamma_{\bar ba}\, \Pi_{\bar b}) -
  \sum_i\, Q^{-1}_r\,
 {{\epsilon_{ari}\, \pi_i^{(\theta )}}\over {Q_r\, Q^{-1}_a -
 Q_a\, Q^{-1}_r}} \Big],
 \end{eqnarray*}

\bea
  {}^3{\tilde K}_{rs} &=& \sgn\, {{4\pi\, G}\over {c^3}}\, {\tilde \phi}^{-1/3}\,
  \Big( \delta_{rs}\, Q^2_r\,
 [2\, \sum_{\bar b}\, \gamma_{\bar br}\, \Pi_{\bar b} -
\tilde \phi\, \pi_{\tilde \phi}] + 2\, Q_r\, Q_s\, {{\sum_i\,
\epsilon_{rsi}\, \pi_i^{(\theta )}}\over { Q_s\,
Q^{-1}_r - Q_r\, Q^{-1}_s }} \Big). \nonumber \\
 &&{}
 \label{II13}
 \eea

\medskip

 The expression of the super-Hamiltonian constraint and of the
 weak ADM energy in the 3-orthogonal gauges is given in Eqs. (\ref{IV2})
 and (\ref{IV3}).

 \bigskip

In Ref.\cite{25} there is the expression for the {\it 3-normal
gauges with respect to the origin of 3-coordinates} and for the
completely fixed {\it ADM 4-coordinate gauge} used for the ADM
post-Newtonian limit in Refs. \cite{27} (it is a CMC gauge).  In the
York canonical basis these gauge fixings are {\it algebraic
equations} for $\pi_{\tilde \phi}(\tau ,\vec \sigma )$ but {\it
first-order elliptic partial differential equations} for the three
Euler angles $\theta^n(\tau ,\vec \sigma )$.

\bigskip

Instead the family of {\it harmonic gauges}, defined by adding the 4
gauge-fixing constraints $\chi^A = \sum_B\, \partial_B\, \Big(N\,
{}^3e\, g^{AB}\Big) = 0$ to the secondary first-class constraints
and used both in theoretical studies \cite{28} and in the
post-Newtonian approximation \cite{29}, belongs to a different class
of gauges at the Hamiltonian level. Their gauge fixings  are neither
algebraic conditions nor elliptic equations defined on a single
instantaneous 3-space $\Sigma_{\tau}$.

By using the first half (\ref{a10}) of the Hamilton equations (the
kinematical connection between velocities and phase space variables)
associated with the Dirac Hamiltonian (\ref{a7}), the gauge-fixing
constraints $\chi_A(\tau ,\vec \sigma ) \approx 0$ can be rewritten
as  four Hamiltonian gauge fixings {\it explicitly depending upon
the four Dirac multipliers} $\lambda_n = \partial_{\tau}\, n$ and
$\lambda_{\vec n\, (a)} = \partial_{\tau}\, {\bar n}_{(a)}$ (see
Eqs.(6.4) of Ref.\cite{25} for their expression in the York
canonical basis). These unconventional Hamiltonian constraints
($\chi^{\tau} \approx 0$ does not define a CMC gauge) are four
coupled equations for $\pi_{\phi}$ and $\theta^i$ in terms of
$\phi$, $R_{\bar a}$, $\Pi_{\bar a}$, $n$, $\lambda_n =
\partial_{\tau}\, n$, ${\bar n}_{(a)}$, $\lambda_{\vec n (a)} =
\partial_{\tau}\, {\bar n}_{(a)}$.

The stability of these gauge fixings requires to impose
$\partial_{\tau}\, {\tilde \chi}_a(\tau ,\vec \sigma ) \approx 0$
and $\partial_{\tau}\, {\tilde \chi}_{\tau}(\tau ,\vec \sigma )
\approx 0$. In this way we get four equations for the determination
of $n$ and ${\bar n}_{(a)}$. But these are not equations of the
"elliptic" type like with ordinary gauge fixings. They are coupled
equations depending upon $n$, $\partial_r\, n$, $\partial_{\tau}\,
n$, $\partial^2_{\tau}\, n$ and ${\bar n}_{(a)}$, $\partial_r\,
{\bar n}_{(a)}$, $\partial_{\tau}\, {\bar n}_{(a)}$,
$\partial^2_{\tau}\, {\bar n}_{(a)}$, namely {\it hyperbolic}
equations like Eq.(\ref{VI3}). As a consequence there is a {\it
problem of initial conditions not only for $R_{\bar a}$ but also for
the lapse and shift functions of the harmonic gauge}. Each possible
set of initial values should correspond to a different completely
fixed harmonic gauge, since once we have a solution for $n$ and
${\bar n}_{(a)}$ the corresponding Dirac multipliers are determined
by taking their $\tau$-derivative.

\section{The Super-Momentum Constraints and their Solution.}

\subsection{The Super-Momentum Constraints.}

By using the results of Ref.\cite{3} for the transformation property
${}^3\omega_{r(a)(b)} = [R\, {}^3{\bar \omega}_r\, R^T + R\,
\partial_r\, R^T]_{(a)(b)}$ of the spin connection (\ref{b17}) under
O(3)- rotations and Eqs. (\ref{II7}) and (\ref{II11}), the
super-momentum constraints (\ref{a4}) in presence of matter, to be
solved in $\pi_i^{(\theta )}(\tau ,\vec \sigma )$, take the form
(${\bar D}_{r(a)(b)}$ is the covariant derivative for
$\alpha_{(a)}(\tau ,\vec \sigma ) = 0$)

\begin{eqnarray*}
 {\cal H}_{(a)} &=& {\cal H}^{(o)}_{(a)} - {}^3e^v_{(a)}\, {\cal
 M}_v = \sum_c\, R_{(a)(c)}\, {\bar {\cal H}}_{(c)} \approx
 \sum_c\, R_{(a)(c)}\, {\tilde {\bar H}}_{(a)} \approx 0,\nonumber \\
 &&{}\nonumber  \\
  {\cal H}^{(o)}_{(a)} &{\buildrel {def}\over =}& \sum_{rb}\,  D_{r(a)(b)}\,
  {}^3{\tilde \pi}^r_{(b)} = \sum_r\, \partial_r\,
 {}^3{\tilde \pi}^r_{(a)} - \sum_{rbc}\, \epsilon_{(a)(b)(c)}\, {}^3\omega_{r(b)}\,
 {}^3{\tilde \pi}^r_{(c)} =\nonumber \\
 &=& \sum_{rc}\, \partial_r\, [R_{(a)(c)}\,
 {}^3{\bar \pi}^r_{(c)}] +
  \sum_{rbc}\, [R\, {}^3{\bar \omega}_r\, R^T + R\, \partial_r\,
 R^T]_{(a)(b)}\, R_{(b)(c)}\, {\bar \pi}^r_{(c)} =\nonumber \\
 &=& \sum_{rc}\, R_{(a)(c)}\, \Big[\partial_r\, {\bar \pi}^r_{(c)} + \sum_d\, {}^3{\bar
 \omega}_{r(c)(d)}\, {\bar \pi}^r_{(d)}\Big] \approx \sum_{vc}\, R_{(a)(c)}\, {}^3{\bar
 e}^v_{(c)}\, {\cal M}_v,\nonumber \\
 &&{}\nonumber \\
 {\bar {\cal H}}_{(a)} &{\buildrel {def}\over =}& \sum_r\, \partial_r\, {}^3{\bar
 \pi}^r_{(a)} + \sum_{rb}\, {}^3{\bar \omega}_{r(a)(b)}\, {}^3{\bar
 \pi}^r_{(b)} - \sum_v\, {}^3{\bar e}^v_{(a)}\, {\cal M}_v
 \approx
 \end{eqnarray*}

\begin{eqnarray*}
 &\approx& {\tilde {\bar {\cal H}}}_{(a)} \, {\buildrel {def}\over
 =}\,\,  \sum_{rb}\, {\bar D}_{r(a)(b)}\, {}^3{\tilde {\bar \pi}}^r_{(b)}
 - \sum_v\, {}^3{\bar e}^v_{(a)}\, {\cal M}_v =\nonumber \\
 &&{}\nonumber \\
 &=& \sum_{rb}\, \Big[\delta_{ab}\, \partial_r +
{1\over 2}\, \sum_{ucd}\, [\delta_{ac}\, \delta_{bd} -
\delta_{ad}\, \delta_{bc}]\,  V_{ud}(\theta^n)\nonumber \\
  && \Big[Q_c\, Q^{-1}_d\, \Big({1\over 3}\,
  [V_{uc}(\theta^n)\, \partial_r\, ln\, \tilde \phi - V_{rc}(\theta^n)\,
  \partial_u\, ln\, \tilde \phi ] +\nonumber \\
  &+& \sum_{\bar b}\, \gamma_{\bar bc}\, [V_{uc}(\theta^n)\,
  \partial_r\, R_{\bar b} - V_{rc}(\theta^n)\, \partial_u\,
  R_{\bar b}] + \partial_r\, V_{uc}(\theta^n) - \partial_u\,
  V_{rc}(\theta^n)\Big) +\nonumber \\
  &+& {1\over 2}\, \sum_{ve}\, Q^2_e\, Q^{-1}_d\, Q^{-1}_c\,
  V_{vc}(\theta^n)\, V_{re}(\theta^n)\nonumber \\
  && \Big({1\over 3}\,[V_{ue}(\theta^n)\, \partial_v\, ln\, \tilde \phi -
  V_{ve}(\theta^n)\, \partial_u\, ln\, \tilde \phi ] +\nonumber \\
  &+&  \sum_{\bar b}\, \gamma_{\bar be}\, [V_{ue}(\theta^n)\,
  \partial_v\, R_{\bar b} - V_{ve}(\theta^n)\, \partial_u\,
  R_{\bar b}] + \partial_v\, V_{ue}(\theta^n) - \partial_u\,
  V_{ve}(\theta^n)\Big)\Big]\,\,\, \Big]
  \end{eqnarray*}

\bea
 &&\Big[{\tilde \phi}^{-1/3}\, \Big( V_{rb}(\theta^n)\, Q^{-1}_b\,
 ( \tilde \phi\, \pi_{\tilde \phi} +  \sum_{\bar c}\, \gamma_{\bar cb}\,
 \Pi_{\bar c}) +\nonumber \\
 &+& \sum_f^{f \not= b}\, \sum_{twi}\, Q^{-1}_f\, {{V_{rf}(\theta^n)\,
 \epsilon_{bft}\, V_{tw}(\theta^n)}\over {Q_f\, Q^{-1}_b - Q_b\,
 Q^{-1}_f}}\, B_{iw}(\theta^n)\, \pi^{(\theta )}_i \Big) \Big] -\nonumber \\
 &-& {\tilde \phi}^{-1/3}\, \sum_v\, V_{va}(\theta^n)\, Q^{-1}_v\, {\cal M}_v
 \quad \approx 0.
 \label{III1}
 \eea

\subsection{Their Solution.}

The solution of Eqs.(\ref{III1}) in terms of the matter mass-current
${\cal M}_r$ is

\begin{eqnarray*}
 {}^3{\tilde {\bar \pi}}^r_{(a)}(\tau ,\vec \sigma ) &\approx&
 g^r_{(a)}(\tau ,\vec \sigma ) - \int d^3\sigma_1\,
 \sum_c\, {\bar \zeta}^r_{(a)(c)}(\vec \sigma ,{\vec \sigma}_1;\tau )\,
 J_{(c)}(\tau ,{\vec \sigma}_1) =\nonumber \\
 &{\buildrel {def}\over =}& \sum_b\, {}^3{\bar e}_{(b)}^r(\tau ,\vec \sigma )\,
 \Big[g_{(a)(b)} + j_{(a)(b)}\Big](\tau ,\vec \sigma ),
 \end{eqnarray*}

 \begin{eqnarray*}
 J_{(a)} &{\buildrel {def}\over =}& \sum_r\, {}^3{\bar e}^r_{(a)}\, {\cal M}_r =
 {\tilde \phi}^{-1/3}\, \sum_r\, V_{ra}(\theta^n)\, Q^{-1}_a\, {\cal
 M}_r =\nonumber \\
 &=& \sum_{sd}\, {\bar D}_{s(a)(d)}\, \sum_b\, {}^3{\bar
  e}^s_{(b)}\, j_{(d)(b)},\nonumber \\
 &&{}\nonumber \\
  j_{(a)(b)}(\tau ,\vec \sigma ) &=& - \sum_{rc}\, {}^3{\bar
 e}_{(b)r}(\tau ,\vec \sigma )\, \int d^3\sigma_1\,
 {\bar \zeta}^r_{(a)(c)}(\vec \sigma ,{\vec \sigma}_1;\tau )\,
 J_{(c)}(\tau ,{\vec \sigma}_1) =\nonumber \\
  &=& - \sum_{rc}\, \Big[{\tilde \phi}^{1/3}\, V_{rb}(\theta^n)\,
  Q_b\Big](\tau ,\vec \sigma )\,
  \int d^3\sigma_1\, {\bar \zeta}^r_{(a)(c)}(\vec \sigma ,{\vec
  \sigma}_1;\tau |\theta^n, \phi ,R_{\bar a}]\nonumber \\
  && \sum_s\, \Big[{\tilde \phi}^{-1/3}\, V_{sc}(\theta^n)\, Q^{-1}_c\,
  {\cal M}_s\Big](\tau ,{\vec  \sigma}_1),
 \end{eqnarray*}

 \begin{eqnarray*}
 &&g^r_{(a)} = \sum_b\, g_{(a)(b)}\, {}^3e^r_{(b)},\qquad
g_{(a)(b)} = \sum_r\, g^r_{(a)}\, {}^3{\bar e}_{(b)r},
 \nonumber \\
  &&{}\nonumber\\
 &&g_{((a)(b))} = {1\over 2}\, (g_{(a)(b)} + g_{(b)(a)}),\qquad
 g_{[(a)(b)]} = {1\over 2}\, (g_{(a)(b)} - g_{(b)(a)}),
 \end{eqnarray*}

\begin{eqnarray*}
 \Rightarrow\,\, \sum_r\, {}^3{\bar e}_{(b)r}\, {}^3{\tilde {\bar \pi}}^r_{(a)} &=&
 g_{(a)(b)} + j_{(a)(b)} = g_{((a)(b))} + j_{((a)(b))} +
 g_{[(a)(b)]} + j_{[(a)(b)]} =\nonumber \\
 &=&  \delta_{ab}\, (\tilde \phi\, \pi_{\tilde \phi} +
 \sum_{\bar b}\, \gamma_{\bar ba}\, \Pi_{\bar b}) +
  \sum_{twi}\, {{\epsilon_{abt}\, V_{tw}(\theta^n)\, B_{iw}(\theta^n)\,
 \pi_i^{(\theta )}}\over {Q_b\, Q^{-1}_a - Q_a\,
 Q^{-1}_b}} = \nonumber \\
 &=& \sum_r\, {}^3{\bar e}_{((a)r}\, {\tilde {\bar
 \pi}}_{(b))}^r,\nonumber \\
 &&{}\nonumber \\
 \Rightarrow && g_{[(a)(b)]} = - j_{[(a)(b)]},
 \end{eqnarray*}

\bea
  \sum_{rb}\, {\bar D}_{r(a)(b)}\, g^r_{(b)}(\tau ,\vec \sigma ) &=&
 \Big[\sum_r\, \partial_r\, g^r_{(a)} + \sum_{rb}\, {}^3{\bar \omega}_{r(a)(b)}\,
 g^r_{(b)}\Big](\tau ,\vec \sigma ) = 0,
 \label{III2}
 \eea

\noindent where the Green function ${\bar \zeta}^r_{(a)(b)}$ of Ref.
\cite{2} is given in Eq.(\ref{d1}) of Appendix D and $g^r_{(a)}$ are
{\it zero modes of the covariant divergence with the covariant
derivative} ${\bar D}_{r(a)(b)}$.

\bigskip

Then Eq.(\ref{II6}), evaluated in the reduced phase space of the
Schwinger time gauge  $\varphi_{(a)}(\tau ,\vec \sigma ) \approx 0$,
$\alpha_{(a)}(\tau ,\vec \sigma ) \approx 0$,  implies

\bea
 \pi^{(\theta )}_i(\tau ,\vec \sigma ) &=&
  - \sum_{lmra}\, \Big[A_{mi}(\theta^n)\, \epsilon_{mlr}\,
 {}^3{\bar e}_{(a)l}\, {}^3{\tilde {\bar \pi}}^r_{(a)}\Big](\tau ,\vec \sigma )
 \approx\nonumber \\
 &&{}\nonumber \\
 &\approx&  - \sum_{lmra}\, \Big[A_{mi}(\theta^n)\, \epsilon_{mlr}\,
 {\tilde \phi}^{1/3}\, V_{la}(\theta^n)\, Q_a\Big](\tau ,\vec \sigma )\,
 \Big[g^r_{(a)}(\tau ,\vec \sigma  ) -\nonumber \\
 &-& \int d^3\sigma_1\,
 \sum_c\, {\bar \zeta}^r_{(a)(c)}(\vec \sigma ,{\vec \sigma}_1;\tau )\,
 J_{(c)}(\tau ,{\vec \sigma}_1)\Big] =\nonumber \\
 &=& - \sum_{lmrab}\, \Big[A_{mi}(\theta^n)\, \epsilon_{mlr}\,
  V_{la}(\theta^n)\, V_{rb}(\theta^n)\, Q_a\, Q^{-1}_b\Big](\tau
  ,\vec \sigma )\, \Big[g_{(a)(b)}
  + j_{(a)(b)}\Big](\tau ,\vec \sigma  ).\nonumber \\
  &&{}
 \label{III3}
 \eea

Let us remark that, in absence of matter ($J_{(a)}(\tau ,\vec \sigma
) = 0$) and with the choice $g^r_{(a)}(\tau ,\vec \sigma ) = 0$ for
the homogeneous solution, we get $\pi_i^{(\theta )}(\tau ,\vec
\sigma ) \approx 0$.

\subsection{The Zero Modes of the Covariant Divergence.}

We have now to see whether we can find the zero modes
$g^r_{(a)}(\tau ,\vec \sigma )$ of the operator ${\bar
D}_{r(a)(b)}$.

\medskip

If we put $g^r_{(a)} = \sum_c\, {}^3{\bar e}^r_{(c)}\, g_{(a)(c)}$
in the second of Eqs.(\ref{II6}), we can determine $g_{(a)(a)}$
since, by using Eqs.(\ref{III2}), we have

\begin{eqnarray*}
 \Lambda_a\, P^a &=& \sum_r\, {}^3{\bar e}_{(a)r}\, \sum_c\,
 {}^3{\bar e}^r_{(c)}\, [g_{(a)(c)} + j_{(a)(c)}] = g_{(a)(a)} +
 j_{(a)(a)}, \nonumber \\
 &&{}\nonumber \\
 g_{(a)(a)}(\tau ,\vec \sigma ) &=& \Big[\Lambda_a\, P^a -
 j_{(a)(a)}\Big](\tau ,\vec \sigma ) =
   \Big[\tilde \phi\, \pi_{\tilde \phi}
 +  \sum_{\bar b}\, \gamma_{\bar ba}\, \Pi_{\bar b}\Big](\tau
 ,\vec \sigma ) +\nonumber \\
 &+& \sum_{sc}\, \Big[{\tilde \phi}^{1/3}\, V_{sa}(\theta^n)\, Q_a\, \int d^3\sigma_1\,
 {\bar \zeta}^s_{(a)(c)}(\vec \sigma ,{\vec \sigma}_1;\tau )
 \nonumber \\
 &&\sum_w\, \Big[{\tilde \phi}^{-1/3}\, V_{wc}(\theta^n)\, Q^{-1}_c\,
 {\cal M}_w\Big](\tau ,{\vec \sigma}_1),\nonumber \\
 &&{}\nonumber \\
 &&and\,\, for\,\, a \not= b,
 \end{eqnarray*}

 \bea
 g_{(a)(b)} &=& g_{((a)(b))} + g_{[(a)(b)]} = g_{((a)(b))} -
 j_{[(a)(b)]} =\nonumber \\
 &=& g_{((a)(b))} + {1\over 2}\, {\tilde \phi}^{1/3}(\tau ,\vec \sigma )\,
 \sum_{rc}\, \int d^3\sigma_1\nonumber \\
 &&\Big(\Big[V_{rb}(\theta^n)\, Q_b\Big](\tau ,\vec \sigma )\,
 {\bar \zeta}^r_{(a)(c)} -
  \Big[V_{ra}(\theta^n)\, Q_a\Big](\tau ,\vec \sigma )\,
 {\bar \zeta}^r_{(b)(c)} \Big)(\vec \sigma ,{\vec \sigma}_1;\tau )\nonumber \\
 &&\sum_s\, \Big[{\tilde \phi}^{-1/3}\, V_{sc}(\theta^n)\, Q^{-1}_c\,
 {\cal M}_s\Big](\tau ,{\vec \sigma}_1).
 \label{III4}
 \eea

As a consequence, the homogeneous equation for $g^r_{(a)}$ in
Eq.(\ref{III2}) gives rise to an inhomogeneous equation for
$g_{((a)(b))}$ with $a \not= b$

\begin{eqnarray*}
 &&\sum_{rb}\, {\bar D}_{r(a)(b)}(\tau ,\vec \sigma )\,
 \sum_c\, \Big[{}^3{\bar e}^r_{(c)}\, g_{(b)(c)}
 \Big](\tau ,\vec \sigma ) = 0,\nonumber \\
 &&{}\nonumber \\
 &&{}\nonumber \\
 &&\sum_{rbc}^{b\not= c}\, {\bar D}_{r(a)(b)}(\tau ,\vec \sigma )\,
 \Big[{}^3{\bar e}^r_{(c)}\, g_{((b)(c))}\Big](\tau ,\vec \sigma ) =\nonumber \\
 &=& - \sum_{rb}\, {\bar D}_{r(a)(b)}(\tau ,\vec \sigma )\,
 \Big[{}^3{\bar e}^r_{(b)}\, g_{(b)(b)} + \sum_c^{c \not= b}\,
 {}^3{\bar e}^r_{(c)}\, g_{[(b)(c)]}\Big](\tau ,\vec \sigma ),\nonumber \\
 &&{}\nonumber \\
 &&\Downarrow
 \end{eqnarray*}

\bea
  &&g_{((a)(b))}(\tau ,\vec \sigma ) =
 g^{hom}{}_{((a)(b))}(\tau,\vec{\sigma})+ \int d^3\sigma_1\, \sum_d\,
 {\bar {\cal G}}_{((a)(b))(d)} (\vec \sigma ,{\vec \sigma}_1;\tau )
 \nonumber \\
 && \sum_{re}\, {\bar D}_{r(d)(e)}(\tau ,{\vec \sigma}_1)\,
 \Big[{}^3{\bar e}^r_{(e)}\, g_{(e)(e)} + \sum_f^{f \not= e}\,
 {}^3{\bar e}^r_{(f)}\, g_{[(e)(f)]}\Big](\tau ,{\vec \sigma}_1),
 \nonumber \\
 &&{}\nonumber \\
 &&{}\nonumber \\
 &&\sum_{r bc}^{b \not= c}\, \Big[{\bar D}_{r(a)((b)}\,\, {}^3{\bar
 e}^r_{(c))}\Big](\tau ,\vec \sigma )\, {\bar {\cal G}}_{((b)(c))
 (d)} (\vec \sigma ,{\vec \sigma}_1;\tau ) = - \delta_{ad}\,
 \delta^3(\vec \sigma ,{\vec \sigma}_1),\nonumber \\
 &&{}\nonumber \\
 &&\sum_{r bc}^{b \not= c}\, \Big[{\bar D}_{r(a)((b)}\,\, {}^3{\bar
 e}^r_{(c))}\Big](\tau ,\vec \sigma )\, g^{hom}_{((b)(c))}(\tau
 ,\vec \sigma ) = 0,
 \label{III5}
 \eea

\noindent where ${\bar {\cal G}}_{((a)(b))(d)} (\vec \sigma ,{\vec
\sigma}_1;\tau )$ is the Green function of the operator $\sum_r\,
\Big[{\bar D}_{r(a)((b)}\,\, {}^3{\bar e}^r_{(c))}\Big]{|}_{b \not=
c}$ [see Eq.(\ref{d2})  and its Minkowski limit  given in
Eq.(\ref{d3})]. In Eq.(\ref{III5}) $g^{hom}_{((a)(b))}(\tau ,\vec
\sigma )$ is an arbitrary zero mode of the operator $\sum_r\,
\Big[{\bar D}_{r(a)((b)}\,\, {}^3{\bar e}^r_{(c))}\Big]{|}_{b \not=
c}$. There are as many independent such zero modes as independent
zero modes $g^r_{(a)}$ of ${\bar D}_{r(a)(b)}$.
\medskip

The presence of this second Green function is a consequence of using
the Darboux canonical basis identified by the York map in the
solution of the super-momentum constraints (so that the Green
function (\ref{d1}) is no more sufficient).

\medskip

Eqs.(\ref{III5}) imply

 \begin{eqnarray*}
 &&\sum_{ca}^{c \not= a}\, \Big[{}^3{\bar e}^r_{(c)}\, g_{((a)(c))}
 \Big](\tau ,\vec \sigma ) = \int d^3\sigma_1\, \sum_d\,
 {\bar \eta}^r_{(a)(d)}(\vec \sigma ,{\vec \sigma}_1;\tau )\times
\nonumber \\
 &&\sum_{se}\, \Big[{\bar D}_{s(d)(e)}\, \Big({}^3{\bar e}^s_{(e)}\,
 g_{(e)(e)} + \sum_c^{c \not= e}\, {}^3{\bar e}^s_{(c)}\,
 g_{[(e)(c)]}\Big)\Big](\tau ,{\vec \sigma}_1) +
 \end{eqnarray*}

 \bea
 &+& \sum_c^{c\not= a}\, \Big[{}^3{\bar e}^r_{(c)}\,
 g^{hom}_{((a)(c))}\Big](\tau ,\vec \sigma ),\qquad with\nonumber \\
 &&{}\nonumber \\
 &&{}\nonumber \\
  &&{\bar \eta}^r_{(a)(d)}(\vec \sigma ,{\vec \sigma}_1;\tau )\,
  {\buildrel {def}\over =}\,\,
 \sum_c^{c \not= a}\, {}^3{\bar e}^r_{(c)}(\tau ,\vec \sigma )\,
{\bar {\cal G}}_{((a)(c))(d)} (\vec \sigma ,{\vec \sigma}_1;\tau ),
 \label{III6}
 \eea

\noindent so that a lengthy calculation, given in Eqs.(C1)-(C4) of
Appendix C of Ref.\cite{25}, and the use of Eq.(\ref{III2}) for
$j_{(a)(b)}$ and of Eq.(\ref{III3}) for $\pi_i^{(\theta )}$ lead to
the following form for the solution of the super-momentum
constraints, which explicitly shows its non-uniqueness being defined
modulo the zero modes of the covariant divergence

\bea
 && \pi_i^{(\theta )}(\tau ,\vec \sigma ) \approx -
\sum_{lmrab}\, \Big[A_{mi}(\theta^n)\, \epsilon_{mlr}\, {}^3{\bar
e}_{(a)l}\,{}^3{\bar e}^r_{(b)}\Big](\tau ,\vec \sigma )\,
\Big[g_{(a)(b)} + j_{(a)(b)}\Big](\tau ,\vec \sigma )=\nonumber\\
&&\nonumber\\
&&{}\nonumber \\
 &=& - \sum_{lmrab}\, \Big[A_{mi}(\theta^n)\,
\epsilon_{mlr}\, V_{la}(\theta^n)\, V_{rb}(\theta^n)\, Q_a\,
Q^{-1}_b\Big](\tau ,\vec \sigma )\nonumber \\
&&\nonumber\\
&&\Big[\,\,\, g^{hom}{}_{((a)(b))}(\tau,\vec{\sigma}) - \sum_d\,
\int d^3\sigma_1\, \bar{\cal G}_{((a)(b))(d)}(\vec{\sigma},
\vec{\sigma}_1;\tau)\, \Big[ {\tilde \phi}^{-1/3}\, \sum_w\,
V_{wd}(\theta^n)\, Q^{-1}_d\, {\cal M}_w -\nonumber \\
 &-&  \sum_{s,e}\, \bar{D}_{s(d)(e)}\, V_{se}(\theta^n )\,
 {\tilde \phi}^{-1/3}\, Q^{-1}_e\, \Big(\tilde \phi\, \pi_{\tilde \phi} + \sum_{\bar{b}}\,
 \gamma_{\bar{b}{e}}\, \Pi_{\bar{b}}
\Big)\Big](\tau ,\vec{\sigma}_1) -\nonumber\\
 &&\nonumber\\
 &-& \sum_{ec}^{c\neq e}\, \int d^3\sigma_1\, \left(
\delta_{ac}\, \delta_{be}\, \delta^3(\vec{\sigma},\vec{\sigma_1}) +
\sum_{d,s}\, \bar{\cal G}_{((a)(b))(d)}(\vec{\sigma},
\vec{\sigma}_1; \tau)\, \Big[\bar{D}_{s(d)(e)}\, V_{sc}(\theta^n
)\, {\tilde \phi}^{-1/3}\, Q^{-1}_c\Big](\tau ,\vec{\sigma}_1)\right)\nonumber\\
 &&\nonumber\\
 &&\sum_{r,f}\,\int d^3\sigma_2\, \frac{1}{2}\, \Big[\,
\Big(V_{re}(\theta^n)\, {\tilde \phi}^{1/3}\, Q_e\Big)(\tau
,\vec{\sigma_1})\,
\bar{\zeta}^r_{(c)(f)}(\vec{\sigma}_1,\vec{\sigma}_2;\tau)\,
+\nonumber\\
 &&\nonumber\\
 &+& \Big(V_{rc}(\theta^n)\, {\tilde \phi}^{1/3}\, Q_c\Big)(\tau ,\vec{\sigma_1})\,
\bar{\zeta}^r_{(e)(f)}(\vec{\sigma}_1,\vec{\sigma}_2;\tau)\,\Big]\,\,
 \Big(\phi^{-2}\, \sum_w\, V_{wf}(\theta^n)\, Q^{-1}_f\, {\cal
M}_w\Big)(\tau ,\vec{\sigma}_2)\,\,\, \Big],
 \label{III7}
 \eea

\noindent Since we have $Q_a(\tau ,\vec \sigma )\, Q^{-1}_b(\tau
,\vec \sigma ) = \Lambda_a(\tau ,\vec \sigma )\, \Lambda_b^{-1}(\tau
,\vec \sigma ) \rightarrow_{r \rightarrow \infty}\,\, 1$, the
leading order of Eq.(\ref{III7}) is $\epsilon_{iab}\, f_{(ab)} = 0$,
consistently with the vanishing of $\pi_i^{(\theta )}(\tau ,\vec
\sigma )$ at spatial infinity.
\medskip

In the final expression of Eq.(\ref{III7}) we made explicit the
symmetry in $e$ and $c$.
\medskip
Let us remark that  the zero modes $g^{hom}_{((a)(b))}(\tau ,\vec
\sigma )$  may be written in the following form

\bea
 g^{hom}_{((a)(b))}(\tau ,\vec \sigma ) &=&
 \sum_{ec}^{c\neq e}\, \int d^3\sigma_1\, \Big( \delta_{c(a}\,
\delta_{b)e}\, \delta^3(\vec{\sigma},\vec{\sigma_1})
 +\nonumber \\
 &+& \sum_{ds} \, \bar{\cal G}_{((a)(b))(d)}(\vec{\sigma},\vec{\sigma}_1;\tau)\,
 {1\over 2}\, \Big[\bar{D}_{s(d)(e)}\, V_{sc}(\theta^n)\, {\tilde \phi}^{-1/3}\,
Q^{-1}_c +\nonumber \\
 &+& \bar{D}_{s(d)(c)}\, V_{se}(\theta^n)\,
{\tilde \phi}^{-1/3}\, Q^{-1}_e\Big](\tau ,\vec{\sigma}_1)\Big)\,
{\tilde g}_{(ec)}(\tau ,{\vec \sigma}_1),
 \label{III8}
 \eea

\noindent with an arbitrary ${\tilde g}_{(ec)}(\tau ,\vec \sigma )$
symmetric in $e$ and $c$.

\bigskip

If we put Eq.(\ref{III7}) into Eq.(\ref{II13}), we get the
expression of ${}^3{\tilde {\bar \pi}}^r_{(a)}(\tau ,\vec \sigma ) =
\Big[\sum_b\, {}^3{\bar e}^r_{(b)}\Big(g_{(a)(b)} + j_{(a)(b)}
\Big)\Big](\tau ,\vec \sigma )$ when restricted to the solution of
the super-momentum constraints (see Eq.(3.10) of Ref.\cite{25}).

 \bigskip

Let us remark that both $\pi_i^{(\theta )}(\tau ,\vec \sigma )$ and
${}^3{\tilde {\bar \pi}}^r_{(a)}(\tau ,\vec \sigma )$ are defined
modulo homogeneous solutions

\noindent i) of Eq.(\ref{d1}): ${\bar \zeta}^r_{(a)(b)} \mapsto
{\bar \zeta}^r_{(a)(b)} + {\bar \zeta}^{(hom)r}_{(a)(b)}$ with
$\sum_{rb}\, {\bar D}_{r(a)(b)}(\tau ,\vec \sigma )\, {\bar
\zeta}^{(hom)r}_{(b)(c)}(\vec  \sigma ,{\vec \sigma}_1; \tau ) = 0$;
\medskip

\noindent ii) of Eq.(\ref{d2}): ${\bar {\cal G}}_{((a)(b))(d)}
\mapsto {\bar {\cal G}}_{((a)(b))(d)} + {\bar {\cal
G}}_{((a)(b))(d)}^{(hom)}$ with $\sum_{rbc}^{b \not= c}\, \Big[{\bar
D}_{r(a)((b)}\, {}^3{\bar e}^r_{(c))}\Big](\tau ,\vec \sigma )\,$
${\bar {\cal G}}_{((a)(b))(d)}^{(hom)}(\vec \sigma ,{\vec \sigma}_1;
\tau ) = 0$.
\medskip

While these freedoms are connected to the {\it choice of the initial
data} \footnote{Like the choice of the retarded, advanced or
symmetric Green functions in the  Lienard-Wiechert solution for an
electro-magnetic field coupled to charged matter.}, Eq.(\ref{III8})
connects the freedom ${\tilde g}_{(ab)}(\tau ,\vec \sigma )$  to the
{\it existence in general relativity of the zero modes}
$g^{hom}_{((a)(b))}(\tau ,\vec \sigma )$ of the operators $\sum_r\,
\Big[{\bar D}_{r(a)((b)}\, {}^3{\bar e}^r_{(c))}\Big]{|}_{b \not=
c}$, see Eq.(\ref{III5}), and $g^r_{(a)}(\tau ,\vec \sigma ) =
\sum_b\, \Big(\Big[g^{hom}_{((a)(b))} + ....\Big]\, {}^3{\bar
e}^r_{(b)}\Big)(\tau ,\vec \sigma )$ of ${\bar D}_{r(a)(b)}$. The
associated residual gauge freedom is connected to the group of
3-diffeomorphisms and not to a Lie group like in Yang-Mills theory
(see the discussion in Section V).
\bigskip

Due to the distributional nature of the Green function ${\bar {\cal
G}}$ (whose flat limit is given in Eq.(E4) of Ref.\cite{25}),
required by the Shanmugadhasan canonical transformations (\ref{II3})
and (\ref{II9}), to avoid distributional problems in the expression
of the super-hamiltonian constraint and in the weak ADM energy we
need a suitable choice  of the arbitrary zero mode
$g^{hom}_{((a)(b))}$, i.e. of ${\tilde g}_{(ce)}$, which will be
done elsewhere when we will solve the theory in the weak field
limit.

\section{The Final Form of the Super-Hamiltonian Constraint and of
the Weak ADM Energy in a Completely Fixed  3-Orthogonal Schwinger
Time Gauge.}

As already said, the gauge fixings $\varphi_{(a)}(\tau ,\vec \sigma
) \approx 0$, $\alpha_{(a)}(\tau ,\vec \sigma ) \approx 0$, whose
$\tau$-constancy implies $\lambda_{\varphi (a)}(\tau ,\vec \sigma )
= 0$ and $\lambda_{\alpha (a)}(\tau ,\vec \sigma ) = 0$, define a
{\it special Schwinger time gauge}. We assume to have eliminated
these variables by going to Dirac brackets (we go on to denote them
as Poisson brackets).

\medskip

Let us now consider the {\it completely fixed 3-orthogonal Schwinger
time gauge} defined by the gauge fixing

\beq
 \theta^i(\tau ,\vec \sigma ) \approx 0,\qquad
 \pi_{\tilde \phi}(\tau ,\vec \sigma ) \approx -
{{c^3}\over {12\pi\, G}}\,  K(\tau ,\vec \sigma ),
 \label{IV1}
 \eeq

\noindent i.e. with ${}^3\tilde K(\tau ,\vec \sigma ) \approx \sgn\,
K(\tau ,\vec \sigma )$. In this way the function $K(\tau ,\vec
\sigma )$ {\it will show explicitly how the dynamics depends on the
shape of $\Sigma_{\tau}$, namely on the convention for the
synchronization of clocks}.

\subsection{The Super-Hamiltonian Constraint.}

In these  completely fixed 3-orthogonal Schwinger time gauge, by
using Eq.(\ref{III8}) (or Eq.(3.10) of Ref.\cite{25}) for the
solution of the super-momentum constraints, the super-Hamiltonian
constraint (\ref{II12}), i.e. the Lichnerowicz equation for $\tilde
\phi (\tau ,\sigma ) = \phi^6(\tau ,\vec \sigma )$, becomes
\footnote{The steps to get Eq.(\ref{IV2}) from Eq.(\ref{II12}) are
described in Eqs. (B6), (B7), (B8), (C5), (C6), (C7), (C9), (D10),
(D11) of Appendices B, C and D of Ref.\cite{25}. We do not give the
final expression (B8) for $Z(\tau ,\vec \sigma )$, because, being
rather complicated, its explicit dependence on $\tilde \phi$ (either
algebraic or under integrals) is irrelevant for the general
discussion.}

\begin{eqnarray*}
{\cal H}(\tau ,\vec \sigma )
 &\approx& \sgn\, \Big[{{c^3}\over {16\, \pi\, G}}\, \Big({\tilde \phi}^{1/6}\,\,
 [- 8\, \hat \triangle [R_{\bar a}] + {}^3\hat R[R_{\bar a}]\,\, ]\, {\tilde
 \phi}^{1/6}\, \Big)(\tau ,\vec \sigma ) -\nonumber \\
 &-& {1\over c}\, {\cal M}(\tau ,\vec \sigma ) + {{c^3}\over {24\,
 \pi\, G}}\, \Big(\tilde \phi\, K^2\Big)(\tau ,\vec \sigma ) - {{4\pi\, G}\over
 {c^3}}\, \Big({\tilde \phi}^{-1}\, \sum_{\bar b}\, \Pi^2_{\bar b}\Big)(\tau
 ,\vec \sigma ) -\nonumber \\
 &-& {{4\pi\, G}\over {c^3}}\, {\tilde \phi}^{-1}(\tau ,\vec \sigma )\,
 Z(\tau ,\vec \sigma )\Big],
\end{eqnarray*}

 \beq
  Z(\tau ,\vec \sigma  ) = Z_{\theta}(\tau ,\vec \sigma ){|}_{\theta =0}
 = \sum_{ab}^{b\neq a}\,{\cal
S}^2_{((a)(b))}\mid_{\theta=0}(\tau ,\vec \sigma ),
 \label{IV2}
 \eeq

\noindent where the $\tilde \phi$-dependent function $Z$ takes into
account the contribution of the $\Gamma$-$\Gamma$ term ${\cal S}$,
containing the inertial potentials present in the non-inertial
frame. In particular $Z$ contains terms linear in $K(\tau ,\vec
\sigma )$.

\bigskip

Its unknown solution is a functional $\tilde \phi (\tau ,\vec \sigma
| R_{\bar a}, \Pi_{\bar a}, {\cal M}, {\cal M}_r, K]$ of the
gravitational tidal degrees of freedom $R_{\bar a}$, $\Pi_{\bar a}$,
of both the mass density ${\cal M}$ and mass-current density ${\cal
M}_r$ of the matter and of the gauge parameter $K$ describing the
shape of the hyper-surfaces $\Sigma_{\tau}$ (the convention for
clock synchronization and for the Cauchy surface) having the given
3-orthogonal 3-coordinate system.

\medskip

Even if Eq.(\ref{IV2}) is a non-linear integro-differential equation
for $\tilde \phi$, the presence of the  Laplace-Beltrami operator on
$\Sigma_{\tau}$ (with its associated theory of harmonic functions)
suggests the  plausibility that the assumed behavior $\tilde \phi
(\tau ,\vec \sigma )\, \rightarrow_{r \rightarrow\infty}\,\, 1 +
O(r^{-1})$  at spatial infinity will identify a unique solution.

\subsection{The Weak ADM Energy.}

In these completely fixed 3-orthogonal Schwinger time gauges, by
using Eq.(C9) of Ref.\cite{25} for the term quadratic in the momenta
and Eq.(\ref{c1}) for the $\Gamma$-$\Gamma$ term ${\cal S}$, the
weak ADM energy (\ref{II12}) becomes [$Z$ has been defined in
Eq.(\ref{IV2})]

\bea
 E_{ADM} &=& \int d^3\sigma\, \Big[ {\cal M} -
{{c^4}\over {24\, \pi\, G}}\, \tilde \phi\, K^2 + {{4\pi\, G}\over
 {c^2}}\, {\tilde \phi}^{-1}\, \Big[\sum_{\bar b}\, \Pi^2_{\bar b} + Z\Big]
-\nonumber \\
 &-& {{c^4}\over {16\pi\, G}}\,
{\tilde \phi}^{1/3}\, \sum_a\, Q^{- 2}_a\, \Big(20\, (\partial_a\,
ln\, {\tilde \phi}^{1/6} )^2 - 4\, \sum_r\, (\partial_r\, ln\,
{\tilde \phi}^{1/6} )^2 + 8\,
\partial_a\, ln\, {\tilde \phi}^{1/6}\, \sum_{\bar b}\, \gamma_{\bar ba}\,
\partial_a\, R_{\bar b} -\nonumber \\
 &-& 2\, \sum_r\, \partial_r\, ln\, {\tilde \phi}^{1/6}\, \sum_{\bar b}\, (\gamma_{\bar ba}
 + \gamma_{\bar br})\, \partial_r\, R_{\bar b} +
 (\sum_{\bar b}\, \gamma_{\bar ba}\, \partial_a\, R_{\bar b})^2
 +\nonumber \\
 &+& \sum_{\bar b}\, (\partial_a\, R_{\bar b})^2 - \sum_r\,
 (\sum_{\bar b}\, \gamma_{\bar br}\, \partial_r\, R_{\bar b})\,
 (\sum_{\bar c}\, \gamma_{\bar ca}\, \partial_r\, R_{\bar c})\Big)
 \Big](\tau ,\vec \sigma ),
 \label{IV3}
  \eea
\medskip

While the terms coming from the $\Gamma$-$\Gamma$ term ${\cal S}$
describe the relativistic version of the standard inertial
potentials in this 3-coordinate system (expressed as functions of
$\tilde \phi (\tau ,\vec \sigma )$ and of the tidal effects $R_{\bar
a}(\tau ,\vec \sigma )$), the first line contains the dependence on
the inertial potential $K(\tau ,\vec \sigma )$ (both explicitly,
$K^2$, and inside $Z$) describing the choice of the instantaneous
3-space.

\subsection{The Rest-Frame Conditions and the Spin of the
3-Universe.}

By using Eqs.(\ref{a6}), the rest-frame conditions for the
3-universe \cite{3} in the York basis and in the completely fixed
gauge (\ref{IV1}) are [${\cal S}_{((a)(b))}{|}_{\theta =0}$ is the
contribution of the $\Gamma$-$\Gamma$ term]

\bea
 {P}^r_{ADM} &\approx&
 P^r_{ADM}\mid_{\theta=0} = \int d^3\sigma\, \Big[{\tilde \phi}^{-2/3}\,
Q^{- 2}_r\,  {\cal M}_r\Big](\tau ,\vec \sigma ) -\nonumber \\
 &-& \int d^3\sigma\, \sum_{uv}\, \Big[{\tilde \phi}^{-2/3}\, \Big(
 \delta_{uv}\, Q^{- 2}_v\, (  \sum_{\bar b}\, \gamma_{\bar bv}\,
 \Pi_{\bar b} - {{c^3}\over {12\pi\, G}}\, \tilde \phi\, K) +\nonumber  \\
 &+& Q_u\, Q_v\,
 {\cal S}_{((u)(v))}\mid_{\theta=0} \Big)\,
 \Big(\delta_{ru}\, ({1\over 3}\, \partial_v\, ln\, \tilde \phi + \sum_{\bar
 c}\, \gamma_{\bar cr}\, \partial_v\, R_{\bar c}) +\nonumber \\
 &+& \delta_{rv}\, ({1\over 3}\, \partial_u\, ln\, \tilde \phi + \sum_{\bar
 c}\, \gamma_{\bar cr}\, \partial_u\, R_{\bar c}) -\nonumber  \\
 &-& \delta_{uv}\, ({1\over 3}\, \partial_r\, ln\, \tilde \phi + \sum_{\bar
 c}\, \gamma_{\bar cr}\, \partial_r\, R_{\bar c})\, Q_u\, Q^{-1}_v
 \Big)\Big](\tau ,\vec \sigma ) \approx 0.
 \label{IV4}
 \eea

Like in special relativity \cite{30}, these 3 first class
constraints imply that 3 variables $q^r_{ADM}[R_{\bar a}, \Pi_{\bar
a},...]$, describing the internal canonical 3-center of mass of the
3-universe, are {\it gauge variables} (they describe the
arbitrariness in the choice of the observer used as origin of the
3-coordinates on $\Sigma_{\tau}$). As shown in Ref.\cite{30}, the
natural gauge fixings to eliminate them is to ask the vanishing of
the boost generators in Eqs.(\ref{a6}), i.e. $J^{\tau r}_{ADM}
\approx 0$. These conditions imply the vanishing of the internal
M$\o$ller 3-center of energy so that it can be shown that then
Eqs.(\ref{IV4}) imply also $q^r_{ADM} \approx 0$. In this way the
observer may be identified with the decoupled 4-center of mass (more
exactly with the covariant non-canonical Fokker-Pryce 4-center of
inertia) of the universe.

\bigskip

See Eq.(4.5) of Ref.\cite{25} for the expression of the spin
(\ref{a6}) of the 3-universe in the rest frame.

\section{The shift and lapse functions}

Let us now determine the lapse and shift functions of the completely
fixed 3-orthogonal Schwinger time gauges of Section IV. In all the
equations of this Section $\tilde \phi (\tau ,\vec \sigma )$ should
be replaced by the  unknown solution $\tilde \phi (\tau ,\vec \sigma
| R_{\bar a}, \Pi_{\bar a}, {\cal M}, {\cal M}_r, K]$ of the
Lichnerowicz equation (\ref{IV2}).

\subsection{The 3-orthogonal Gauges, the Shift Functions and a
Generalized Gribov Ambiguity.}

If we use Eqs. (\ref{II12}) for the super-Hamiltonian constraint and
the weak ADM energy in Eqs.(\ref{a7}) and (\ref{a9}),  the Dirac
Hamiltonian in the York basis can be written in the form

\bea
 H_D &=&  \int d^3\sigma\, \Big[(1 + n)\, {\cal M}\Big](\tau ,\vec
 \sigma ) -\nonumber \\
 &-& {{c^4}\over {16\pi\, G}}\, \int d^3\sigma\, \Big[{\cal S} + n\, {\tilde \phi}^{1/6}\,
 \Big( - 8\,   {\hat \triangle} + {}^3{\hat R}\,  \Big)\, {\tilde
 \phi}^{1/6}
 \Big](\tau ,\vec \sigma ) +\nonumber \\
 &+& {{2\pi\, G}\over {c^2}}\, \int d^3\sigma\, \Big[
 (1 + n)\, {\tilde \phi}^{-1}\, \Big(- 3\, (\tilde \phi\, \pi_{\tilde \phi})^2
 + 2\, \sum_{\bar b}\, \Pi^2_{\bar b} +\nonumber \\
 &+& 2\, \sum_{abtwiuvj}\, {{\epsilon_{abt}\, \epsilon_{abu}\, V_{tw}(\theta^n)\,
 B_{iw}(\theta^n)\, V_{uv}(\theta^n)\, B_{jv}(\theta^n)\, \pi_i^{(\theta )}\,
 \pi_j^{(\theta )}}\over {\Big[Q_a\, Q^{-1}_b - Q_b\,
 Q^{-1}_a \Big]^2}} \Big) \Big](\tau ,\vec \sigma ) +\nonumber \\
 &+& \int d^3\sigma\, \Big[\sum_a\, {\bar n}_{(a)}\,
 {\tilde {\bar {\cal H}}}_{(a)} +
 \lambda_n\, \pi_n + \sum_a\, \lambda_{\vec n(a)}\,
 \pi_{\vec n(a)} \Big](\tau ,\vec \sigma ),
 \label{V1}
 \eea

\noindent where the super-momentum constraints ${\tilde {\bar {\cal
H}}}_{(a)}(\tau ,\vec \sigma ) \approx 0$ are given by
Eqs.(\ref{III1}), ${\cal S}$ in Eq.(B1) and ${}^3\hat R[\theta^n,
R_{\bar a}]$ in Eq.(B2) of Appendix B of Ref.\cite{25} [for
$\theta^n(\tau ,\vec \sigma ) \approx 0$ see Eqs. (\ref{c1}) and
(\ref{c2})]. The Hamilton equations of motion must be evaluated with
this Dirac Hamiltonian (see Section VI) and the solution
(\ref{III8}) [or Eq.(3.10) of Ref.\cite{25}] of the super-momentum
constraints {\it can be used only after having evaluated the Poisson
brackets}.

\bigskip

With the Hamiltonian ({\ref{V1}) the time-constancy of the gauge
fixings $\theta^n(\tau ,\vec \sigma ) \approx 0$, determining the
shift functions, implies [here $\approx$ means by using these gauge
fixings, the one of Eq.(\ref{IV1}) and the solution (\ref{c3}) of
the super-momentum constraints]

\begin{eqnarray*}
 \partial_{\tau}\, \theta^i(\tau ,\vec \sigma
 ) &=& \{ \theta^i(\tau ,\vec \sigma ), H_D\} \approx\nonumber \\
 &=& \sum_a\, \int d^3\sigma_1\, {\bar n}_{(a)}(\tau ,{\vec
 \sigma}_1)\, \{ \theta^i(\tau ,\vec \sigma ), {\tilde {\bar {\cal
 H}}}_{(a)}(\tau ,{\vec \sigma}_1 )\} + \nonumber \\
 &+& \{ \theta^i(\tau ,\vec \sigma ), E_{ADM} - \sgn\, c\, \int
 d^3\sigma_1\, n(\tau ,{\vec \sigma}_1 )\, {\cal H}(\tau
 ,{\vec \sigma}_1 ) \} \approx \nonumber \\
 &\approx& \sum_a\, \int d^3\sigma_1\, {\bar n}_{(a)}(\tau ,{\vec \sigma}_1)\,
 {\tilde Z}_{(a)i}(\tau ,{\vec \sigma}_1)\, \delta^3(\vec \sigma
 ,{\vec \sigma}_1) - W_i(\tau ,\vec \sigma ) =
 \end{eqnarray*}

\begin{eqnarray*}
 &=& \sum_a\, Z_{(a)i}(\tau ,\vec \sigma )\, {\bar n}_{(a)}(\tau
 ,\vec \sigma ) - W_i(\tau ,\vec \sigma ) \approx 0,
 \end{eqnarray*}

\bea
 {\tilde Z}_{(a)i}(\tau ,\vec \sigma ) &{\buildrel {def}\over =}&
 \sum_{rb}\, \Big[\Big(\delta_{ab}\, \partial_{1r} +
  \epsilon_{(a)(b)(c)}\, {}^3{\bar \omega}_{r(c)}\Big)\,
 G^{(o)r}_{(b)i}\Big](\tau ,\vec \sigma ) =
  \sum_{rb}\, \Big[{\bar D}_{r(a)(b)}\, G^{(o)r}_{(b)i}\Big](\tau
 ,\vec \sigma ),\nonumber \\
 Z_{(a)i}(\tau ,\vec \sigma ) &{\buildrel {def}\over =}&
  - \sum_{arb} \Big[ \,G^{(o)r}_{(b)i}\,
 {\bar D}_{r(b)(a)}\,\Big](\tau ,\vec \sigma ),\nonumber \\
 W_i(\tau,\vec{\sigma}) &{\buildrel {def}\over =}&
 - \Big[{{\delta}\over {\delta\, \pi^{(\theta )}_i(\tau ,\vec
 \sigma )}}\, \Big(E_{ADM} + \int d^3\sigma_1\,
 \Big(- \sgn\, c\, n\, {\cal H}\Big)(\tau
 ,{\vec \sigma}_1)\Big)\Big],
   \label{V2}
  \eea

\noindent where we used Eq.(\ref{III1}) and Eq.(\ref{II13}) but with
the notation  $G^{(o)r}_{(a)i} = - {\tilde \phi}^{-1/3}\, Q^{-1}_r\,
{{\epsilon_{rai}}\over {Q_r\, Q^{-1}_a - Q_a\, Q^{-1}_r}}$ of
Eq.(\ref{b16}).

\medskip

Eqs.(\ref{III8}) and (\ref{c3}) imply  the following  expression for
$W_i(\tau,\vec{\sigma})$ [with the substitution $ \tilde \phi\,
\pi_{\tilde \phi} \mapsto - {{c^3}\over {12\pi\, G}}\, \tilde \phi\,
K$ into the functions $F_{(ab)}$ of Eq.(\ref{c3}) in accord with
Eq.(\ref{IV1}): let us remark that these functions have a linear
dependence on $\pi_{\tilde \phi}$, namely they {\it know the sign}
of $K(\tau ,\vec \sigma )$]

\bea
 &&W_i(\tau,\vec{\sigma}) = \Big[- {{8\pi\, G}\over {c^2}}\,
 {\tilde \phi}^{-1}\, \Big(1+n\Big)\,
\sum_{abj}\, {{\epsilon_{abi}\, \epsilon_{abj}\, \pi_j^{(\theta
)}}\over {[Q_a\, Q^{-1}_b - Q_b\, Q^{-1}_a]^2}}
 \Big](\tau,\vec{\sigma}) \approx\nonumber\\
 &&\nonumber\\
 &\approx& - {{16\pi\, G}\over {c^2}}\, {\tilde \phi}^{-1}(\tau,\vec{\sigma})\,
\Big(1+n(\tau,\vec{\sigma})\Big)\, \sum_{ab}\,
{{\epsilon_{iab}}\over {Q_a\, Q^{-1}_b - Q_b\, Q^{-1}_a}}(\tau ,\vec
\sigma )\,
F_{(ab)}(\tau ,\vec \sigma ).\nonumber\\
 &&{}
 \label{V3}
 \eea

\medskip

Since Eqs.(\ref{b16}) with $H^{(o)}_{(b)ri} =  {1\over 2}\,
\epsilon_{bri}\, {\tilde \phi}^{1/3}\, Q_r\, (Q_r\, Q^{-1}_b - Q_b\,
Q^{-1}_r)$, imply $\sum_{ar}\, H^{(o)}_{(a)rj}\,\, G^{(o)r}_{(a)i} =
\delta_{ij}$ and $\sum_{ri}\, H^{(o)}_{(b)ri}\,\, G^{(o)r}_{(a)i}\,
= \delta_{ab}$, we get

\begin{eqnarray*}
 &&\sum_a\, Z_{(a)i}\, {\bar n}_{(a)} =
 \sum_{ar}\, \Big[ \sum_b\, G^{(o)r}_{(b)i}\, (\partial_r\,
 \delta_{ba} + \sum_c\, \epsilon_{(b)(a)(c)}\, {}^3{\bar \omega}_{r(c)})\Big]\,
 {\bar n}_{(a)} =\nonumber \\
 &&= \sum_s\, \partial_s\, \Big(\sum_a\, G^{(o)s}_{(a)i}\, {\bar n}_{(a)}\Big)
 -\end{eqnarray*}

 \bea
 &-& \sum_{sd}\, \Big(\sum_{bc}\, \epsilon_{(b)(c)(d)}\, G^{(o)s}_{(b)i}\,
 {}^3{\bar \omega}_{s(c)} + \partial_s\,
 G^{(o)s}_{(d)i}\Big)\,\,
 \sum_a\, \Big(\sum_{rj}\, H^{(o)}_{(d)rj}\, G^{(o)r}_{(a)j}\Big)\,
 {\bar n}_{(a)} =\nonumber \\
 &&{}\nonumber \\
 && {\buildrel {def}\over =}\, \sum_{jr}\, {\tilde D}_{rij}\,
 \Big(\sum_a\, G^{(o)r}_{(a)j}\, {\bar n}_{(a)}\Big) \approx W_i,
 \label{V4}
 \eea

\noindent where we have introduced the modified covariant derivative
operator

 \bea
 &&{\tilde D}_{rij} = \delta_{ij}\, \partial_r - T_{rij},\qquad
 Z_{(a)i} = - \sum_{arb}\,G^{(o)r}_{(b)i}\,
 {\bar D}_{r(b)(a)} = \sum_{jr}\, {\tilde D}_{rij}\,
 G^{(o)r}_{(a)j}, \nonumber \\
 &&{}\nonumber \\
 &&T_{rij} = \sum_{sd}\, \Big(\sum_{bc}\, \epsilon_{(b)(c)(d)}\,
 G^{(o)s}_{(b)i}\,\, {}^3{\bar \omega}_{s(c)} +
 \partial_s\, G^{(o)s}_{(d)i}\Big)\, H^{(o)}_{(d)rj}
 =\nonumber \\
 &&= - \sum_{abcds}\, \epsilon_{abc}\, \epsilon_{asi}\,
 \epsilon_{crj}\, \epsilon_{sbd}\, Q_r\, Q^{-1}_d\, \partial_d\,
 ({1\over 3}\, ln\, \tilde \phi + \Gamma^{(1)}_s)\, {{Q_r\, Q^{-1}_c - Q_c\, Q^{-1}_r}
 \over {Q_s\, Q^{-1}_a - Q_a\, Q^{-1}_s}} -\nonumber \\
 &-& \sum_{as}\, \epsilon_{asi}\, \epsilon_{arj}\, Q_r\,
 Q^{-1}_s\, {{Q_r\, Q^{-1}_a - Q_a\, Q^{-1}_r}\over {Q_s\, Q^{-1}_a
  - Q_a\, Q^{-1}_s}}\, \Big[ \partial_s\, ({1\over 3}\, ln\, \tilde \phi +
  \Gamma^{(1)}_s) +\nonumber \\
 &+& {{Q_r\, Q^{-1}_a + Q_a\, Q^{-1}_r}\over {Q_s\, Q^{-1}_a
 - Q_a\, Q^{-1}_s}}\, \partial_s\, (\Gamma^{(1)}_s -
 \Gamma^{(1)}_a)\Big],\nonumber \\
 &&{}\nonumber \\
 &&Q_a = e^{\Gamma^{(1)}_a},\qquad  \Gamma^{(1)}_a = \sum_{\bar
 a}\, \gamma_{\bar aa}\, R_{\bar a},\qquad \sum_a\,
 \Gamma^{(1)}_a = 0.
 \label{V5}
 \eea

By using the Green function of the operator ${\tilde D}_{rij}$,
defined in Eq.(\ref{d4}), the shift functions can be determined and
have the following expression as functions of the lapse function and
of the dynamical variables (it is linear in $n$)

 \bea
 {\bar n}_{(a)}(\tau ,\vec \sigma ) &=& {\cal N}_{(a)}(\tau ,\vec
 \sigma | \tilde \phi , n, R_{\bar a}, \Pi_{\bar a}, {\cal M}, {\cal M}_r, K]
 =\nonumber \\
 &=& f_{(a)}(\tau ,\vec \sigma )
 + \sum_{ri}\, H^{(o)}_{(a)ri}(\tau ,\vec \sigma )\, \sum_j\, \int d^3\sigma_1\,
 {\tilde \zeta}^r_{ij}(\vec \sigma ,{\vec \sigma}_1;\tau )\,
 W_j(\tau ,{\vec \sigma}_1),\nonumber \\
 &&{}\nonumber \\
 &&{}\nonumber \\
 \sum_a \Big[Z_{(a)i}\,
 f_{(a)}\Big](\tau ,\vec \sigma ) &{\buildrel {def}\over =}& -
 \sum_{abr}\, \Big[G^{(o)r}_{(b)i}\, {\bar D}_{r(b)(a)}\, f_{(a)}\Big](\tau
 ,\vec  \sigma ) = \sum_{rj}\, \Big[{\tilde D}_{rij}\, \sum_a\,
 G^{(o)r}_{(a)j}\, f_{(a)}\Big](\tau ,\vec \sigma ) =\nonumber \\
 &&{\buildrel {def}\over =}\, \sum_{jr}\, \Big[{\tilde D}_{rij}\,
 {\tilde f}^r_j\Big](\tau ,\vec \sigma ) = 0,
  \label{V6}
  \eea

\noindent with $f_{(a)}(\tau ,\vec \sigma ) = \sum_{rj}\,
\Big[H^{(o)}_{(a)rj}\, {\tilde f}^r_j\Big](\tau ,\vec \sigma )$ {\it
zero modes} of $Z_{(a)i}(\tau ,\vec \sigma )$, namely with ${\tilde
f}^r_j(\tau ,\vec \sigma ) = \Big[\sum_a\, G^{(o)r}_{(a)j}\,
f_{(a)}\Big](\tau ,\vec \sigma )$ {\it zero modes} of the operator
${\tilde D}_{rij}(\tau ,\vec \sigma )$.
\medskip

Naturally the shift functions are defined modulo homogeneous
solution of Eqs.(\ref{d4}): ${\tilde \zeta}^r_{ij} \mapsto {\tilde
\zeta}^r_{ij} + {\tilde \zeta}^{(hom)r}_{ij}$ with $\sum_{rj}\,
{\tilde D}_{rij}(\tau ,\vec \sigma )\, {\tilde
\zeta}^{(hom)r}_{jk}(\vec \sigma ,{\vec \sigma}_1; \tau ) = 0$.
Again this is a problem of choice of the initial conditions.

\medskip

When the operator $Z_{(a)i}$ has zero modes, $\sum_a\,
\Big[Z_{(a)i}\, f_{(a)}\Big](\tau ,\vec \sigma ) = 0$, also its
adjoint operator ${\tilde Z}_{(a)i} = \sum_{rb}\, {\bar
D}_{r(a)(b)}\, G^{(o)r}_{(b)i}$, appearing in Eq.(\ref{V2}), has
zero modes $h_i(\tau ,\vec \sigma )$, i.e. $\sum_i\, \Big[{\tilde
Z}_{(a)i}\, h_i\Big](\tau ,\vec \sigma ) = 0$. Then Eq.(\ref{V2})
imposes the following restriction on $W_i$

\beq
 \int d^3\sigma\,
\sum_i\, W_i(\tau ,\vec \sigma )\, h_i(\tau ,\vec \sigma ) = 0.
 \label{V7}
 \eeq

\bigskip

Therefore, in the 3-orthogonal gauges there is a {\it residual gauge
freedom or generalized Gribov ambiguity} in the determination of the
shift functions associated to the zero modes of the operator
$Z_{(a)i}$ (or of ${\tilde D}_{rij}$). Since in general relativity
the Gauss law constraints ${\tilde {\bar {\cal H}}}_{(a)}(\tau ,\vec
\sigma ) \approx 0$ are a suitable reformulation of the constraints
$\Theta_r(\tau ,\vec \sigma ) \approx 0$ generating
3-diffeomorphisms [see after Eq.(\ref{a4})], this extra ambiguity is
connected to the group of 3-diffeomorphisms.

\medskip

Given $\theta^i(\tau ,\vec \sigma )$ and its modification
$\theta^i(\tau ,\vec \sigma ) + \sum_a\, Z^{'}_{ai}(\tau ,\vec
\sigma )\, \beta_a(\tau ,\vec \sigma )$ ($Z_{ai} =
Z^{'}_{ai}{|}_{\theta^r = 0}$) induced by a (modified)
3-diffeomorphism [generated by $\int d^3\sigma\, \sum_a\,
\beta_a(\tau ,\vec \sigma )\, {\tilde  {\bar H}}_{(a)}(\tau ,\vec
\sigma )$], we have that the vanishing of the first as a gauge
fixing, $\theta^i(\tau ,\vec \sigma ) \approx 0$, implies the
vanishing also of the modified one $\theta^i(\tau ,\vec \sigma ) +
\sum_a\, Z_{ai}(\tau ,\vec \sigma )\, \beta_a(\tau ,\vec \sigma )
\approx 0$ when $\beta_a$ coincides with one  of the zero modes
$f_{(a)}$. The same happens in Yang-Mills (YM) theory: for certain
gauge potentials arising from special connections the gauge fixing
$\vec \partial \cdot {\vec A}_a \approx 0$ (so that ${\vec A}_a =
{\vec A}_{a\perp}$) implies that there are transformed gauge
potentials ${\vec A}^U_a = {\vec A}_a + U^{-1}\, {\vec D}^{(\vec
A)}\, U$, $U = e^{i\alpha}$ also satisfying $\vec
\partial \cdot {\vec A}_a^U \approx 0$ if $K({\vec A}_{\perp})\,
\alpha = 0$ [$K({\vec A}_{\perp}) = - \vec
\partial \cdot {\vec D}^{({\vec A}_{\perp})}$]. In these cases the
connection originating the gauge potential ${\vec A}_a$ has {\it
gauge symmetries} (stability subgroup of gauge transformations)
implying the existence of zero modes of the Faddeev-Popov operator
$K({\vec A}_{\perp})$ and of the operator $\triangle ({\vec
A}_{\perp}) = {\vec D}^{({\vec A}_{\perp})} \cdot {\vec D}^{({\vec
A}_{\perp})}$. This leads  to the Gribov ambiguity (see
Ref.\cite{12} for a review) \footnote{In the YM case the canonical
variables are $A^o_a$, $\pi^o_a$, ${\vec A}_a$, ${\vec \pi}_a$ and
the Gauss laws $\vec \partial \cdot {\vec \pi}_a \approx 0$ are
secondary constraints implied by the primary ones $\pi^o_a \approx
0$. In ordinary Sobolev spaces the Gribov ambiguity creates problems
in the analogue of Eq.(\ref{V2}), namely $\partial_{\tau}\, \vec
\partial \cdot {\vec A}_a \approx 0$, needed for the determination
of the gauge variables $A^o_a$'s. The YM constraint manifold is a
stratification of Gribov copies labeled by a winding number with the
different sectors  separated by Gribov horizons. In suitable
weighted Sobolev spaces \cite{11} the Faddeev-Popov does not have
zero modes, and there is no Gribov ambiguity (the connections with
gauge symmetries have a constant limit at spatial infinity not
allowed in these spaces) and we have that the only solution of
$\sum_b\,D_{rab}\, f_b = 0$ is $f_a = 0$. In these spaces is also
absent the other aspect of the Gribov ambiguity, i.e. the existence
of special field strengths stable ($F = F^U$) under a subgroup of
gauge transformations (the problem of {\it gauge copies}). However
also in YM, the absence of zero modes of the Faddeev-Popov operator
does not fix the Green function appearing in the solution of the
Gauss laws (the non-Abelian generalization of ${\bar
\zeta}^r_{(a)(b)}$ of Eq.(\ref{d1}) in flat space-time): there is
the usual freedom (connected to the choice of the initial data) in
the choice of homogeneous solutions.\hfill\break See Refs.
\cite{11,31} for the known results on the zero modes of operators
like $Z_{(a)i}$ and ${\tilde Z}_{(a)i}$ in the case of Yang-Mills
and Einstein equations.}.
\bigskip

Since the shift functions determine which points on different
$\Sigma_{\tau}$'s have the same numerical value of the chosen
3-orthogonal 3-coordinates $\vec \sigma$ (and then the inertial
gravito-magnetic potential), we see that, when $Z_{(a)i}$ has zero
modes, there are as many independent 3-orthogonal gauges, and,
therefore, non-inertial frames, as zero modes (each one with the
gauge freedom in the choice of $\pi_{\tilde \phi}$, i.e. in the form
of $\Sigma_{\tau}$). It is an open problem whether this generalized
Gribov ambiguity (gauge symmetries of the gauge variables
$\theta^i(\tau ,\vec \sigma )$, i.e. existence of a stability
subgroup of passive 3-diffeomorphisms, whose group-manifold has a
mathematical structure not yet under control), whose possibility was
noted in Ref.\cite{8} (see page 765), can be eliminated by a
suitable restriction of the function space like it happens in the YM
case (but here the gauge group is a Lie group with a well understood
theory of associated principal bundles) with the restriction to
suitable weighted Sobolev spaces \cite{11}. Since our class of
non-compact space-times does not admit \cite{10} asymptotically
vanishing Killing vectors, the only known result (see the first of
Refs.\cite{10}, pages 133-136) is that in weighted Sobolev spaces
suitable elliptic operators acting on the simultaneity surfaces
$\Sigma_{\tau}$ have no zero modes. If it is possible to apply these
results to the covariant divergence ${\bar D}_{r(a)(b)}$ and to the
operator $Z_{(a)i}$, given the assumed behavior ${\bar n}_{(a)}(\tau
,\vec \sigma )\, \rightarrow_{r \rightarrow \infty}\,\,
O(r^{-\epsilon})$, $\epsilon > 0$, at spatial infinity, also the
zero modes (\ref{III8}) would be absent in these function spaces.

\bigskip

The time-constancy of Eq.(\ref{V6}), i.e. the time-constancy of the
induced gauge fixings, determining the shift functions, determines
the Dirac multiplier $\lambda_{\vec n\, (a)}(\tau ,\vec \sigma )$,
since we have

\bea
 &&\partial_{\tau}\, \Big[{\bar n}_{(a)}(\tau ,\vec \sigma ) -
 f_{(a)}(\tau ,\vec \sigma ) - \sum_{ri}\, H^{(o)}_{(a)ri}(\tau ,\vec \sigma )\,
 \sum_j\, \int d^3\sigma_1\, {\tilde \zeta}^r_{ij}(\vec \sigma ,
 {\vec \sigma}_1;\tau )\, W_j(\tau ,{\vec \sigma}_1)\Big] \approx 0,\nonumber \\
 &&{}\nonumber \\
 &&{}\nonumber \\
  &&\lambda_{\vec n\, (a)}(\tau ,\vec \sigma ) \approx
{{\partial\, f_{(a)}(\tau ,\vec \sigma )}\over {\partial\, \tau}} +
\{ \sum_{ri}\, H^{(o)}_{(a)ri}(\tau ,\vec \sigma )\,
 \sum_j\, \int d^3\sigma_1\, {\tilde \zeta}^r_{ij}(\vec \sigma ,
 {\vec \sigma}_1;\tau )\, W_j(\tau ,{\vec \sigma}_1),
 H_D\}.\nonumber \\
 &&{}
 \label{V8}
 \eea

Therefore $\lambda_{\vec n(a)}(\tau ,\vec \sigma )$ inherits the
arbitrariness of ${\bar n}_{(a)}(\tau ,\vec \sigma )$.

\subsection{The Lapse Function in the  3-Orthogonal Schwinger
Time Gauges.}

The time constancy of the  other gauge fixing (\ref{IV1}), evaluated
with the Dirac Hamiltonian (\ref{V1}) determines the lapse function
[the calculations can be found in  Eqs.(D8) and (D9) of Appendix D
of Ref.\cite{25}]

\bea
 \partial_{\tau}&& \Big[\pi_{\tilde \phi}(\tau ,\vec \sigma
 ) + {{c^3}\over {12\pi\, G}}\, K(\tau ,\vec \sigma )\Big]=\nonumber \\
 &=& {{c^3}\over {12\pi\, G}}\,
 {{\partial\, K(\tau ,\vec \sigma )}\over  {\partial\, \tau}}
 + \{ \pi_{\tilde \phi}(\tau ,\vec \sigma ), H_D\} =\nonumber \\
 &&\nonumber\\
  &=& {{c^3}\over {12\pi\, G}}\,
 {{\partial\, K(\tau ,\vec \sigma )}\over  {\partial\, \tau}}
 -  {{\delta}\over {\delta\, \phi (\tau ,\vec \sigma )}}\nonumber \\
 &&\Big(E_{ADM} + \int d^3\sigma_1\, \Big[- \sgn\, c\, n\, {\cal H}
 + \sum_a\, {\cal N}_{(a)}[\tilde \phi ,n, R_{\bar a}, \Pi_{\bar a},
 {\cal M}, {\cal M}_r, K]\, {\tilde {\bar {\cal H}}}_{(a)}\Big](\tau
 ,{\vec \sigma}_1)\Big) \approx 0,\nonumber \\
 &&{}\nonumber \\
 &&\Downarrow\nonumber \\
  &&{}\nonumber \\
 n(\tau ,\vec \sigma ) &\approx& {\cal N}(\tau ,\vec \sigma |\tilde \phi
 , R_{\bar a}, \Pi_{\bar a}, {\cal M}, {\cal M}_r, K].
 \label{V9}
\eea

\bigskip
The explicit expression for this linear  integro-differential
equation for the lapse function is given in Eq.(5.9) of
Ref.\cite{25}.\medskip

It is impossible to judge whether Eq.(\ref{V9}), with the assumed
behavior $n(\tau ,\vec \sigma )\, \rightarrow_{r \rightarrow
\infty}\,\, O(r^{-(2 + \epsilon)})$, $\epsilon > 0$ at spatial
infinity, admits a further residual gauge freedom (ambiguity in the
determination of the proper time element $n(\tau ,\vec \sigma )\,
d\tau$ in each point of $\Sigma_{\tau}$, giving the packing of the
$\Sigma_{\tau}$'s  in the foliation), besides the generalized Gribov
ambiguity for the shift functions.

\bigskip

The time-constancy of the induced gauge fixing (\ref{V9}) determines
the Dirac multiplier $\lambda_n(\tau ,\vec \sigma )$

\bea
 &&{{\partial}\over {\partial \tau}}\, [n(\tau ,\vec \sigma ) -
 {\cal N}(\tau ,\vec \sigma |\tilde \phi
 , R_{\bar a}, \Pi_{\bar a}, {\cal M}, {\cal M}_r, K]\,\, ]
 \approx 0,\nonumber \\
 &&{}\nonumber \\
 &&\downarrow\nonumber \\
 &&{}\nonumber \\
 && \lambda_n(\tau ,\vec \sigma ) \approx \{{\cal N}(\tau ,\vec \sigma |\tilde \phi
 , R_{\bar a}, \Pi_{\bar a}, {\cal M}, {\cal M}_r, K], H_D \}.
 \label{V10}
 \eea

\section{Equations of Motion for the Tidal Effects $R_{\bar a}$
and $\Pi_{\bar a}$ in Schwinger Time Gauges.}

In this Section we shall consider the Hamilton equations in the York
basis both in arbitrary Schwinger time gauges and in a completely
fixed 3-orthogonal Schwinger time gauge.

\subsection{Equations of Motion in the York Basis.}

In the York canonical basis in an arbitrary Schwinger time gauge the
effective Dirac Hamiltonian is given in Eq.(\ref{V1}). As a
consequence the first half of Hamilton equations becomes [the
equations for $\partial_{\tau}\, \theta^i$ are written using the
results in Eq.(\ref{V2})]

\begin{eqnarray*}
 \partial_{\tau}\, n(\tau ,\vec \sigma ) &=& \lambda_n(\tau ,\vec
 \sigma ),\nonumber \\
  \partial_{\tau}\, {\bar n}_{(a)}(\tau ,\vec \sigma ) &=&
  \lambda_{\vec n (a)}(\tau ,\vec \sigma ),\nonumber \\
   \partial_{\tau}\, {\tilde \phi}^{1/6} (\tau ,\vec \sigma ) &=& \Big[-{{2\pi\,
   G}\over { c^2}}\, (1 + n)\, {\tilde \phi}^{-1/6}\, \pi_{\tilde \phi}
   -\nonumber \\
   &-&{1\over 6}\, {\tilde \phi}^{-1/6}\,  \sum_{rb}\,  V_{rb}(\theta^n)\,
  Q^{-1}_b\, \sum_a\, {\bar D}_{r(b)(a)}\, {\bar n}_{(a)}
 \Big](\tau ,\vec \sigma ), \nonumber \\
  \partial_{\tau}\, \theta^i(\tau ,\vec \sigma ) &=& \Big[{{8\pi\, G}\over {c^2}}\,
  (1 + n)\, {\tilde \phi}^{-1}\, \sum_{abtwuvj}\, {{\epsilon_{abt}\,
  \epsilon_{abu}\, V_{tw}(\theta^n)\, B_{iw}(\theta^n)\, V_{uv}(\theta^n)\,
  B_{jv}(\theta^n)\, \pi_j^{(\theta )}}\over {[Q_a\, Q^{-1}_b -
  Q_b\, Q^{-1}_b]^2}}  +\nonumber \\
  &+&\sum_a\,  Z_{(a)i}\, {\bar n}_{(a)}\Big](\tau ,\vec \sigma ),\nonumber \\
  \partial_{\tau}\, R_{\bar a}(\tau ,\vec \sigma ) &=&
  \Big[{{4\pi\, G}\over {c^2}}\, (1 + n)\, {\tilde \phi}^{-1}\, \Pi_{\bar a}
  -\nonumber \\
  &-& {\tilde \phi}^{-1/3}\, \sum_{rb}\, \gamma_{\bar ab}\, V_{rb}(\theta^n)\,
  Q^{-1}_b\, \sum_a\, {\bar D}_{r(b)(a)}\, {\bar n}_{(a)}\Big](\tau
  ,\vec \sigma ),
  \end{eqnarray*}

  \bea
  &&\Downarrow\nonumber \\
  &&{}\nonumber \\
  \Pi_{\bar a}(\tau ,\vec  \sigma ) &=& \Big[{{c^2}\over {4\pi\, G}}\,
  {{{\tilde \phi}^{2/3}}\over {1 + n}}\, \Big(\partial_{\tau}\, R_{\bar a} +
  {\tilde \phi}^{-1/3}\, \sum_{rb}\, \gamma_{\bar ab}\, V_{rb}(\theta^n)\,
  Q^{-1}_b\, \sum_a\, {\bar D}_{r(b)(a)}\, {\bar n}_{(a)}
  \Big)\Big](\tau ,\vec \sigma ),\nonumber \\
  \pi_{\tilde \phi}(\tau ,\vec \sigma ) &=& \Big[- {{ c^2}\over {2\pi\, G}}\,
 {{{\tilde \phi}^{-1/6}}\over {1 + n}}\, \Big(\partial_{\tau}\, {\tilde \phi}^{1/6} +
 {1\over 6}\, {\tilde \phi}^{-1/6}\, \sum_{rb}\, V_{rb}(\theta^n)\,
  Q^{-1}_b\, \sum_a\, {\bar D}_{r(b)(a)}\, {\bar n}_{(a)}
 \Big)  \Big](\tau ,\vec \sigma ).\nonumber \\
 &&{}
 \label{VI1}
 \eea

\noindent In the last two lines we have given $\Pi_{\bar a}$ and
$\pi_{\tilde \phi}$ in terms of the velocities and of the
configuration variables. Also $\pi_i^{(\theta )}$ could be expressed
in the same way, so that the solution (\ref{III8}) of the
super-momentum constraints could be transformed on a statement about
the velocities $\partial_{\tau}\, \theta^i$. As said in the previous
Section, the vanishing of these velocities become the equations for
the shift functions of 3-orthogonal gauges.

\medskip

The second half of Hamilton equations, to which the unsolved first
class constraints have to be added, is

\bea
 \partial_{\tau}\, \pi_{\tilde \phi}(\tau ,\vec \sigma ) &=& - {{\delta\,
 H_D}\over {\delta\, \tilde \phi (\tau ,\vec \sigma )}},\qquad
 {\cal H}(\tau ,\vec \sigma ) \approx 0, \nonumber \\
  \partial_{\tau}\, \pi_i^{(\theta )}(\tau ,\vec \sigma ) &=& - {{\delta\,
 H_D}\over {\delta\, \theta^i (\tau ,\vec \sigma )}},\qquad
 {\tilde {\bar {\cal H}}}_{(a)}(\tau ,\vec \sigma ) \approx 0, \nonumber \\
 \partial_{\tau}\, \Pi_{\bar a}(\tau ,\vec \sigma ) &=&
 - {{\delta\, H_D}\over {\delta\, R_{\bar a}(\tau ,\vec \sigma )}},
 \label{VI2}
 \eea

\noindent where the super-Hamiltonian and super-momentum constraints
have the forms given in  Eqs.(\ref{II12}) and (\ref{III1}),
respectively. The equation for $\partial_{\tau}\, \pi_{\tilde \phi}$
may be obtained by using Eq.(\ref{V9}).
\medskip

The content of the equations for $\partial_{\tau}\, \tilde \phi$ in
Eqs.(\ref{VI1}) and for $\partial_{\tau}\, \pi_i^{(\theta )}$ in
Eqs.(\ref{VI2}) is the preservation in time of the super-Hamiltonian
and super-momentum constraints, respectively.

\medskip

By using the expression of $\Pi_{\bar a}(\tau ,\vec \sigma )$ given
in Eqs.(\ref{VI1}) and Eq.(D12) of Appendix D  of Ref.\cite{25} for
$\delta\, H_D / \delta\, R_{\bar a}(\tau ,\vec \sigma )$, the
equations $\partial_{\tau}\, \Pi_{\bar a}(\tau ,\vec \sigma ) = -
\delta\, H_D / \delta\, R_{\bar a}(\tau ,\vec \sigma )$ assume the
following form [$Q_a = e^{\sum_{\bar a}\, \gamma_{\bar aa}\, R_{\bar
a}}$, $Q_1\, Q_2\, Q_3 = 1$]

\begin{eqnarray*}
 && \Big[\partial_{\tau}^2\, R_{\bar a} + \sum_{rs\bar b}\, A_{rs\bar a\bar b}\,
 \partial_r\, \partial_s\, R_{\bar b}\Big](\tau ,\vec \sigma )
 = \Big[\sum_{\bar b}\, B_{\bar a\bar b}\, \partial_{\tau}\, R_{\bar b} +
 \sum_{r\bar b\bar c}\, B_{r\bar a\bar b\bar c}\, \partial_{\tau}\, R_{\bar b}\,
 \partial_r\, R_{\bar c} +\nonumber \\
 &+& \sum_{rs\bar b\bar c}\, C_{rs\bar a\bar b\bar c}\,
 \partial_r\, R_{\bar b}\, \partial_s\, R_{\bar c} + \sum_{r\bar b}\,
 C_{r\bar a\bar b}\, \partial_r\, R_{\bar b} +
 F_{\bar a}\Big](\tau ,\vec \sigma ),
 \end{eqnarray*}

\bea
 && A_{rs\bar a\bar b}\qquad\qquad functions\, of\quad
  Q_a,\, \tilde \phi,\, n,\, \theta^i,\, \partial_u\, \tilde \phi,\, \partial_u\,
 n,\, \partial_u\, \theta^i,\, \partial_u\, \partial_v\, \tilde \phi,\,
 \partial_u\, \partial_v\, n,\, \partial_u\, \partial_v\,
 \theta^i,\nonumber \\
 &&{}\nonumber \\
 &&B_{\bar a\bar b},\, B_{r\bar a\bar b\bar c},\, C_{r\bar a\bar
 b},\, C_{rs\bar a\bar b\bar c}\qquad functions\, of\, the\,
 same\, variables\, and\, of
 \qquad \pi_{\tilde \phi}, \pi_i^{(\theta )}, {\bar n}_{(a)},
 \partial_u\, {\bar n}_{(a)},\nonumber \\
 &&{}\nonumber \\
 &&F_{\bar a}\qquad\qquad functions\, of\, the\, previous\,
 variables\, and\, of\quad  {\cal M},\, {\cal M}_v.
 \label{VI3}
 \eea

\medskip

The hyperbolic equations (\ref{VI3}) show explicitly that the
equations of motion for the two tidal degrees of freedom $R_{\bar
a}(\tau ,\vec \sigma )$ of the gravitational field {\it depend upon
the arbitrary gauge variables (the inertial effects)} $n$, ${\bar
n}_{(a)}$, $\theta^i$, $\pi_{\tilde \phi}$, and on the unknowns
$\tilde \phi$ and $\pi_i^{(\theta )}$ in the super-Hamiltonian and
super-momentum constraints. In particular the term in $\delta\, H_D
/ \delta\, R_{\bar a}(\tau ,\vec \sigma )$ coming from the
super-momentum constraints (\ref{III1}) (see Eq.(D12) of
Ref.\cite{25} for its expression in the 3-orthogonal gauges) {\it
depends linearly on} $\pi_{\tilde \phi}$: since its {\it sign} (i.e.
the sign of the trace of the extrinsic curvature of the simultaneity
surface) is not fixed, $\pi_{\tilde \phi}$ {\it describes a
relativistic inertial force which may vary from attractive to
repulsive} from a region of $\Sigma_{\tau}$ to another one with an
opposite sign of ${}^3K(\tau ,\vec \sigma )$.\medskip

Therefore, {\it to get a deterministic evolution we must go to a
completely fixed gauge}. The same holds for Einstein's equations,
but only at the Hamiltonian level it can be made explicit.

\bigskip

A naive background-independent linearization of Eqs.(\ref{VI3})
along the lines of Ref.\cite{16} could be attempted by requiring
$|R_{\bar a}(\tau ,\vec \sigma )| << 1$ [so that $Q_a \approx 1 +
\sum_{\bar a}\, \gamma_{\bar aa}\, R_{\bar a}$] \footnote{The
presence of the denominators $(Q_a\, Q^{-1}_b - Q_b\,
Q^{-1}_a)^{-k}$, $k = 1,2,3$, in Eqs. (\ref{VI1}) and Eq.(D12) of
Appendix D of Ref.\cite{25} suggests the necessity of a point
canonical transformation from the tidal variables $R_{\bar a}$ to
new variables more suitable for the linearization. This problem will
be studied elsewhere.} producing equations of the type
$\Big[\partial_{\tau}^2\, R_{\bar a} + A^{(o)}_{rs}\, \partial_r\,
\partial_s\, R_{\bar a} +..... + M^{(o)}\, R_{\bar a} +
F^{(o)}\Big](\tau ,\vec \sigma ) = 0$ (the quantities
$A^{(o)}_{rs}$, ..., are evaluated for $Q_a \rightarrow 1$) with a
{\it pseudo-squared-mass term} $M^{(o)}$ depending upon ${\cal
M}^{(o)}$ (the metric-independent part of ${\cal M}$), ${\cal M}_r$
and the gauge variables, i.e. upon the inertial effects
\footnote{Since the sign of the non-inertial-frame-dependent term
$M^{(o)}(\tau ,\vec \sigma )$ is unknown and may vary from a region
of $\Sigma_{\tau}$ to another one, we have not used a notation like
in the Klein-Gordon equation $(\Box + m^2)\, \phi = 0$.}. This type
of term will appear also in completely fixed gauges. Let us remark
that in the standard linearization on a background one ignores the
dependence of the matter energy-momentum tensor $T^{\mu\nu}$ on the
4-metric: it too would generate a similar pseudo-squared-mass term.
\medskip

In conclusion Eqs.(\ref{VI3}) show that the refusal of particle
physicists \footnote{See Feynman's statement \cite{32} that {\it the
geometrical interpretation is not really necessary or essential to
physics}.} to accept the geometrical view of the gravitational field
with its reduction to a spin 2 (massless graviton) theory in an
inertial frame of the background Minkowski (or DeSitter) space-time,
is not acceptable as already noticed long time ago in Ref.\cite{33}.
Inertial effects and the coupling to matter give a {\it
non-inertial-frame-dependent} description of the tidal degrees of
freedom even in the limit of the relativistic linearized theory,
which has to be defined and understood before going to the
post-Newtonian limit, the only one required till now by the solar
system tests of general relativity.

\subsection{Equations of Motion in the  3-Orthogonal Schwinger Time
Gauges.}

Let us now look at the Hamilton equations in the completely fixed
3-orthogonal time gauge.

Let us remark that it is not convenient to use the Dirac brackets
eliminating the super-Hamiltonian and super-momentum constraints and
their respective gauge fixings (\ref{IV1}), because otherwise the
tidal variables $R_{\bar a}$ and $\Pi_{\bar a}$ would not be any
more canonical and the search of the final canonical Dirac
observables ${\tilde R}_{\bar a}$, ${\tilde \Pi}_{\bar a}$, would be
extremely difficult. Therefore we can use these constraints and the
final gauge fixing for $\pi_{\tilde \phi}(\tau ,\vec \sigma )$ and
$\theta^i(\tau ,\vec \sigma )$ only after the evaluation of the
Poisson brackets.

Therefore the Hamilton equations of motion with the Dirac
Hamiltonian (\ref{V1}) are

\bea
 \partial_{\tau}\, R_{\bar a}(\tau ,\vec \sigma ) &=& \{ R_{\bar
 a}(\tau ,\vec \sigma ), H_D \} = {{\delta\, H_D}\over {\delta\,
 \Pi_{\bar a}(\tau ,\vec \sigma )}} =\nonumber \\
 &=& \Big[{{4\pi\, G}\over {c^2}}\, (1 + n)\, {\tilde \phi}^{-1}\, \Pi_{\bar a}
  - {\tilde \phi}^{-1/3}\, \sum_{rb}\, \gamma_{\bar ab}\, V_{rb}(\theta^n)\,
  Q^{-1}_b\, \sum_a\, {\bar D}_{r(b)(a)}\, {\bar n}_{(a)}\Big](\tau
  ,\vec \sigma ),\nonumber \\
  &&{}\nonumber \\
 \partial_{\tau}\, \Pi_{\bar a}(\tau ,\vec \sigma ) &=& \{ \Pi_{\bar
 a}(\tau ,\vec \sigma ), H_D \} = - {{\delta\, H_D}\over {\delta\,
 R_{\bar a}(\tau ,\vec \sigma )}}.
 \label{VI4}
 \eea

\noindent where now the functional derivatives, given by Eqs.(D12)
of Appendix D of Ref.\cite{25}, are evaluated by using the gauge
fixings (\ref{IV1}) and the solution (\ref{V6}) for the shift
functions ${\bar n}_{(a)}$ (after a choice for the residual gauge
freedom). To these equations we must add:

i) the coupled Hamilton equations for the matter;

ii) the Lichnerowicz equation (\ref{IV2}) for $\tilde \phi$;

iii) the equation (\ref{V9}) for the lapse function $n$.

\medskip

All these equations depend on the solution (\ref{III8}) (or
Eq.(3.10) of Ref.\cite{25}) of the super-momentum constraints, on
the choice of the zero modes (\ref{III9}), (\ref{V6}) and on the
choice of the three Green functions (\ref{d1}), (\ref{d2}) and
(\ref{d4}).

\bigskip

Having completely fixed the gauge, we have chosen a well defined
non-inertial frame and  a well defined pattern of inertial
potentials in the density of the weak ADM energy (the
$\Gamma$-$\Gamma$ term ${\cal S}$), in terms of the generalized
tidal effects $R_{\bar a}(\tau ,\vec \sigma )$, $\Pi_{\bar a}(\tau
,\vec \sigma )$. As a consequence in Eqs.(\ref{VI4}) there are {\it
relativistic inertial forces} associated to the chosen gauge and a
well defined deterministic evolution.

\medskip

Modulo the ambiguities in the shift functions and in the solution of
the equations ii) and iii), the resulting Hamilton equations
(\ref{VI4}) and i) are a hyperbolic system of partial differential
equations ensuring a deterministic evolution for $\tau \geq \tau_o$
of the tidal effects $R_{\bar a}(\tau ,\vec \sigma )$, $\Pi_{\bar
a}(\tau ,\vec \sigma )$ and of the matter from a set of Cauchy data
for $R_{\bar a}(\tau_o, \vec \sigma )$, $\Pi_{\bar a}(\tau_o,\vec
\sigma )$ and the matter on a Cauchy surface $\Sigma_{\tau_o}$.

\medskip

The solution of all these equations is equivalent to a solution
${}^4g_{\mu\nu}$ of Einstein's equations written in the radar
4-coordinate system associated to the chosen 3-orthogonal
non-inertial frame. This leads to an Einstein space-time, whose {\it
chrono-geometrical structure $ds^2 = {}^4g_{\mu\nu}(x)\, dx^{\mu}\,
dx^{\nu}$ is dynamically determined by the solution}. In particular,
there is a {\it dynamical emergence of 3-space} \cite{4}: the leaves
of the 3+1 splitting determined by the solution in the adapted radar
4-coordinates (i.e. the dynamically selected non-inertial frame
centered on some time-like observer) are the instantaneous 3-spaces
(the 3-universe) corresponding to a dynamical convention for the
synchronization of distant clocks. One of the leaves is the Cauchy
surface of the solution.

\bigskip

Since Eqs.(\ref{IV2}), (\ref{V6}) and (\ref{V9}) imply that both $n$
and ${\bar n}_{(a)}$ depend upon the momenta $\Pi_{\bar a}$, it
becomes non trivial to re-express them in terms of the velocities
$\partial_{\tau}\, R_{\bar a}$, of $R_{\bar a}$ and of the matter,
like it was possible in Eqs.(\ref{VI1}) before fixing the gauge.
This is the price to be paid to have deterministic evolution. As a
consequence the analogue of Eq.(\ref{VI3}) becomes extremely
complicated and much more non-linear. However the
background-independent linearization of Eqs.(\ref{VI4}) will lead to
a linearized equation with the same type of behavior as the
linearization of Eq.(\ref{VI3}).

\section{Conclusions}

As shown in Ref.\cite{4}, ADM canonical gravity is sufficiently
developed on  both the theoretical and interpretational levels so
that it is now possible to see which are the implications of a
coherent and systematic use of constraint theory. We can finally
give the Hamiltonian re-interpretation of all the procedures
developed till now in the covariant Lagrangian approach, even if for
some of them are understood only at the theoretical level without
suitable approximation schemes for practical calculations (for
instance a weak field background-independent Post-Minkowskian
approximation with relativistic matter motion is now under
investigation). In this paper we give an alternative formulation of
the York-Lichnerowicz conformal approach clarifying all its aspects
like the elusive York map and which is the natural scheme for gauge
fixing. Regarding this last point, so relevant for numerical
gravity, we show that the determination of the lapse and shift
functions is implied by the gauge fixing constraints for the
super-Hamiltonian and super-momentum secondary constraints and
should not be given independently as it happens in most of the
treatments of numerical gravity \footnote{A priori any set of gauge
fixing constraints, satisfying an orbit condition, is possible.
However, as it happens using coordinates not adapted to the existing
structures, in this way there the risk that {\it coordinate
singularities} will develop in the time evolution, as often happens
in numerical gravity.}. Also the harmonic gauges, so relevant for
the covariant approach and its Post-Newtonian approximations, have
been shown to belong to a peculiar family of Hamiltonian gauge
fixings without analog in finite-dimensional constrained systems.

\medskip

As a consequence, we now have a good understanding of the
Hamiltonian framework and we can try to face concrete problems
ranging from ${1\over {c^3}}$ relativistic effects near the geoid
\cite{34} (inertial effects, clock synchronization) to the notion of
simultaneity to be used in astrophysics and cosmology (with the
associated problem of which is the 1-way propagation velocity of
electromagnetic and gravitational signals) and to the weak field
approximation but with relativistic motion (fast binaries and
relativistic quadrupole emission formula). \bigskip

The rest-frame instant form of tetrad gravity developed in
Refs.\cite{2,3,16} for the canonical treatment of vacuum Einstein's
equations in Christodoulou-Klainermann space-times and emphasizing
the role of the non-inertial frames (the only one allowed by the
equivalence principle), has been modified in this paper so to allow
the inclusion of matter. A new parametrization of the 3-metric has
made possible the explicit construction of a York map as a partial
Shanmugadhasan canonical transformation. This map,  end point of the
Lichnerowicz-York conformal approach \cite{20,23}, had been shown to
correspond to a canonical transformation \cite{24}, but no-one had
been able to build it.

\medskip

In the York canonical basis we have  the identification of three
groups of variables (all of them have a well defined expression in
terms of the original variables, differently from the canonical
basis of Refs.\cite{3,16} for which only the inverse canonical
transformation was explicitly known):

i) The conformal factor $\phi(\tau ,\vec \sigma )$ of the 3-metric
or, better, the volume element $\tilde \phi = \phi^6$ on
$\Sigma_{\tau}$ (the unknown in the super-Hamiltonian constraint,
namely the Lichnerowicz equation), and three momenta $\pi_i^{(\theta
)}(\tau ,\vec \sigma )$ (the unknowns in the super-hamiltonian
constraints).

ii) The 14 gauge variables describing {\it generalized inertial
effects} in the non-inertial frames identified by the admissible 3+1
splittings of space-time. They are 13 configurational variables plus
the momentum $\pi_{\tilde \phi}(\tau ,\vec \sigma )$ proportional to
the York time ${}^3K(\tau ,\vec \sigma)$, whose fixation amounts to
a convention for the synchronization of distant clocks and to the
identification of the instantaneous 3-space. The meaning of the
other 13 gauge variables $\alpha_{(a)}(\tau ,\vec \sigma )$,
$\varphi_{(a)}(\tau ,\vec \sigma )$, $\theta^i(\tau ,\vec \sigma )$,
$N(\tau ,\vec \sigma ) = 1 + n(\tau ,\vec \sigma )$, $n_{(a)}(\tau
,\vec \sigma )$ has been clarified in Subsection C of Section II.

iii) two pairs of canonical (in general non-covariant) variables
describing the genuine physical degrees of freedom of the
gravitational field ({\it generalized tidal effects}). The two
configurational ones are determined by the eigen-values of the
3-metric. Since the Shanmugadhasan canonical transformation is
adapted only to 10 of the 14 first-class constraints, they are not
the final Dirac observables. However, if we fix completely the gauge
freedom and we go to Dirac brackets, they become 4 functions of the
Dirac observables of that gauge, whose identification amounts to
find a Darboux basis for the Dirac brackets.

\bigskip

In the rest-frame instant form of tetrad gravity \cite{1,3} the
Dirac Hamiltonian contains also the weak ADM energy $E_{ADM} = \int
d^3\sigma\, {\cal E}_{ADM}(\tau ,\vec \sigma )$, besides all the
first class constraints. The ADM energy density ${\cal E}_{ADM}(\tau
,\vec \sigma )$ depends on all the gauge variables. Since a
completely fixed Hamiltonian gauge corresponds to the choice of a
global non-inertial frame, in which the observers have fixed
metrological conventions, it is natural that the ADM energy density
is a gauge-dependent quantity (the problem of energy in general
relativity): it contains the inertial potentials generating the
inertial effects (for instance the $\Gamma$-$\Gamma$ term ${\cal S}$
give rise to the coordinate-dependent pattern of relativistic
Coriolis, centrifugal ... forces).

\bigskip

We have given the general solution of the super-momentum constraints
in the York canonical basis and the explicit form of the
super-Hamiltonian constraint, of the weak ADM energy and of the
Hamilton equations for the tidal degrees of freedom of the
gravitational field in a  {\it family of completely fixed
3-orthogonal Schwinger time gauges (the 3-metric is diagonal;
$\theta^i(\tau ,\vec \sigma ) \approx 0$) parametrized by the gauge
function} ${}^3K(\tau ,\vec \sigma )$ (so that the convention for
clock synchronization varies smoothly from one gauge to another
one). Unfortunately till now we do not yet know how to make
calculations in the family of Hamiltonian harmonic gauges, so that
it is not possible to compare the results with those in harmonic
coordinates.

\bigskip

The study of the equations for the shift functions, emerging from
the preservation in time of the gauge fixings $\theta^i(\tau ,\vec
\sigma ) \approx 0$ of the 3-orthogonal gauges, shows the appearance
of a generalized Gribov ambiguity connected to the gauge freedom in
the choice of the 3-coordinates on the simultaneity surfaces
$\Sigma_{\tau}$ (the 3-diffeomorphism subgroup of the gauge
transformations), like the ordinary Gribov ambiguity of Yang-Mills
theory is connected to the freedom of non-abelian gauge
transformations. It is connected to the existence of {\it zero
modes} of the covariant divergence, which imply the non-uniqueness
of the momenta $\pi_i^{(\theta )}(\tau ,\vec \sigma )$ given by the
solution (\ref{III7}) of the super-momentum constraints [see the
$g^{hom}_{((a)(b))}(\tau ,\vec \sigma )$'s  in Eq.(\ref{III8})].

The possibility of such an ambiguity in general relativity is
pointed out in Ref.\cite{8} (see page 765). As a consequence, the
3-orthogonal 3-coordinate system is identified on the Cauchy surface
$\Sigma_{\tau_o}$ not only by the gauge fixings $\theta^i(\tau ,\vec
\sigma ) \approx 0$, but also by gauge fixings modified by the
addition of zero mode terms as shown in Subsection A of Section V.
Since to each such gauge fixing are associated different shift
functions \footnote{This is a byproduct of the natural scheme for
the gauge fixings implied by constraint theory.} (\ref{V6}), whose
difference is connected with the the zero modes $f_{(a)}(\tau ,\vec
\sigma )$ of the operator $Z_{(a)i}$ of Eqs.(\ref{V2}), (\ref{V4}),
and since this ambiguity is inherited by the lapse function, it
turns out that there are inequivalent 3+1 splittings (i.e.
non-inertial frames) of $M^4$ ({\it Gribov copies}) with the same
3-orthogonal 3-coordinates. In the copies with $f_{(a)}(\tau ,\vec
\sigma ) \not= 0$ there are the restrictions (\ref{V7}) on the
dynamical variables.

Till now we were able to identify the generalized Gribov ambiguity
only in the 3-orthogonal gauges. It would be important to check
whether it arises also in other gauges, to exclude the possibility
that the 3-orthogonal gauges are globally ill-defined due to some
unknown pathology.

\medskip

While in Yang-Mills theory the choice \cite{11} of suitable weighted
Sobolev spaces eliminates the ordinary Gribov ambiguity , it is not
clear if the assumed direction-independent behavior  of the various
fields at spatial infinity (required by the absence of
super-translations) is enough to eliminate the generalized Gribov
ambiguity of canonical gravity. In canonical gravity, where there
are no asymptotically vanishing Killing vectors \cite{10} in our
class of space-times, the use of weighted Sobolev spaces (see the
first paper in Ref.\cite{10}, pages 133-136) implies the absence of
zero modes for suitable elliptic operators acting on the
simultaneity surfaces $\Sigma_{\tau}$. It is an open problem whether
there are weighted Sobolev spaces compatible with the assumed
direction-independent behavior at spatial infinity for the cotriads
and the lapse and shift functions and such that the zero modes
$g^r_{(a)}(\tau ,\vec \sigma )$ of the covariant divergence and the
asymptotically vanishing ones $f_{(a)}(\tau ,\vec \sigma )$ of the
operator $Z_{(a)i}$ are expelled from the function space.

To find the suitable function space for gravity plus the standard
model of elementary particles, in which there are neither ordinary
or generalized Gribov ambiguities nor Killing vectors, could be a
difficult task, since the group manifold in large of the
diffeomorphisms is not under mathematical control and we do not have
the well understood topological properties of the principal fiber
bundles of Yang-Mills theory.

\medskip

Three independent Green functions\footnote{Two are needed for the
solution of the super-momentum constraint and one for the
determination of the shift functions in the 3-orthogonal gauges},
each one defined modulo solutions of the corresponding homogeneous
equation, appear inside the Hamilton equations, we will have to
specify not only the initial data for the dynamical variables but
also which type of conditions we have to assume on the gravitational
fields at $\tau\, \rightarrow\, - \infty$ (in the linearized theory
in harmonic coordinates one usually uses {\it retarded} conditions
on the incoming radiation at minus null infinity).
\medskip

\bigskip

Then we have written the Hamilton equations for the tidal variables
in the York canonical basis in arbitrary Schwinger time gauges for
tetrad gravity and explicitly shown that to get a deterministic
evolution we must completely fix the gauge, i.e. we must choose  a
well defined non-inertial frame with its pattern of inertial forces.
Given Cauchy data for the tidal variables (and matter, if present)
on an instantaneous 3-space $\Sigma_{\tau_o}$, one Einstein
space-time is identified by solving these equations.

There are strong indications that in a generic gauge the
background-independent linearization along the lines of
Ref.\cite{16} will lead to the appearance of a gauge-dependent
pseudo-square-mass term. This result, joined with Ref.\cite{33},
makes the refusal of the geometrical view of the gravitational
field, with its replacement with a linear spin 2 theory in an {\it
inertial frame} of a flat background space-time, unacceptable. This
refusal is induced by the fact that till now we are able to define
the creation and annihilation operators for quantum fields only in
such inertial frames, where there is a well posed notion of Fourier
transform. The first step towards a better approximation, even if
still with a background, would be the definition of quantum fields
in non-inertial frames  in Minkowski space-time. But this is still
an unsolved problem due to the Torre-Varadarajan no-go theorem
\cite{35}, showing that in general there is no unitary evolution in
the Tomonaga-Schwinger formalism. The open problem, discussed in
Ref.\cite{14} treating the quantization of particles in non-inertial
frames, is to identify the family of non-inertial frames admitting a
unitary evolution. But then the effective non-inertial Hamiltonian
density will be frame-dependent also in flat space-time, due to the
inertial potentials like it happens in general relativity, with the
same interpretational problems.

\medskip

As shown in Refs.\cite{4}, each independent solution of Einstein
equations corresponds to an equivalence class of gauge  equivalent
Cauchy data on simultaneity surfaces leaves of the 3+1 splittings
connected by the gauge transformations admitted by the solution.
Therefore each solution admits {\it preferred dynamical non-inertial
frames corresponding to the dynamical chrono-geometrical structure
of the solution} (including dynamically determined conventions for
the synchronization of clocks implying a {\it dynamical emergence of
a  notion of instantaneous 3-space}, absent  in special relativity).

\bigskip

Let us add a final remark on the dependence of the Hamilton
equations (\ref{VI4}) on the gauge function $K(\tau ,\vec \sigma )$,
both explicitly as a consequence of Eq.(\ref{III1}) and implicitly
through the shift and lapse functions, in the family of 3-coordinate
systems where  ${}^3K(\tau ,\vec \sigma ) = \sgn\, K(\tau ,\vec
\sigma )$ [see Eqs.(\ref{IV1})]. Each choice of ${}^3K(\tau ,\vec
\sigma )$ corresponds to the presence of {\it inertial forces,
attractive or repulsive according to the sign of ${}^3K(\tau ,\vec
\sigma )$,  dictated by the convention chosen for the
synchronization of distant clocks, i.e. for the identification of
the instantaneous 3-space}. This {\it non-local} effect has no
non-relativistic counterpart, because Newtonian physics in Galilei
space-time makes use of an absolute notion of simultaneity and there
is an {\it absolute Euclidean 3-space}. Since the post-Newtonian
calculations in harmonic coordinates \cite{29} agree with the ADM
ones \cite{27} using ${}^3K(\tau ,\vec \sigma ) = 0$ at the 3PN
order and since there are no calculations at fixed 3-coordinates but
with varying ${}^3K(\tau ,\vec \sigma )$, we do not know the
influence of this inertial effect on the gravitational dynamics.

\medskip

It is important to find a relativistic solution of the Hamilton
equations in these gauges, for instance in the weak field
approximation but with relativistic motion, so to be able to
understand this effect. Such a solution would allow to study the
motion of a test particle along a time-like geodesic spiralling
around a compact mass distribution visualized in an instantaneous
3-space $\Sigma_{\tau}$ in a family of completely fixed gauges like
the ones of Eqs.(\ref{IV1}) depending in a continuous way on the
function $K(\tau ,\vec \sigma )$. As a consequence, we would find
how the velocity of the test particle depends on the instantaneous
distance inside $\Sigma_{\tau}$ (along a space-like 3-geodesic) of
the test particle from the center of mass of the matter distribution
and how this dependence changes as a function of $K(\tau ,\vec
\sigma )$, i.e. of the definition of the instantaneous 3-space (the
clock synchronization convention). Therefore, for the first time we
could explicitly check which are the (weak field) general
relativistic deviations from the Kepler virial theorem, which is
used in the interpretation of the observational data about the
rotation curves of galaxies \cite{36}. Furthermore, in this
calculation one should replace the instantaneous distance in
$\Sigma_{\tau}$, the general relativistic alternative to the
absolute Euclidean distance of Newton theory, with the {\it
luminosity distance} (a property of a congruence of light rays), the
only one definable from the observed electro-magnetic signals. Do
the general relativistic deviations go in the direction of reducing
the quantity named {\it dark matter}? Or does it (or part of it)
correspond to an inertial appearance, as it happens for the
gravito-magnetic frame dragging? Which is the dependence of these
deviations depend on the clock synchronization convention, i.e. on
the definition of the instantaneous 3-spaces and of the associated
pattern of inertial forces? Are we sure that the till now undetected
WIMPs are the explaination of dark matter or, in other words, are we
sure that the prevailing interpretation of the observational data is
the correct one? If the quoted deviations will turn out to be
negligible, this would reduce the strength of points of view like
the relativistic version \cite{37} of the non-relativistic MOND
model \footnote{It is based on a modification of Newton's law in an
inertial frame of the absolute Euclidean 3-space of Newton physics.
While in the MOND model one modifies the acceleration side of the
equations of motion, in general relativity it is the force side to
be modified by the inertial effects.} \cite{38} or like the
gravito-magnetic relativistic inertial effect of Ref.\cite{39}.
Otherwise there should be some coordinate-independent signature of
dark matter (for instance an effective mass higher of the rest mass
for ordinary matter), like it happens with the Lense-Thirring
effect, a consequence of the gravito-magnetic gauge variables in
presence of matter.

\bigskip

Moreover, the gauge dependence [including a dependence upon
${}^3K(\tau ,\vec \sigma )$] of the ADM energy  density ${\cal
E}_{ADM}(\tau ,\vec \sigma )$, namely its dependence on the chosen
non-inertial frame \footnote{Let us remark that this already happens
for the Hamiltonian describing the evolution of both relativistic
and non-relativistic particles in non-inertial frames \cite{14}.},
should play some role in the understanding of what is the {\it dark
energy}, which in some way has to take into account the
gravitational energy. We have to understand whether our results may
help to clarify the kinematical back-reaction effects appearing in
the scenario of Ref.\cite{40}, where cosmology is seen as an
effective description emerging from a coarse graining starting from
the gravitational field at small scales and going to larger and
larger scales.
\bigskip

A general open problem in the astrophysical and cosmological
contexts is what has to be understood with the word "{\it
observable}": usually it is said that it must be {\it 4-coordinate
independent} (see the description of quantities connected with
obseved light rays). In the context of general relativity this means
independent from the 4-diffeomorphisms at the Lagrangian level, i.e.
independent from the Hamiltonian gauge transformations (namely
independent from the inertial effects) at the canonical level. But,
apart from Einstein's {\it point-coincidence quantities} (what do
they mean in cosmology?), we do not yet have control on this
subject: also the coordinate-independent Weyl scalars of the
Newman-Penrose approach \cite{41} (used in the framework of
gravitational waves) are gauge-dependent on the chice of the null
tetrads. We are just beginning to understand the non-covariant
coordinate-dependent Dirac observables, invariant under Hamiltonian
gauge transformations, but we are still far away from identifying
the coordinate- and Hamiltonian -gauge- transformation independent
Bergmann observables (see Refs.\cite{4,6} about what is known on the
four eigenvalues of the Weyl tensor). As a conclusion, it is not
clear to us how many interpretational problems are hidden behind the
empirical notions of dark matter and dark energy.

\appendix

\section{Notations for Tetrad Gravity.}

\subsection{Tetrads, Cotetrads and the 4- and 3- Metric.}

We shall use the signature $\sgn\, (+---)$  for the 4-metric, with
$\sgn = \pm$, according to particle physics and general relativity
conventions respectively.

After an admissible 3+1 splitting of space-time with space-like
hyper-surfaces $\Sigma_{\tau}$, we introduce adapted coordinates,
namely the {\it radar 4-coordinates} $\sigma^A = (\tau ; \sigma^r)$
\footnote{For the sake of simplicity we shall use the notation $\vec
\sigma$ for $\{ \sigma^r\}$. $(\alpha )$ and $(a)$ are flat 4- and
3-indices, respectively; $\mu$ is a world 4-index; $A$ is a
$\Sigma_{\tau}$-adapted world 4-index.} adapted to the 3+1 splitting
and centered on an arbitrary time-like observer (they define a {\it
non-inertial frame} centered on the observer, so that they are {\it
observer and frame- dependent}).\medskip

Namely, instead of local coordinates $x^{\mu}$ for $M^4$, we use
local coordinates $\sigma^A$ on $R\times \Sigma \approx M^4$ with
$\Sigma \approx R^3\,\,\,$ [$x^{\mu}=z^{\mu}(\sigma )$ with inverse
$\sigma^A= \sigma^A(x)$], i.e. a {\it $\Sigma_{\tau}$-adapted
holonomic coordinate basis} for vector fields
$\partial_A={{\partial}\over {\partial \sigma^A}}\in T(R\times
\Sigma ) \mapsto b^{\mu}_A(\sigma )
\partial_{\mu} ={{\partial z ^{\mu}(\sigma )}\over {\partial
\sigma^A}} \partial_{\mu} \in TM^4$, and for differential one-forms
$dx^{\mu}\in T^{*}M^4 \mapsto d\sigma^A=b^A _{\mu}(\sigma
)dx^{\mu}={{\partial \sigma^A(z)}\over {\partial z^{\mu}}} dx ^{\mu}
\in T^{*}(R\times \Sigma )$.

\bigskip

As shown in Ref.\cite{2}, the general cotetrads ${}^4E^{(\alpha
)}_{\mu}$ (dual to the tetrads ${}^4E^{\mu}_{(\alpha )}$), appearing
in the 4-metric of the ADM action principle, are connected to the
cotetrads $\eo^{(\alpha )}_A$ (and tetrads) adapted to the 3+1
splitting (the time-like tetrad is normal to $\Sigma_{\tau}$) by a
point-dependent standard Lorentz boost for time-like orbits acting
on the flat indices (it sends the unit future-pointing time-like
vector ${\buildrel o\over V}^{(\alpha )} = (1; 0)$ into the unit
time-like vector $V^{(\alpha )} = l^A\, {}^4E^{(\alpha )}_A =
\Big(\sqrt{1 + \sum_a\, \varphi^2_{(a)}}; \varphi^{(a)} = - \sgn\,
\varphi_{(a)}\Big)$, where $l^A$ is the unit future-pointing normal
to $\Sigma_{\tau}$)

\medskip

\bea
 {}^4E^{(\alpha )}_A &=& L^{(\alpha )}{}_{(\beta )}(\varphi_{(a)})\, \eo^{(\beta
 )}_A, \qquad
 {}^4E^A_{(\alpha )} = \eo^A_{(\beta )}\, L^{(\beta )}{}_{(\alpha
 )}(\varphi_{(a)}),\nonumber \\
 &&{}\nonumber \\
 g_{AB} &=& {}^4g_{AB} = {}^4E^{(\alpha )}_A\, {}^4\eta_{(\alpha
 )(\beta )}\, {}^4E^{(\beta )}_B = \eo^{(\alpha )}_A\, {}^4\eta_{(\alpha
 )(\beta )}\, \eo^{(\beta )}_B
 \label{a1}
 \eea
\bigskip

The adapted tetrads and cotetrads  (corresponding to the {\it
Schwinger time gauge} of tetrad gravity) are expressed at the
Hamiltonian level in terms the lapse $N = 1 + n > 0$ (so that $N\,
d\tau$ is positive from $\Sigma_{\tau}$ to $\Sigma_{\tau + d\tau}$)
and shift $n_{(a)} = {}^3e_{(a)r}\, N^r = {}^3e^r_{(a)}\, N_r$
functions and of cotriads ${}^3e_{(a)r}$ (dual to the triads
${}^3e^r_{(a)}$) on $\Sigma_{\tau}$

\bea
 \eo^A_{(o)} &=& {1\over {N}}\, (1; - n_{(a)}\,
 {}^3e^r_{(a)}) = l^A,\qquad \eo^A_{(a)} = (0; {}^3e^r_{(a)}), \nonumber \\
 &&{}\nonumber  \\
 \eo^{(o)}_A &=& N\, (1; 0) = \sgn\, l_A,\qquad \eo^{(a)}_A
= (n_{(a)}; {}^3e_{(a)r}),
 \label{a2}
 \eea
\medskip

As a consequence, our configuration variables for tetrad gravity are
$n$, $n_{(a)}$, ${}^3e_{(a)r}$ and the boost parameters
$\varphi_{(a)}$. The future-oriented unit normal to $\Sigma_{\tau}$
is $l_A = \sgn\, N\, (1; 0)$ ($ g^{AB}\, l_A\, l_B = \sgn$), $l^A =
\sgn\, N\, g^{A\tau} = {1\over {N}}\, (1; - n^r) = {1\over {N}}\,
(1; - n_{(a)}\, {}^3e_{(a)}^r)$.

\bigskip

As shown in Ref.\cite{2}, the induced 4-metric ${}^4g_{AB}$ and the
inverse 4-metric ${}^4g^{AB}$ become in the adapted basis
\footnote{If ${}^4\gamma^{rs}$ is the inverse of the spatial part of
the 4-metric (${}^4\gamma^{ru}\, {}^4g_{us}=\delta^r_s$), the
inverse of the 3-metric is ${}^3g^{rs}=-\epsilon \, {}^4\gamma^{rs}$
(${}^3g^{ru}\, {}^3g_{us}=\delta^r_s$).}

 \bea
 g_{\tau\tau} &=&
 {}^4g_{\tau\tau} = \sgn\, [N^2 - {}^3g^{rs}\, n_r\,
 n_s] = \sgn\, [N^2 - n_{(a)}\, n_{(a)}],\nonumber \\
 g_{\tau r} &=& {}^4g_{\tau r} = - \sgn\, n_r = -\sgn\, n_{(a)}\,
 {}^3e_{(a)r},\nonumber \\
  g_{rs} &=& {}^4g_{rs} = -\sgn\, {}^3g_{rs} = - \sgn\, {}^3e_{(a)r}\,
  {}^3e_{(a)s},\nonumber \\
 &&{}\nonumber \\
 g^{\tau\tau} &=& {}^4g^{\tau\tau} = {{\sgn}\over {N^2}},\qquad
 g^{\tau r} = {}^4g^{\tau r} = -\sgn\, {{n^r}\over {N^2}} = -\sgn\, {{{}^3e^r_{(a)}\,
 n_{(a)}}\over {N^2}},\nonumber \\
 g^{rs} &=& {}^4g^{rs} = -\sgn\, ({}^3g^{rs} - {{n^r\, n^s}\over
 {N^2}}) = -\sgn\, {}^3e^r_{(a)}\, {}^3e^s_{(b)}\, (\delta_{(a)(b)} -
 {{n_{(a)}\, n_{(b)}}\over {N^2}}).\nonumber \\
 &&{}
 \label{a3}
 \eea

\bigskip

The 3-metric ${}^3g_{rs}$ has signature $(+++)$, so that we will put
all the flat 3-indices {\it down}.

\subsection{The Constraints and the Dirac Hamiltonian.}

As shown in Refs.\cite{2,3,16,17,18}, in presence of matter with
Hamiltonian mass-energy density ${\cal M}(\tau ,\sigma )$ and
3-momentum density ${\cal M}_r(\tau ,\vec \sigma )$ (${\cal M}$
depends on the 4-metric but not on its gradients) the primary and
secondary constraints are \footnote{ ${}^3G_{o(a)(b)(c)(d)} =
\delta_{(a)(c)}\, \delta_{(b)(d)} + \delta_{(a)(d)}\,
\delta_{(b)(c)} - \delta_{(a)(b)}\, \delta_{(c)(d)}$ is the flat
Wheeler-DeWitt super-metric. The covariant derivative is
$D_{r(a)(b)} = \delta_{ab}\, \partial_r + \epsilon_{(a)(b)(c)}\,
{}^3\omega_{r(c)}$, where ${}^3\omega_{r(a)}$ is the (cotriad
dependent) 3-spin connection.}

\bea
 \pi_n(\tau ,\vec \sigma ) &\approx& 0,\qquad
 \pi_{\vec n (a)}(\tau ,\vec \sigma ) \approx 0,\qquad
 \pi_{\vec \varphi (a)}(\tau ,\vec \sigma ) \approx 0,\nonumber \\
 M_{(a)}(\tau ,\vec \sigma ) &=& \sum_{bcr}\,\epsilon_{(a)(b)(c)}\,
 [{}^3e_{(b)r}\, {}^3{\tilde \pi}^r_{(c)}](\tau ,\vec \sigma ) \approx 0,\nonumber \\
 &&{}\nonumber \\
 {\cal H}(\tau ,\vec \sigma ) &=& \sgn \Big[{{c^3}\over {16\pi\, G}}\, {}^3e\,
 {}^3R - {1\over c}\, {\cal M} -\nonumber \\
 &&-{{2\pi\, G}\over {c^3\,\, {}^3e}}
{}^3G_{o(a)(b)(c)(d)}\, {}^3e_{(a)r}\, {}^3{\tilde \pi}^r_{(b)}\,
{}^3e_{(c)s}\, {}^3{\tilde \pi}^s_{(d)}
 \Big](\tau ,\vec \sigma )\approx 0,\nonumber \\
 {\cal H}_{(a)}(\tau ,\vec \sigma ) &=& \Big[\sum_{rb}\,
 D_{r(a)(b)}\, {}^3{\tilde \pi}^r_{(b)} - {}^3e^v_{(a)}\, {\cal M}_v
 \Big](\tau ,\vec \sigma ) \approx 0,
 \label{a4}
 \eea

In Ref.\cite{2} it is shown that the super-momentum constraints
${\cal H}_{(a)}(\tau ,\vec \sigma ) \approx 0$ are not the
Hamiltonian generators of passive 3-diffeomorphims: the actual
generators are $\Theta_r(\tau ,\vec \sigma ) = \sum_{as}\,
\Big[{}^3{\tilde \pi}^s_{(a)}\,
\partial_r\, {}^3e_{(a)s} -
\partial_s\, ({}^3e_{(a)r}\, {}^3{\tilde \pi}^s_{(a)})\Big](\tau
,\vec \sigma )\approx 0$ and that we have ${\cal H}_{(a)}(\tau ,\vec
\sigma ) = - \Big[{}^3e^r_{(a)}\, \Big(\Theta_r + \sum_b\,
{}^3\omega_{r(b)}\, M_{(b)}\Big)\Big](\tau ,\vec \sigma )$.

The constraints $M_{(a)}(\tau ,\vec \sigma ) \approx 0$ generate
O(3)- rotations, which vary the angles $\alpha_{(a)}(\tau ,\vec
\sigma )$ hidden inside the cotriads. The boost parameters
$\varphi_{(a)}(\tau ,\vec \sigma )$ and the angles
$\alpha_{(a)}(\tau ,\vec \sigma )$ describe the O(3,1) gauge freedom
of the tetrads in their flat indices $(\alpha )$ in each point of
$\Sigma_{\tau}$. The constraints $M_{(a)}(\tau ,\vec \sigma )
\approx 0$ and $\pi_{\vec \varphi\, (a)}(\tau ,\vec \sigma ) \approx
0$ replace the standard generators of the O(3,1) (proper Lorentz
group) gauge transformations ($\varphi_{(a)}$ and $\alpha_{(a)}$ are
our parametrization of the 6 gauge variables also appearing in the
Newman-Penrose formalism, where they label the arbitrariness in the
choice of the null tetrads).

\medskip

Therefore, with our parametrization  the independent configuration
variables and the conjugate momenta of our canonical basis  are

\bea
 &&\begin{minipage}[t]{3cm}
\begin{tabular}{|l|l|l|l|} \hline
$\varphi_{(a)}$ & $n$ & $n_{(a)}$ & ${}^3e_{(a)r}$ \\ \hline
$\approx 0$ & $\approx 0$ & $  \approx 0 $ & ${}^3{\tilde
\pi}^r_{(a)}$ \\ \hline
\end{tabular}
\end{minipage}
 \label{a5}
  \eea

\noindent This is a Shanmugadhasan canonical basis already naturally
adapted to seven of the primary constraints. See Refs. \cite{1,3}
for the assumed (direction independent) behavior at spatial infinity
of these variables: the basic information is ${}^3e_{(a)r}(\tau
,\vec \sigma )\, \rightarrow_{r\, \rightarrow \infty }\, (1+{M\over
{2r}}) \delta_{(a)r}+ O(r^{-3/2})$, $N(\tau ,\vec \sigma ) = 1 +
n(\tau ,\vec \sigma )\, \rightarrow_{r\, \rightarrow \infty }\, 1 +
O(r^{- (2+\epsilon )})$ ($\epsilon > 0$), $n_r(\tau ,\vec \sigma ) =
n_{(a)}(\tau ,\vec \sigma )\, {}^3e_{(a)r}(\tau ,\vec \sigma )\,
\rightarrow_{r\, \rightarrow \infty }\, O(r^{- \epsilon})$ ($r =
|\vec \sigma |$).
\medskip

From Eqs.(25) of Ref.\cite{3} the weak or volume form of the ADM
Poincar\'e charges of metric gravity is [$\gamma = |det\,
{}^3g_{rs}| = ({}^3e)^2 = \phi^{12}$, ${}^3e = det\, {}^3e_{(a)r}$]

\begin{eqnarray*}
 E_{ADM} &=& - \sgn\, c\,
{P}^{\tau}_{ADM}=  \int d^3\sigma \Big[{\cal M} -
 {{c^4}\over {16\pi\, G}}\, \sqrt{\gamma}\,\, \sum_{rsuv}\,
 {}^3g^{rs}({}^3\Gamma^{u} _{rv}\,
{}^3\Gamma^{v}_{su}-{}^3\Gamma^{u}_{rs}\, {}^3\Gamma^{
v} _{vu}) +\nonumber \\
 &+&  {{8\pi\, G}\over {c^2\, \sqrt{\gamma}
}}\, \sum_{rsuv}\, {}^3G_{rsuv}\, {}^3{\tilde \Pi}^{rs}\,
{}^3{\tilde \Pi}^{ uv}\Big] (\tau ,\vec \sigma ),\nonumber \\
 {P}^{r}_{ADM}&=&- 2\int d^3\sigma \, \Big[\sum_{su}\, {}^3\Gamma
^{r}_{su}(\tau ,\vec \sigma )\, {}^3{\tilde \Pi}^{su}- {1\over 2}\,
\sum_s\, {}^3g^{rs} {\cal M}_{s}\Big](\tau ,\vec \sigma ),
 \end{eqnarray*}

 \bea
{J}^{\tau r}_{ADM}&=&-{ J}^{r\tau}_{ADM}=
 \int d^3\sigma   \Big( \sigma^{ r}\,\,
\Big[ {{c^3}\over {16\pi\, G}}\, \sqrt{\gamma}\,\, \sum_{nsuv}\,
{}^3g^{ns}({}^3\Gamma^{ u}_{ nv}\,
{}^3\Gamma^{v}_{su}-{}^3\Gamma^{u} _{ns}\, {}^3\Gamma^{ v}_{ vu})
-\nonumber \\
 &-& {{8\pi\, G}\over {c^3\, \sqrt{\gamma}}}\,
\sum_{nsuv}\, {}^3G_{nsuv}\, {}^3{\tilde \Pi}^{ns}\, {}^3{\tilde
\Pi}^{
uv} - {1\over c}\, {\cal M}\Big] +\nonumber \\
 &+& {{c^3}\over {16\pi\, G}}\, \sum_{nsuv}\,  \delta^{
r}_{u}({}^3g_{vs}-\delta _{vs})
\partial_{n}\Big[ \sqrt{\gamma}({}^3g^{ns} \,
{}^3g^{uv}-{}^3g^{nu}\, {}^3g^{
sv})\Big] \Big) (\tau ,\vec \sigma ),\nonumber \\
 {J}^{rs}_{ADM}&=&
\int d^3\sigma  \Big[\sum_{uv}\, (\sigma^{ r}\, {}^3\Gamma^{s} _{u
v} - \sigma^{s}\, {}^3\Gamma^{r}_{uv})\, {}^3{\tilde \Pi}^{uv} -
{1\over 2}\, \sum_u\, (\sigma^{r}\, {}^3g^{su} - \sigma^{s}\,
 {}^3g^{ ru})\, {\cal M}_{u}\Big] (\tau ,\vec \sigma ). \nonumber \\
 &&{}
 \label{a6}
\end{eqnarray}

\noindent  These  weak Poincar\'e charges are expressed in terms of
cotriads ${}^3e_{(a)r}$ and their conjugate momenta ${}^3{\tilde
\pi}^r_{(a)}$, by using ${}^3g_{rs}=\sum_a\, {}^3e_{(a)r}\,
{}^3e_{(a)s}$, ${}^3{\tilde \Pi}^{rs}= {1\over 4}\, \sum_a\,
[{}^3e^r_{(a)}\, {}^3{\tilde \pi}^s_{(a)}+{}^3e^s_{(a)}\,
{}^3{\tilde \pi}^r_{(a)}]$ (see Eq.(12) of  Ref.\cite{3}).

\bigskip

The Dirac Hamiltonian is (the $\lambda$'s are arbitrary Dirac
multipliers \footnote{In canonical metric gravity they are only 4
(not 8), namely the Hamiltonian gauge group has 8 generators (both
the primary and secondary constraints) but the same number of
parameters (i.e. arbitrary functions) like the 4-diffeomorphism
group of the covariant Lagrangian approach. The configurational
lapse and shift variables in front of the secondary constraints are
the effective parameters, because the kinematical part of the
Hamilton equations implies that the Dirac multipliers are their
$\tau$-derivatives.})

\bea
 H_D &=& E_{ADM} + \int d^3\sigma\, \Big[- \sgn\, c\, n\, {\cal H} + n_{(a)}\,
 {\cal H}_{(a)}\Big](\tau ,\vec \sigma ) +\nonumber \\
 &+& \int d^3\sigma\, \Big[\lambda_n\, \pi_n + \lambda_{\vec n(a)}\,
 \pi_{\vec n(a)} + \lambda_{\vec \varphi (a)}\, \pi_{\vec \varphi (a)}
 + \lambda_{(a)}\, M_{(a)} \Big](\tau ,\vec \sigma ),
 \label{a7}
 \eea

\noindent where the explicit form of the weak ADM energy in tetrad
gravity is

\bea
 E_{ADM} &=&  \int d^3\sigma\, \Big[{\cal M} -
  {{c^4}\over {16\pi\, G}}\, {\cal S} +
   {{2\pi\, G}\over {c^2\,\, {}^3e}}\, \sum_{abcd}\,
 {}^3G_{o(a)(b)(c)(d)}\, {}^3e_{(a)r}\, {}^3{\tilde \pi}^r_{(b)}\, {}^3e_{(c)s}\,
 {}^3{\tilde \pi}^r_{(d)}\Big](\tau ,\vec \sigma ),\nonumber \\
 &&{}\nonumber \\
 {\cal S}(\tau ,\vec \sigma ) &=& \Big[{}^3e\, \sum_{rsuv}\,
 {}^3e^r_{(a)}\, {}^3e^s_{(a)}\, \Big({}^3\Gamma^u_{rv}\, {}^3\Gamma^v_{su} -
 {}^3\Gamma^u_{rs}\, {}^3\Gamma^v_{uv}\Big)\Big](\tau ,\vec \sigma ).
 \label{a8}
 \eea

\noindent It  is the sum of the matter mass density, of the
$\Gamma$-$\Gamma$ potential term $-  {{c^4}\over {16\pi\, G}}\, \int
d^3\sigma\, {\cal S}(\tau ,\vec \sigma )$ and of the kinetic term
quadratic in the momenta.

\medskip

As a consequence we have

\bea
 &&E_{ADM} + \int d^3\sigma\, \Big[- \sgn\, c\, n\, {\cal H}\Big](\tau ,\vec \sigma )
 =  \int d^3\sigma\, \Big[(1 + n)\, {\cal
 M}\Big](\tau ,\vec \sigma ) -\nonumber \\
 &&{}\nonumber \\
 &&-  {{c^4}\over {16\pi\, G}}\, \int d^3\sigma\, \Big({\cal S} + {}^3e\,
 n\, {}^3R\,\, \Big) (\tau ,\vec \sigma ) +\nonumber \\
 &&+  {{2\pi\, G}\over {c^2}}\, \int d^3\sigma\, \Big[{1\over {{}^3e}}\, (1 + n)\,
 \sum_{abcd}\, {}^3G_{o(a)(b)(c)(d)}\,
 {}^3e_{(a)r}\, {}^3{\tilde \pi}^r_{(b)}\, {}^3e_{(c)s}\,
 {}^3{\tilde  \pi}^r_{(d)}\Big](\tau ,\vec \sigma ).
 \label{a9}
 \eea

\bigskip

The extrinsic curvature of the hyper-surface $\Sigma_{\tau}$ and the
first half of Hamilton equations are \cite{2} (see Section II for
the definition of the $\alpha_{(a)}$ -independent "barred"
variables)

\begin{eqnarray*}
  {}^3K_{rs} &=& \sgn\, {{4\pi\, G}\over {c^3\,\, {}^3{\bar e}}}\,
  \sum_{abu}\, \Big[\Big({}^3{\bar
 e}_{(a)r}\, {}^3{\bar e}_{(b)s} + {}^3{\bar e}_{(a)s}\, {}^3{\bar
 e}_{(b)r}\Big)\, {}^3{\bar e}_{(a)u}\, {\bar \pi}^u_{(b)} -
 {}^3{\bar e}_{(a)r}\, {}^3{\bar e}_{(a)s}\, {}^3{\bar e}_{(b)u}\,
 {\bar \pi}^u_{(b)}\Big],\nonumber \\
 &&{}\nonumber \\
 {}^3K &=& - \sgn\, {{4\pi\, G}\over {c^3}}\, {{\sum_{ar}\,
{}^3{\bar e}_{(a)r}\, {\bar \pi}^r_{(a)}}\over {{}^3{\bar e}}},\nonumber \\
 &&{}\nonumber \\
 \partial_{\tau}\, n(\tau ,\vec \sigma ) &=& \{ n(\tau ,\vec
 \sigma ), H_D\} = \lambda_n(\tau ,\vec \sigma ),\nonumber \\
 \partial_{\tau}\, n_{(a)}(\tau ,\vec \sigma ) &=& \{ n_{(a)}(\tau ,\vec
 \sigma ), H_D\} = \lambda_{\vec n (a)}(\tau ,\vec \sigma ),\nonumber \\
 \partial_{\tau}\, \varphi_{(a)}(\tau ,\vec \sigma ) &=& \{
 \varphi_{(a)}(\tau ,\vec \sigma ), H_D\} = \lambda_{\vec \varphi (a)}(\tau
 ,\vec \sigma ),
 \end{eqnarray*}

 \bea
 \partial_{\tau}\, {}^3e_{(a)r}(\tau ,\vec \sigma ) &=& \{
 {}^3e_{(a)r}(\tau ,\vec \sigma ), H_D\} =
 \Big[\epsilon_{(a)(b)(c)}\, \lambda_{(b)}\, {}^3e_{(c)r} - N\,
 \sum_s\, {}^3K_{rs}\, {}^3e^s_{(a)} + \partial_r\, n_{(a)}
 +\nonumber \\
 &+& \sum_{bs}\, n_{(b)}\, {}^3e^s_{(b)}\, (\partial_s\, {}^3e_{(a)r} -
 \partial_r\, {}^3e_{(a)s}) \Big](\tau ,\vec \sigma ),\nonumber \\
 \Rightarrow&& \partial_{\tau}\, {}^3g_{rs}(\tau ,\vec \sigma ) =
 \Big[n_{r|s} + n_{s|r} - 2\, N\, {}^3K_{rs}\Big](\tau ,\vec \sigma
 ).
 \label{a10}
 \eea
\bigskip

The gauge-fixing procedure illustrated in the Introduction implies
that at the end the Dirac multipliers are consistently determined by
the preservation in time of the gauge-fixing constraints \cite{3}.

\section{The Canonical Transformation (\ref{II3}).}

\subsection{Its Determination}

By putting Eq.(\ref{II4}) into Eqs.(\ref{II6}) we get the following
three sets of equations for the kernels $K$, $G$, $F$

\begin{eqnarray*}
 &&\sum_{sha}\, \epsilon_{(k)(h)(a)}\, {}^3e_{(h)s}(\tau ,\vec \sigma
 )\, K^s_{(a)b}(\tau ,\vec \sigma ) = 0,\nonumber \\
 &&\sum_{sha}\, \epsilon_{(k)(h)(a)}\, {}^3e_{(h)s}(\tau ,\vec \sigma
 )\, G^s_{(a)b}(\tau ,\vec \sigma ) = 0,\nonumber \\
 &&\sum_{khsa}\, \Big[A_{(k)(c)}(\alpha_{(e)})\, \epsilon_{(k)(h)(a)}\,
 {}^3e_{(h)s}\, F^s_{(a)(b)}\Big](\tau ,\vec \sigma ) = -
 \delta_{(b)(c)},\end{eqnarray*}

 \begin{eqnarray*}
 &&\sum_{ar}\, K^r_{(a)c}(\tau ,\vec \sigma )\,
 R_{(a)(b)}(\alpha_{(e)}(\tau ,\vec \sigma ))\, V_{rb}(\theta^n(\tau ,\vec \sigma
 )) = \delta_{cb},\nonumber \\
 &&\sum_{ar}\, G^r_{(a)i}(\tau ,\vec \sigma )\,
 R_{(a)(b)}(\alpha_{(e)}(\tau ,\vec \sigma ))\, V_{rb}(\theta^n(\tau ,\vec \sigma
 )) = 0,\nonumber \\
 &&\sum_{ar}\, F^r_{(a)(b)}(\tau ,\vec \sigma )\,
 R_{(a)(b)}(\alpha_{(e)}(\tau ,\vec \sigma ))\, V_{rb}(\theta^n(\tau ,\vec \sigma
 )) = 0,\end{eqnarray*}

 \bea
 &&\sum_{rla}\, \epsilon_{mlr}\, {}^3e_{(a)l}(\tau ,\vec \sigma
 )\, K^r_{(a)c}(\tau ,\vec \sigma ) = 0,\nonumber \\
 &&\sum_{rla}\, \epsilon_{mlr}\, {}^3e_{(a)l}(\tau ,\vec \sigma
 )\, G^r_{(a)i}(\tau ,\vec \sigma ) = - B_{mi}(\theta^n(\tau ,\vec \sigma
 )),\nonumber \\
 &&\sum_{rla}\, \epsilon_{mlr}\, {}^3e_{(a)l}(\tau ,\vec \sigma
 )\, F^r_{(a)(c)}(\tau ,\vec \sigma ) = 0.
 \label{b1}
 \eea

The solutions of the first three equations (\ref{b1}) are

\bea
 K^s_{(a)b} &=& \sum_h\, {}^3e^s_{(h)}\, {\tilde
 K}_{(h)(a)b},\qquad {\tilde K}_{(h)(a)b} = {\tilde K}_{(a)(h)b} =
 {\tilde K}_{((h)(a))b},
 \nonumber \\
 &&{}\nonumber \\
 G^s_{(a)i} &=& \sum_h\, {}^3e^s_{(h)}\, {\tilde
 G}_{(h)(a)i},\qquad {\tilde G}_{(h)(a)i} = {\tilde G}_{(a)(h)i}
 = {\tilde G}_{((h)(a))i},
 \nonumber \\
 &&{}\nonumber \\
 F^s_{(a)(b)} &=& - {1\over 2}\, \sum_{kh}\,
 \epsilon_{(h)(a)(k)}\, B_{(b)(k)}(\alpha_{(e)})\, {}^3e^s_{(h)} +
 \sum_h\, {\tilde \Lambda}_{(h)(a)(b)}\, {}^3e^s_{(h)} =\nonumber \\
 &{\buildrel {def}\over =}& \sum_h\, {}^3e^s_{(h)}\, \Big[Z_{(h)(a)(b)}
 + {\tilde \Lambda}_{(h)(a)(b)}\Big],\nonumber \\
 && {\tilde \Lambda}_{(h)(a)(b)} = {\tilde \Lambda}_{(a)(h)(b)}
 = {\tilde \Lambda}_{((h)(a))(b)},
 \qquad Z_{(h)(a)(b)} = - Z_{(a)(h)(b)} = Z_{[(h)(a)](b)}.
 \label{b2}
 \eea

The second set of three equations (\ref{b1}) may be rewritten in the
form

\bea
 &&\sum_{ha}\, {\cal M}_{(h)(a)b}\, {\tilde K}_{((h)(a))c} =
 \delta_{bc},\nonumber \\
 &&\sum_{ha}\, {\cal M}_{(h)(a)b}\, {\tilde G}_{((h)(a))i} =
 0,\nonumber \\
 &&\sum_{ha}\, {\cal M}_{(h)(a)b}\, \Big[Z_{[(h)(a)](c)} + {\tilde
 \Lambda}_{((h)(a))(c)}\Big] = 0,\nonumber \\
 &&{}\nonumber \\
 &&with\nonumber \\
 &&{}\nonumber \\
 {\cal M}_{(h)(a)b} &=& \sum_r\, {}^3e^r_{(h)}\,
 R_{(a)(b)}(\alpha_{(e)})\, V_{rb}(\theta^n) =
 {{R_{(a)(b)}(\alpha_{(e)})\, R_{(h)(b)}(\alpha_{(e)})}
 \over {\Lambda^b}} = {\cal M}_{((h)(a))b},\nonumber \\
 &&{}
 \label{b3}
 \eea

\noindent and has the following solutions

\begin{eqnarray*}
 {\tilde K}_{((h)(a))c} &=& \sum_{lm}\,
 R_{(h)(l)}(\alpha_{(e)})\, R_{(a)(m)}(\alpha_{(e)})\,
 K^{'}_{((l)(m))c} =\nonumber \\
 &=& \sum_{lm}^{l \not= m}\,
 R_{(h)(l)}(\alpha_{(e)})\, R_{(a)(m)}(\alpha_{(e)})\,
 K^{'}_{((l)(m))c} + R_{(h)(c)}(\alpha_{(e)})\,
 R_{(a)(c)}(\alpha_{(e)})\, \Lambda^c, \nonumber \\
 &&\qquad K^{'}_{(l)(l)c} = \delta_{lc}\, \Lambda^c,
 \end{eqnarray*}

 \begin{eqnarray*}
 {\tilde G}_{((h)(a))i} &=& \sum_{lm}^{l \not= m}\,
 R_{(h)(l)}(\alpha_{(e)})\, R_{(a)(m)}(\alpha_{(e)})\,
 G^{'}_{((l)(m))i},\qquad G^{'}_{(l)(l)i} = 0,\nonumber \\
  {\tilde \Lambda}_{((h)(a))c} &=& \sum_{lm}^{l \not= m}\,
 R_{(h)(l)}(\alpha_{(e)})\, R_{(a)(m)}(\alpha_{(e)})\,
 \Lambda^{'}_{((l)(m))c},\qquad \Lambda^{'}_{(l)(l)c} = 0,
 \end{eqnarray*}

 \beq
 Z_{[(h)(a)](c)} = - {1\over 2}\, \epsilon_{(h)(a)(k)}\,
 B_{(c)(k)}(\alpha_{(e)}) =
  \sum_{lm}^{l \not= m}\,
 R_{(h)(l)}(\alpha_{(e)})\, R_{(a)(m)}(\alpha_{(e)})\,
 Z^{'}_{[(l)(m)](c)}.
 \label{b4}
 \eeq

By defining

\bea
 {\cal N}^m_{(h)(a)} &=& \sum_{rl}\, \epsilon_{mlr}\,
 {}^3e_{(a)l}\, {}^3e^r_{(h)} =
   \sum_{bk}\, R_{(a)(b)}(\alpha_{(e)})\,
  R_{(h)(k)}(\alpha_{(e)})\, {\cal N}^{'\, m}_{(b)(k)},
  \nonumber \\
  &&{}\nonumber \\
  &&with\nonumber \\
  &&{}\nonumber \\
  {\cal N}^{'\, m}_{(b)(k)} &=& \Big[\sum_{rl}\, \epsilon_{mlr}\,
  V_{lb}(\theta^n)\, V_{rk}(\theta^n)\Big]\, {{\Lambda_b}\over
  {\Lambda_k}} {\buildrel {def}\over =}\, Q^m_{bk}\,
  {{\Lambda_b}\over {\Lambda_k}},\qquad
  Q^m_{bk} = - Q^m_{kb} = Q^m_{[bk]},
 \label{b5}
 \eea
\medskip

\noindent the third set of three equations (\ref{b1}) may be written
in the form

\bea
 &&\sum_{bk}^{b \not= k}\, {\cal N}^{'\, m}_{(b)(k)}\, K^{'}_{((k)(b))c}
 = 0,\nonumber \\
 &&\sum_{bk}^{b \not= k}\, {\cal N}^{'\, m}_{(b)(k)}\, G^{'}_{((k)(b))i}
 = - B_{mi}(\theta^n),\nonumber \\
 &&\sum_{bk}^{b \not= k}\, {\cal N}^{'\, m}_{(b)(k)}\, \Big[Z^{'}_{[(k)(b)](c)}
 + \Lambda^{'}_{((k)(b))(c)}\Big] = 0,
 \label{b6}
 \eea

\noindent which does not contain the already known components
$K^{'}_{(b)(b)c} = \delta_{bc}\, \Lambda^b$, $G^{'}_{(b)(b)i} = 0$,
$\Lambda^{'}_{(b)(b)(c)} = 0$.

The solution of Eqs.(\ref{b6}) are

\begin{eqnarray*}
 k \not= b&&{}\nonumber \\
 &&{}\nonumber \\
 Z^{'}_{[(k)(b)](c)} &=& - {1\over 2}\, \sum_m\,
 B_{(c)(m)}(\alpha_{(e)})\, \sum_{ha}\, \epsilon_{(h)(a)(m)}\,
 R_{(h)(k)}(\alpha_{(e)})\, R_{(a)(b)}(\alpha_{(e)}),\nonumber \\
 W^{'}_{(k)(b)i} &=& - {1\over 2}\, \sum_{tuw}\, \epsilon_{tuw}\,
 B_{iw}(\theta^n)\, {{\Lambda_k}\over {\Lambda_b}}\,
 V_{uk}(\theta^n)\, V_{tb}(\theta^n),\quad W^{'}_{(b)(b)i} =
 0,\nonumber \\
 &&{}\nonumber \\
 K^{'}_{((k)(b))c} &=& 0,\qquad
 \Lambda^{'}_{((k)(b))(c)} = {{{{\Lambda_k}\over {\Lambda_b}} +
 {{\Lambda_b}\over {\Lambda_k}}}\over {{{\Lambda_k}\over {\Lambda_b}} -
 {{\Lambda_b}\over {\Lambda_k}}}}\, Z^{'}_{[(k)(b)](c)},
 \end{eqnarray*}

\bea
 G^{'}_{((k)(b))i} &=& {{2\, {{\Lambda_b}\over {\Lambda_k}}}\over
 {{{\Lambda_k}\over {\Lambda_b}} - {{\Lambda_b}\over {\Lambda_k}}}}\,
 W^{'}_{(k)(b)i} =  \sum_{tw}\, {{\epsilon_{bkt}\, V_{tw}(\theta^n)\,
 B_{iw}(\theta^n)}\over {{{\Lambda_k}\over {\Lambda_b}} -
 {{\Lambda_b}\over {\Lambda_k}}}}.
 \label{b7}
 \eea

Therefore the kernels in Eq.(\ref{II4}) are

\bea
 K^r_{(a)b} &=& \sum_h\, {}^3e^r_{(h)}\,
 R_{(h)(b)}(\alpha_{(e)})\, R_{(a)(b)}(\alpha_{(e)})\, \Lambda_b =
 R_{(a)(b)}(\alpha_{(e)})\, V_{rb}(\theta^n), \nonumber \\
 G^r_{(a)i} &=&  \sum_{ml}^{m \not= l}\, \sum_{htuw}\, {}^3e^r_{(h)}\,
 R_{(h)(l)}(\alpha_{(e)})\, R_{(a)(m)}(\alpha_{(e)})\,
 {{\epsilon_{tuw}\, B_{iw}(\theta^n)\, V_{ul}(\theta^n)\, V_{tm}(\theta^n)}
 \over {{{\Lambda_l}\over {\Lambda_m}} - {{\Lambda_m}\over
 {\Lambda_l}}}} =\nonumber \\
 &=&  \sum_{lb}^{l \not= b}\, \sum_{tw}\, R_{(a)(b)}(\alpha_{(e)})
 {{V_{rl}(\theta^n)\, \epsilon_{blt}\, V_{tw}(\theta^n)\,
 B_{iw}(\theta^n)}\over {\Lambda_l\, \Big({{\Lambda_l}\over {\Lambda_b}} -
 {{\Lambda_b}\over {\Lambda_l}}\Big)}},\nonumber \\
 F^r_{(a)(c)} &=& - \sum_{lm}^{l \not= m}\, \sum_{huvk}\,
 {}^3e^r_{(h)}\, {{{{\Lambda_l}\over {\Lambda_m}}}\over
 {{{\Lambda_l}\over {\Lambda_m}} - {{\Lambda_m}\over {\Lambda_l}}}}\,
 R_{(h)(l)}(\alpha_{(e)})\, R_{(a)(m)}(\alpha_{(e)})\,
 R_{(u)(l)}(\alpha_{(e)})\, R_{(v)(m)}(\alpha_{(e)})\nonumber \\
 &&\epsilon_{(u)(v)(k)}\, B_{(c)(k)}(\alpha_{(e)}) =\nonumber \\
 &=& - \sum_{lb}^{l \not= b}\, R_{(a)(b)}(\alpha_{(e)})\, \sum_t\,
 {{V_{rl}(\theta^n)\, \epsilon_{(l)(b)(t)}\, R_{(t)(k)}(\alpha_{(e)})\,
 B_{(c)(k)}(\alpha_{(e)})}\over {\Lambda_l\, \Big({{\Lambda_l}\over
 {\Lambda_b}} - {{\Lambda_b}\over {\Lambda_l}}\Big)}}.
 \label{b8}
 \eea

\medskip

\subsection{Inversion of $G^r_{(a)i}$}

Let us look for a kernel $H_{(a)rj}$, which is an inverse of
$G^r_{(a)i}$ in the following sense (Eqs.(\ref{b2}) and (\ref{b4})
are used)

\bea
 \delta_{ij} &=& \sum_{ar}\, H_{(a)rj}\, G^r_{(a)i} =
   \sum_{ar}\, H_{(a)rj}\, \sum_h\, {}^3e^r_{(h)}\, \sum_{lm}^{l
  \not= m}\, R_{(h)(l)}(\alpha_{(e)})\, R_{(a)(m)}(\alpha_{(e)})\,
  G^{'}_{((l)(m))i} =\nonumber \\
  &=& \sum_{lm}^{l \not= m}\, H^{'}_{(l)(m)j}\, G^{'}_{((l)(m))i},
 \label{b9}
 \eea

\noindent where we have introduced the kernel  $H^{'}_{(l)(m)j}$

\beq
 H^{'}_{(l)(l)j} = 0,\qquad
 H^{'}_{(l)(m)j} = \sum_{arh}\, H_{(a)rj}\, {}^3{\bar
 e}^r_{(l)}\, R_{(a)(m)}(\alpha_{(e)}),\qquad for\,\, l \not= m.
 \label{b10}
 \eeq

Since $G^{'}_{(l)(l)i} = 0$, we can define the following $3 \times
3$ matrix

\beq
 {\cal G}_{li} = G^{'}_{((b)(k))i},\qquad l \not= b, l \not= k, b
 \not= k.
 \label{b11}
 \eeq

Since the first two lines of Eqs.(\ref{b6}) suggest the following
ansatz

\beq
 H^{'}_{(l)(m)j} = H^{'}_{((l)(m))i} = - {1\over 2}\, \sum_t\,
 A_{jt}(\theta^n)\, \Big[{\cal N}^{{'}\, t}_{(l)(m)} +
 {\cal N}^{{'}\, t}_{(m)(l)}\Big],
 \label{b12}
 \eeq

\noindent we can also define the $3 \times 3$ matrix

\beq
 {\cal H}_{jl} = H^{'}_{((b)(k))j},\qquad l \not= b, l \not= k, b
 \not= k.
 \label{b13}
 \eeq

 As a consequence Eqs.(\ref{b9}) are satisfied because the second
 of Eqs.(\ref{b6}) implies

 \beq
  \sum_l\, {\cal H}_{jl}\, {\cal G}_{li} = - \sum_{tbk}\,
  A_{jt}(\theta^n)\, {\cal N}^{{'}\, t}_{(b)(k)}\, G^{'}_{((b)(k))i}
  = \sum_t\, A_{jt}(\theta^n)\, B_{ti}(\theta^n) = \delta_{ij}.
  \label{b14}
  \eeq

  Then the first of Eqs.(\ref{b6}) implies

  \beq
   \sum_j\, v_j\, {\cal H}_{jl} = 0 \Rightarrow v_j = 0
   \Rightarrow det\, ({\cal H}_{jl}) \not= 0,
   \label{b15}
   \eeq

\noindent so that we have also $det\, ({\cal G}_{li}) \not= 0$, i.e.
${\cal G}_{li}\, v_i = 0$ implies $v_i = 0$.

Therefore we get (also the expressions in the 3-orthogonal gauges
$\theta^i(\tau ,\vec \sigma ) \approx 0$ are given)

\begin{eqnarray*}
 H_{(a)rj} &=& \sum_{lm}^{l \not= m}\, R_{(a)(m)}(\alpha_{(e)})\,
 {}^3{\bar e}_{(l)r}\, H^{'}_{((l)(m))j} =\nonumber \\
 &=& - {1\over 2}\, \sum_{lm}^{l \not= m}\, R_{(a)(m)}(\alpha_{(e)})\,
 {}^3{\bar e}_{(l)r}\, \sum_t\, A_{jt}(\theta^n)\, \Big[{\cal N}^{{'}\,
 t}_{(l)(m)} + {\cal N}^{{'}\, t}_{(m)(l)}\Big] =\nonumber \\
 &=& - {1\over 2}\, \sum_{lm}^{l \not= m}\, R_{(a)(m)}(\alpha_{(e)})\,
 V_{rl}(\theta^n)\, \Lambda_l\, \sum_t\, A_{jt}(\theta^n)\nonumber \\
 &&\sum_{uv}\, \epsilon_{tvu}\, \Big[{{\Lambda_l}\over {\Lambda_m}}\,
 V_{um}(\theta^n)\, V_{vl}(\theta^n) + {{\Lambda_m}\over {\Lambda_l}}\,
 V_{ul}(\theta^n)\, V_{vm}(\theta^n)\Big]\nonumber \\
 &&{}\nonumber \\
 &\rightarrow_{\alpha_{(e)}, \theta^n \rightarrow 0}& H^{(o)}_{(a)rj}
 = {1\over 2}\, \epsilon_{arj}\, \Lambda_r\,
 \Big({{\Lambda_r}\over {\Lambda_a}} - {{\Lambda_a}\over
 {\Lambda_r}}\Big) = \Lambda_r\, H^{'}_{((r)(a))j}{|}_{\alpha_{(e)}=
 \theta^n=0},
 \end{eqnarray*}

\bea
 G^r_{(a)i} &\rightarrow_{\alpha_{(e)}, \theta^n \rightarrow 0}&
 G^{(o)r}_{(a)i} = {{\epsilon_{ari}}\over {\Lambda_r\,
 \Big({{\Lambda_r}\over {\Lambda_a}} - {{\Lambda_a}\over
 {\Lambda_r}}\Big)}},\nonumber \\
 &&{}\nonumber \\
 &&{}\nonumber \\
 \sum_{ar}\, H_{(a)rj}{|}_{\theta^n = 0}\,\,&&
 G^r_{(a)i}{|}_{\theta^n = 0} = \delta_{ij},\qquad
 \sum_{ri}\, H_{(b)ri}{|}_{\theta^n = 0}\,\,
 G^r_{(a)i}{|}_{\theta^n = 0}\, = \delta_{ab}.\nonumber \\
 &&{}
 \label{b16}
 \eea

\subsection{The Spin Connection in the York Basis.}

When $\alpha{(a)}(\tau ,\vec \sigma ) = 0$, the spin connection on
$\Sigma_{\tau}$ is given by (also its expression in the 3-orthogonal
gauges is given)

\begin{eqnarray*}
 {}^3{\bar \omega}_{r(a)}&=& {1\over 2}\, \sum_{bc}\, \epsilon_{(a)(b)(c)}\,
 {}^3{\bar \omega}_{r(b)(c)} =
 {1\over 2} \epsilon_{(a)(b)(c)} \sum_u\, \Big[{}^3{\bar e}^u_{(b)}
(\partial_r\, {}^3{\bar e}_{(c)u}-\partial_u\, {}^3{\bar e}_{(c)r})+\nonumber \\
&+&{1\over 2}\, \sum_v\, {}^3{\bar e}^u_{(b)}\, {}^3{\bar
e}^v_{(c)}\, {}^3{\bar e}_{(d)r}(\partial_v\, {}^3{\bar
e}_{(d)u}-\partial_u\, {}^3{\bar e}_{(d)v})\Big] =\nonumber \\
  &=& {1\over 2}\, \sum_{bcu}\, \epsilon_{(a)(b)(c)}\,
  V_{ub}(\theta^n)\,
   \Big[Q_c\, Q^{-1}_b\, \Big({1\over 3}\,
  [V_{uc}(\theta^n)\, \partial_r\, ln\, \tilde \phi - V_{rc}(\theta^n)\,
  \partial_u\, ln\, \tilde \phi ] +\nonumber \\
  &+& \sum_{\bar b}\, \gamma_{\bar bc}\, [V_{uc}(\theta^n)\,
  \partial_r\, R_{\bar b} - V_{rc}(\theta^n)\, \partial_u\,
  R_{\bar b}] + \partial_r\, V_{uc}(\theta^n) - \partial_u\,
  V_{rc}(\theta^n)\Big) +
  \end{eqnarray*}

\bea
  &+& {1\over 2}\, \sum_{vd}\, Q^2_d\, Q^{-1}_b\, Q^{-1}_c\,
  V_{vc}(\theta^n)\, V_{rd}(\theta^n)\nonumber \\
  && \Big({1\over 3}\,[V_{ud}(\theta^n)\, \partial_v\, ln\, \tilde \phi -
  V_{vd}(\theta^n)\, \partial_u\, ln\, \tilde \phi ] +\nonumber \\
  &+&  \sum_{\bar b}\, \gamma_{\bar bd}\, [V_{ud}(\theta^n)\,
  \partial_v\, R_{\bar b} - V_{vd}(\theta^n)\, \partial_u\,
  R_{\bar b}] + \partial_v\, V_{ud}(\theta^n) - \partial_u\,
  V_{vd}(\theta^n)\Big)\Big]\nonumber \\
  &&{}\nonumber \\
  &\rightarrow_{\theta^n \rightarrow 0}&  - \sum_b\,
  \epsilon_{rab}\, Q_r\, Q^{-1}_b\, \partial_b\,
 \Big({1\over 3}\, ln\, \tilde \phi + \sum_{\bar b}\,
 \gamma_{\bar br}\,  R_{\bar b}\Big).
 \label{b17}
 \eea

\section{The 3-Geometry in the 3-Orthogonal Gauges.}

As shown in Appendices B, C and D of Ref.\cite{25}, in the
3-orthogonal gauges we have the following expression for the
3-Christoffel symbols and the $\Gamma$-$\Gamma$ potential (in this
appendix we use $\phi = {\tilde \phi}^{1/6}$)

\begin{eqnarray*}
  {}^3\Gamma^r_{uv}
 &&\rightarrow_{\theta^n \rightarrow 0}\, \delta_{ru}\, (2\,
 \partial_v\, ln\, \phi + \sum_{\bar a}\, \gamma_{\bar ar}\,
 \partial_v\, R_{\bar a}) + \delta_{rv}\, (2\,
 \partial_u\, ln\, \phi + \sum_{\bar a}\, \gamma_{\bar ar}\,
 \partial_u\, R_{\bar a}) -\nonumber \\
 &-& \delta_{uv}\, (2\, \partial_r\, ln\, \phi + \sum_{\bar a}\,
 \gamma_{\bar ar}\, \partial_r\, R_{\bar a})\, Q_u\, Q^{-1}_r,\nonumber \\
 &&{}\nonumber \\
  \sum_v\, {}^3\Gamma^v_{uv}
  &&\rightarrow_{\theta^n \rightarrow 0}\, 6\, \partial_u\, ln\,
 \phi ,
 \end{eqnarray*}

 \bea
 {\cal S}   && \rightarrow_{\theta^n \rightarrow 0}\nonumber \\
  &&{}\nonumber \\
 &&\phi^2\, \sum_a\, Q^{- 2}_a\, \Big(20\, (\partial_a\, ln\, \phi
)^2 - 4\, \sum_r\, (\partial_r\, ln\, \phi )^2 + 8\, \partial_a\,
ln\, \phi\, \sum_{\bar b}\, \gamma_{\bar ba}\, \partial_a\,
R_{\bar b} -\nonumber \\
 &-& 2\, \sum_r\, \partial_r\, ln\, \phi\, \sum_{\bar b}\, (\gamma_{\bar ba}
 + \gamma_{\bar br})\, \partial_r\, R_{\bar b} +
 (\sum_{\bar b}\, \gamma_{\bar ba}\, \partial_a\, R_{\bar b})^2
 +\nonumber \\
 &+& \sum_{\bar b}\, (\partial_a\, R_{\bar b})^2 - \sum_r\,
 (\sum_{\bar b}\, \gamma_{\bar br}\, \partial_r\, R_{\bar b})\,
 (\sum_{\bar c}\, \gamma_{\bar ca}\, \partial_r\, R_{\bar
 c})\Big).
 \label{c1}
 \eea

\bigskip

From Eqs.(223) of Appendix A of  Ref.\cite{3}   for $\theta^n = 0$
we get \footnote{The conformal decomposition ${}^3g_{rs} = \phi^4\,
{}^3{\hat g}_{rs}$ implies (see Eqs. (189)-(190) of Ref.\cite{3})
 ${}^3\Gamma^u_{rs} = {}^3{\hat \Gamma}^u_{rs} + 2\, \Big(\delta_{ur}\,
 \partial_s\, ln\, \phi + \delta_{us}\, \partial_r\, ln\, \phi
 - \sum_{abv}\, V_{ra}(\theta^n)\, V_{sa}(\theta^n)\,
 V_{ub}(\theta^n)\, V_{vb}(\theta^n)\, Q^2_b\, Q^{-2}_a\,
 \partial_v\, ln\, \phi \Big)$ and ${}^3R[\theta^n, \phi ,R_{\bar a}]
 =  \phi^{-5}\, \Big[- 8\,  {\hat \triangle}\, \phi +
 {}^3{\hat R}\,\, \phi \Big]$, where ${}^3\hat R = {}^3\hat R[\theta^n, R_{\bar a}]$
and $\hat \triangle   = \partial_r\, ({}^3{\hat g}^{rs}\,
\partial_s)$ are the scalar curvature and the Laplace-Beltrami
operator associated with the 3-metric ${}^3{\hat g}_{rs}$,
respectively. $\hat \triangle - {1\over 8}\, {}^3\hat R$ is a
conformally invariant operator.}

\begin{eqnarray*}
 {}^3{\hat g}_{rs} &=& Q^2_r\, \delta_{rs},\quad
 det\, {}^3{\hat g}_{rs} = 1,\quad
\Rightarrow \sum_r \, {}^3{\hat \Gamma}^r_{rs}=0,\nonumber \\
 &&{}\nonumber \\
 {}^3R[\phi ,R_{\bar a}]&=& {}^3R[\theta^n = 0, \phi ,R_{\bar a}]
 = \phi^{-5}\, [- 8\, {\hat \triangle}[R_{\bar a}]\, \phi +
{}^3{\hat R}[R_{\bar a}] \phi ] =\nonumber \\
 &&{}\nonumber \\
 &=& - \sum_{uv}\, \Big( (2\,
\partial_v\, ln\, \phi + \sum_{\bar a}\, \gamma_{\bar
au}\, \partial_v\, R_{\bar a})\, (4\, \partial_v\, ln\, \phi -
\sum_{\bar b}\, \gamma
_{\bar bu}\, \partial_v\, R_{\bar b}) +\nonumber \\
 &+& \phi^{-4}\, Q^2_v\, [2\, \partial^2_v\, ln\, \phi + \sum_{\bar
a}\, \gamma_{\bar au}\, \partial_v^2\, R_{\bar a} +\nonumber \\
 &+& 2\, (2\, \partial_v\, ln\, \phi + \sum_{\bar a}\, \gamma
_{\bar au}\, \partial_v\, R_{\bar a})\, \sum_{\bar b}\,
(\gamma_{\bar bu} - \gamma_{\bar bv})\,
\partial_v\, R_{\bar b} -\nonumber \\
 &-& (2\, \partial_v\, ln\, \phi + \sum_{\bar
a}\, \gamma_{\bar av}\, \partial_v\, R_{\bar a})\, (2\,
\partial_v\, ln\, \phi + \sum_{\bar b}\, \gamma_{\bar
bu}\, \partial_v\, R_{\bar b})] \Big) +\nonumber \\
 &+& \phi^{-4}\, \sum_u\, Q^2_u\, [- 2\, \partial^2_u\, ln\, \phi + 2\,
\sum_{\bar a}\, \gamma_{\bar au}\, \partial^2_u\, R_{\bar a} +\nonumber \\
 &+& (2\, \partial_u\, ln\, \phi + \sum_{\bar
a}\, \gamma_{\bar au}\, \partial_u\, R_{\bar a})\, (2\,
\partial_u\, ln\, \phi - 2\, \sum_{\bar b}\, \gamma_{\bar
bu}\, \partial_u\, R_{\bar b})],
 \end{eqnarray*}

\bea
 {}^3\hat R[R_{\bar a}] &=& lim_{\phi\, \rightarrow 1}\,\,
 {}^3R[\phi ,R_{\bar a}] =
  \sum_u\, \Big(1 - 2\, Q^{- 2}_u\, \sum_{\bar b}\, (\partial_u\,
 R_{\bar b})^2\Big) +\nonumber \\
 &+& 2\, \sum_u\, Q^{- 2}_u\, \sum_{\bar a}\, \gamma_{\bar au}\,
 [\partial^2_u\, R_{\bar a} + \sum_{\bar b}\, \gamma_{\bar bu}\,
\partial_u\, R_{\bar a}\, \partial_u\, R_{\bar b}],\nonumber \\
&&{}\nonumber \\
\hat \triangle [R_{\bar a}] &=& \partial_r\, [{}^3{\hat g}^{rs}\,
\partial_s] = {}^3{\hat g}^{rs}\, {}^3{\hat \nabla}_r\,
{}^3{\hat \nabla}_s =
 \sum_r\, Q^{- 2}_r\, [\partial_r^2 - 2\, \sum_{\bar b}\,
\gamma_{\bar br}\, \partial_r\, R_{\bar b}\, \partial_r].
\nonumber \\
 &&{}
 \label{c2}
\end{eqnarray}

Let us remark that we have ${}^3R[\phi ,R_{\bar a} = 0] = - 24\,
\sum_u\, (\partial_u\, ln\, \phi )^2 - 8\, \phi^{-4}\, \sum_u\,
[\partial^2_u\, ln\, \phi - 2\, (\partial_u\, ln\, \phi )^2]\,
\rightarrow_{\phi\, \rightarrow 1}\, {}^3R[1,0] = 0$.

 \bigskip

The solution (\ref{III8}) of the super-momentum constraints becomes

\begin{eqnarray*}
 \pi^{(\theta )}_i(\tau ,\vec \sigma )
 &&\rightarrow_{\theta^n \rightarrow 0}
 \sum_{ab}\, \Big[\epsilon_{iab}\, Q_a\, Q^{-1}_b
\Big](\tau ,\vec \sigma )\times
 \end{eqnarray*}

 \bea
 &&\Big[ \sum_d\, \int d^3\sigma_1\, \bar{\cal G}_{((a)(b))(d)}(\vec{\sigma},
\vec{\sigma}_1;\tau)\, \Big[ {\tilde \phi}^{-1/3}\, Q^{-1}_d\, {\cal
M}_d -\nonumber \\
 &-&  \sum_{e}\, \bar{D}_{e(d)(e)}\, {\tilde \phi}^{-1/3} \,
Q^{-1}_e\, \Big(\tilde \phi\, \pi_{\tilde \phi} +  \sum_{\bar{b}}\,
\gamma_{\bar{b}{e}}\, \Pi_{\bar{b}}
\Big)\Big](\tau ,\vec{\sigma}_1) +\nonumber\\
 &&\nonumber\\
 &+& \sum_{ec}^{c\neq e}\, \int d^3\sigma_1\, \Big(
\delta_{c(a}\,\delta_{b)e}\, \delta^3(\vec{\sigma},\vec{\sigma_1})
+\nonumber \\
 &+& \sum_{d} \, \bar{\cal G}_{((a)(b))(d)}(\vec{\sigma},\vec{\sigma}_1;\tau)\,
 {1\over 2}\, \Big[\bar{D}_{c(d)(e)}\, {\tilde \phi}^{-1/3}\, Q^{-1}_c
 + \bar{D}_{e(d)(c)}\, {\tilde \phi}^{-1/3}\, Q^{-1}_e \Big](\tau
,\vec{\sigma}_1)\Big)\nonumber\\
 &&\nonumber\\
 &&\Big(- {\tilde g}_{ce}(\tau ,{\vec \sigma}_1) +
 \sum_{f}\,\int d^3\sigma_2\,\frac{1}{2}\, \Big[\,\Big(\,{\tilde \phi}^{1/3}\,
 Q_e\Big)(\tau ,\vec{\sigma_1})\,
\bar{\zeta}^e_{(c)(f)}(\vec{\sigma}_1,\vec{\sigma}_2;\tau)\,
+\nonumber\\
 &&\nonumber\\
 &+&\Big(\,{\tilde \phi}^{1/3}\, Q_c\Big)(\tau ,\vec{\sigma_1})\,
\bar{\zeta}^c_{(e)(f)}(\vec{\sigma}_1,\vec{\sigma}_2;\tau)\,\Big]\,\,
 \Big({\tilde \phi}^{-1/3}\, Q^{-1}_f\, {\cal M}_f\Big)(\tau
,\vec{\sigma}_2)\Big)\,\,\, \Big] =\nonumber \\
 &&{}\nonumber \\
 &{\buildrel {def}\over =}& \sum_{ab}\, \Big[\epsilon_{iab}\, Q_a\, Q^{-1}_b
\, F_{(ab)}\Big](\tau ,\vec \sigma ),
 \label{c3}
 \eea

\noindent where the last line defines the function $F_{(a)(b)}$.

\bigskip

The expression in the York basis of the extrinsic curvature
(\ref{II12}) (after having used the solution of the super-momentum
constraints) is given in Eq.(C8) of Appendix C of Ref.\cite{25}.
\medskip

\section{The Green Functions.}

The Green function ${\bar \zeta}^r_{(a)(b)}$ of Ref. \cite{3} in the
3-orthogonal gauges is

\begin{eqnarray*}
 {\bar \zeta}^r_{(a)(b)}(\vec \sigma ,{\vec \sigma}_1;\tau )
  &=& {\bar \zeta}^r_{(a)(b)}(\vec \sigma ,{\vec \sigma}_1;\tau |
  \theta^n, \phi , R_{\bar a}] =\nonumber \\
  &=& d^r_{\gamma_{PP_1}}(\vec
\sigma ,{\vec \sigma}_1)\, \Big( P_{\gamma_{PP_1}}\, e^{\int ^{\vec
\sigma}_{{\vec \sigma}_1}d\sigma_2^w\, {}^3{ \bar
\omega}_{w(c)}(\tau , {\vec \sigma}_2){\hat R}^{(c)} }\,
\Big)_{(a)(b)} =\nonumber \\
 &=&d^r_{\gamma_{PP_1}}(\vec \sigma ,{\vec \sigma}_1)\,
 \sum_{n=0}^{\infty}\, \int_0^1ds_n\, \cdots \nonumber \\
 &&\int_0^1 ds_1\,
 {{d \sigma_2^{i_n}(s_n)}\over {d s_n}}\, {}^3{\bar \omega}_{i_n(c_n)}(\tau
 ,{\vec \sigma}_2(s_n))\, \cdots\nonumber \\
 &&{{d \sigma_2^{i_1}(s_1)}\over {d s_1}}\, {}^3{\bar \omega}_{i_1(c_1)}(\tau
 ,{\vec \sigma}_2(s_1))\, \Big({\hat R}^{(c_n)}\, \cdots {\hat
 R}^{(c_1)}\Big)_{(a)(b)},
 \end{eqnarray*}

 \begin{eqnarray*}
 \Big(R^{(c)}\Big)_{(a)(b)} = \epsilon_{(a)(b)(c)},
 &&\qquad {\vec \sigma}_2(s)\,\, geodesics\,\, \gamma_{PP_1},\,\,
 {\vec \sigma}_2(s=0) = {\vec \sigma}_1,\,\, {\vec \sigma}_2(s=1)
 = \vec \sigma ,\nonumber \\
 &&{}\nonumber  \\
 \theta^n = R_{\bar a} = 0&& \rightarrow\,\,
d^r_{\gamma_{PP_1}}(\vec \sigma ,{\vec \sigma}_1){|}_{R_{\bar
a}=0} \nonumber \\
 &&\Big( P_{\gamma_{PP_1}}\, e^{2\int^{\vec
\sigma}_{{\vec \sigma}_1} d\sigma_2^w\,
\epsilon_{(c)(m)(n)}\delta_{(m)w}\sum_u\delta_{(n)u}
\partial_uln\, \phi (\tau , {\vec \sigma}_2){\hat R}^{(c)} }\,
\Big)_{(a)(b)},
 \end{eqnarray*}

\bea
 \theta^n = R_{\bar a} = 0, \phi =1&& \rightarrow\,\,
 \zeta^{(o)r}_{(a)(b)}(\vec \sigma ,{\vec \sigma}_1)
 = - \delta_{(a)(b)}\, c^r(\vec \sigma - {\vec
 \sigma}_1),\nonumber \\
  \sum_r\, \partial_r\, c^r(\vec \sigma ) = -
 \delta^3(\vec \sigma ),&&  c^r(\vec \sigma ) = - {{\sigma^r}\over
 {4\pi\, |\vec \sigma|^3}},\nonumber \\
 &&{}\nonumber \\
 &&{}\nonumber \\
 \sum_{rb}\, {\bar D}_{r(a)(b)}(\tau ,\vec \sigma )\, {\bar \zeta}^r
 _{(b)(c)}(\vec \sigma, {\vec \sigma}_1; \tau ) &=&
   - \delta_{(a)(c)}\, \delta^3(\vec \sigma , {\vec \sigma}_1 ),
 \label{d1}
 \eea

\noindent where $d^r$ is the Synge bitensor tangent to the geodesics
$\gamma_{PP_1}$, joining the point $\vec \sigma$ to a generic point
${\vec \sigma}_1$ on the same $\Sigma_{\tau}$. The Green function is
defined modulo solutions of the homogeneous equation $\sum_{rb}\,
{\bar D}_{r(a)(b)}(\tau ,\vec \sigma )\, {\bar
\zeta}^{(hom)r}_{(b)(c)}(\vec \sigma, {\vec \sigma}_1; \tau )  = 0$.

\bigskip

The Green function appearing in Eqs.(\ref{III5}) satisfies the
following equation

\bea
 && \sum_{r bc}^{b \not= c}\, \Big[{\bar D}_{r(a)((b)}\,\, {}^3{\bar
 e}^r_{(c))}\Big](\tau ,\vec \sigma )\, {\bar {\cal G}}_{((b)(c))(d)}
 (\vec \sigma ,{\vec \sigma}_1;\tau ) =
  {\tilde \phi}^{-1/3}(\tau ,\vec \sigma )\, \sum_{r bc}^{b \not= c}\,
 \Big[{\hat D}_{r(a)((b)} -\nonumber \\
 &-& 2\, \Big(\delta_{(a)((b)}\,
 \partial_r\, ln\, \phi + \sum_u\, ({}^3{\hat e}_{(a)r}\,
 {}^3{\hat e}^u_{((b)} - {}^3{\hat e}_{((b) r}\, {}^3{\hat e}^u_{(a)})\,
 \partial_u\, ln\, \phi\Big)
 \Big](\tau ,\vec \sigma )\, {}^3{\hat e}^r_{(c))}(\tau ,\vec \sigma )\nonumber \\
 && {\bar {\cal G}}_{((b)(c)) (d)}(\vec \sigma ,{\vec \sigma}_1;\tau )
 = - \delta_{ad}\, \delta^3(\vec \sigma ,{\vec
 \sigma}_1).
 \label{d2}
 \eea

Its explicit form is not yet known. However we know an inhomogeneous
solution $d_{((b)(c))(d)}(\vec \sigma ,{\vec \sigma}_1)$ (see
Eq.(E4) of Ref.\cite{25}) in the flat Minkowski limit, where
Eq.(\ref{d2}) becomes

\beq
 \sum_{bc}^{b\not= c}\, (\delta_{ab}\, \partial_c + \delta_{ac}\,
 \partial_b)\, d_{((b)(c))(d)}(\vec \sigma ,{\vec
 \sigma}_1) = - 2\, \delta_{ad}\, \delta^3(\vec \sigma ,{\vec
 \sigma}_1).
 \label{d3}
 \eeq
\medskip

The Green function of the modified covariant derivative operator
${\tilde D}_{rij}$ of Eq,(\ref{V5}) is

\bea
 &&\sum_{rk}\, {\tilde D}_{rik}(\tau ,\vec \sigma )\,
 {\tilde \zeta}^r_{kj}(\vec \sigma ,{\vec \sigma}_1;\tau )
 = \delta_{ij}\, \delta^3(\vec \sigma ,{\vec \sigma}_1),\nonumber \\
 {\tilde \zeta}^r_{ij}(\vec \sigma ,{\vec \sigma}_1;\tau )
 &=& d^r_{\gamma_{PP_1}}(\vec \sigma ,{\vec \sigma}_1)\,
 \sum_{n=0}^{\infty}\, \int_0^1ds_n\, \int_0^1 ds_{n-1}\cdots \int_0^1 ds_1\,
 {{d \sigma_2^{r_n}(s_n)}\over {d s_n}}\, T_{r_nij_n}(\tau
 ,{\vec \sigma}_2(s_n))\nonumber \\
 &&{{d \sigma_2^{r_{n-1}}(s_{n-1})}\over {d s_{n-1}}}\, T_{r_{n-1}j_nj_{n-1}}(\tau
 ,{\vec \sigma}_2(s_n))\cdots\,
 {{d \sigma_2^{r_1}(s_1)}\over {d s_1}}\, T_{r_1j_1j}(\tau
 ,{\vec \sigma}_2(s_1)),
 \label{d4}
 \eea

\vfill\eject

\end{document}